\documentclass[10pt]{article}

\usepackage[utf8]{inputenc}
\usepackage[dvipsnames]{xcolor}
\usepackage{tikz}
\usepackage{standalone}
\usepackage{amsmath,mathtools}
\usepackage{amsfonts}
\usepackage{amssymb}
\usepackage{lscape}
\usepackage{afterpage}
\usepackage{setspace}
\usepackage{multicol}
\usepackage{changepage}
\usepackage{simplewick}
\usepackage{youngtab}
\usepackage{pgfmath}
\usepackage{pdflscape}
\usepackage{tensor}
\usepackage{blindtext}
\usepackage{comment}
\usepackage{jheppub}
\usepackage{mathbbol}
\usepackage{subfiles}

\allowdisplaybreaks

\usepackage[backgroundcolor=white!95!red, linecolor=purple, bordercolor=purple, textsize=footnotesize]{todonotes}

\title{Branes, fermions, and superspace dualities}

\author[a]{Ander Retolaza,}
\author[b]{Jamie Rogers,}
\author[b]{Radu Tatar,}
\author[b]{Flavio Tonioni}

\affiliation[a]{Institut de Physique Th\'eorique, Universit\'e Paris Saclay, CEA, CNRS, Orme des Merisiers, 91191 Gif-sur-Yvette CEDEX, France.}
\affiliation[b]{Department of Mathematical Sciences, University of Liverpool, Liverpool, L69 7ZL, United Kingdom}

\emailAdd{anderreto@gmail.com}
\emailAdd{jamie.rogers@liverpool.ac.uk}
\emailAdd{rtatar@liverpool.ac.uk}
\emailAdd{flavio.tonioni@liverpool.ac.uk}

\abstract{We use the superspace formulation of supergravity in eleven and ten dimensions to compute fermion couplings on the M2-brane and on D$p$-branes. In this formulation fermionic couplings arise naturally from the $\theta$-expansion of the superfields from which the brane actions are constructed. The techniques we use and develop can in principle be applied to determine the fermionic couplings to general background fields up to arbitrary order. Starting with the superspace formulation of 11-dimensional supergravity, we use a geometric technique known as the `normal coordinate' method to obtain the $\theta$-expansion of the M2-brane action. 
We then present a method which allows us to translate the knowledge of fermionic couplings on the M2-brane to knowledge of such couplings on the D2-brane, and then to any D$p$-brane. This method is based on superspace generalizations of both the compactification taking 11-dimensional supergravity to type IIA supergravity and the T-duality rules connecting the type IIA and type IIB supergravities.

}

\arxivnumber{2106.02090}

\def\bs{\boldsymbol}

\def\cl{\mathcal}
\def\ul{\underline}
\def\sf{\mathsf}

\def\T[#1,#2]{\cl{T}_{#1,#2}}
\def\esix{e^{-\frac{\phi}{6}}}
\def\y{{\sf{y}}}
\def\T{\check{T}}

\def\ten{\indices}

\def\ulv{\clty} 
\def\epsilonconv{\varepsilon^{ijk}}
\def\clty{*}
\def\cmttr{K} 
\def\dual{\tilde}
\def\td{9} 
\def\flux{R} 
\def\F{F}
\def\ds{\mathbb}

\newcommand{\beq}{\begin{equation}}
\newcommand{\eeq}{\end{equation}}
\newcommand{\bea}{\begin{eqnarray}}
\newcommand{\eea}{\end{eqnarray}}

\newcommand{\de}{\mathrm{d}}
\newcommand{\der}{\partial}

\newcommand{\A}{\text{A}}
\newcommand{\B}{\text{B}}
\newcommand{\II}{\text{II}}
\newcommand{\vect}[2]{\left(
\begin{array}{c}
    #1 \\
    #2
\end{array}\right)}
\newcommand{\matr}[4]{\left( \begin{array}{cc}
    #1 \, & \, #2 \\
    #3 \, & \, #4
\end{array}\right)}

\begin{document}

\maketitle

\section{Introduction}
String Theory concerns itself not only with objects extended in a single dimension, the eponymous strings, but also with objects extended in many dimensions, namely \textit{branes}. These extended objects, as well as the quantum fields that live on them, are deeply consequential to modern String Theory research in both its most formal and most phenomenological aspects. Despite the ubiquity of branes in String Theory and the prominent position of fermions in physics, the fermionic fields living on branes are often less well understood than their bosonic counterparts due in no small part to their inherent technical complexities. Nevertheless, many phenomena in high-energy physics involve fermions, and in a large variety of string theoretic scenarios branes are crucial tools, therefore a detailed understanding of fermions on branes is of paramount importance.

Ever since the discovery that branes are objects intrinsic to string theories \cite{Polchinski:1995mt}, they have been extensively studied in a multitude of contexts. In type II theories, D-branes provide string theoretic realizations of gauge theories, supersymmetry breaking, and inflation, among others. In many of these studies their worldvolume fermions play central roles in the mechanisms under investigation. Of particular interest recently is the KKLT scenario \cite{Kachru:2003aw}, a proposal to generate de Sitter vacua in String Theory, where branes are crucial for multiple purposes.  The KKLT construction was originally described at an effective 4-dimensional level and so the viability of the proposal has now got to be scrutinized at the 10-dimensional level. This has been done from many perspectives (see e.g. \cite{Koerber:2007xk, Koerber:2008sx, Bena:2009xk, Baumann:2010sx, Heidenreich:2010ad, Dymarsky:2010mf, Blaback:2011pn, Bena:2014jaa, Bena:2016fqp, Sethi:2017phn, Moritz:2017xto, Bena:2018fqc, Randall:2019ent, Blumenhagen:2019qcg, Bena:2019mte, Demirtas:2019sip, Grana:2020hyu, Gao:2020xqh, Carta:2021lqg}). Initially, KKLT-related works considering fermions on branes focused on counting zero modes of brane instantons (see e.g. \cite{Kallosh:2005gs,Tripathy:2005hv,Kallosh:2005yu}). More recently new developments in this sector have lead to an interest in higher order fermion terms on brane actions \cite{Gautason:2018gln, Hamada:2018qef, Kallosh:2019oxv, Hamada:2019ack,Gautason:2019jwq, Carta:2019rhx, Kachru:2019dvo, Hamada:2021ryq}, bringing to this context open questions first posed by Ho\v{r}ava and Witten  \cite{Horava:1995qa, Horava:1996ma,Horava:1996vs}. In the well-understood case of non-localized gauginos, supersymmetry gives rise to a `perfect square' structure in the action \cite{Dine:1985rz}, and it is not currently known how this structure extends to the case of localized gauginos. Shedding light on these terms has been one of the main motivations that has led us to study higher-order fermionic couplings in D$p$-brane actions. Another feature that makes branes extremely promising tools for model building resides in the fact that they break half of the bulk supersymmetries (this was first observed in \cite{Hughes:1986dn, Hughes:1986fa}). Supersymmetry breaking is still not completely understood in String Theory proposals, but D$p$-branes are good candidates to provide ways to achieve it without spoiling the solution to the Hierarchy Problem since their fermionic degrees of freedom can realize supersymmetry non-linearly \cite{Hughes:1986dn, Hughes:1986fa,Kallosh:2014wsa, Kallosh:2016aep, Cribiori:2020bgt}.   This is a key reason for devoting our interest to the topic from a very generic point of view. In \cite{deWit:1998tk, Grana:2002tu, Marolf:2003ye, Marolf:2003vf, Martucci:2005rb, Kallosh:2005yu}, the worldvolume action of D$p$- and M$p$-branes in an arbitrary bosonic background has been determined up to quadratic terms in fermions. Our aim is to understand more deeply the mathematical structure underlying the action of a D$p$-brane, independently of the fermionic order of interest, and to set the stage for a concrete determination of the order-4 fermionic terms in the imminent future. A fundamental feature will be the structure inherited by the D$p$-branes from the more fundamental underlying theory, the M2-brane theory, as part of the web of string dualities. It would also be possible to inherit the structure from the M5-brane action, but the simplicity of the M2-brane action makes this choice more  practical.  \\

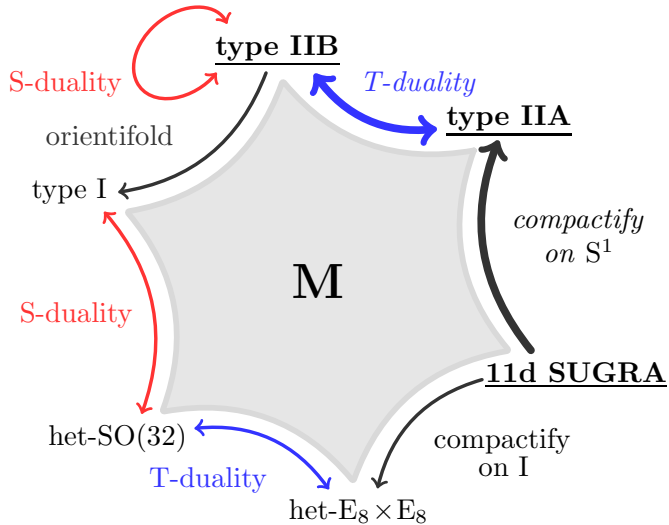
\begin{figure}  \begin{center}       \begin{tikzpicture}
\def\fullscale{1.25} 
\def\linwidth{2pt} 
\def\lincol{gray!30} 
\def\fillcol{gray!20} 
\def\arrowwidth{1.2pt} 
\def\arrowwidthemph{2.7pt} 
\def\compcol{black!80} 
\def\sdualcol{red!80} 
\def\tdualcol{blue!80} 

\def\fullrotate{400}
\begin{scope}[rotate=\fullrotate, scale=\fullscale]
  \draw[line width=\linwidth, \lincol, rounded corners = 1pt, cap=round, fill = \fillcol] (0:2) to[bend left] (60:2) to[bend left] (120:2) to[bend left] (180:2) to[bend left] (240:2) to[bend left] (300:2) to[bend left] (0:2) ;

\draw[->, line width=\arrowwidthemph, \compcol, rounded corners = 1pt, cap=round] (302:2.2) to[bend left]  (-2:2.25);
\draw[->, line width=\arrowwidth, \compcol, rounded corners = 1pt, cap=round] (289:1.88) to[bend right] (246:2.12);
\draw[<->, line width=\arrowwidth, \tdualcol, rounded corners = 1pt, cap=round] (233:2.05) to[bend right] (190:1.89);
\draw[<->, line width=\arrowwidth, \sdualcol, rounded corners = 1pt, cap=round] (177:2.17) to[bend right] (122:2.21);
\draw[<->, line width=\arrowwidthemph, \tdualcol, rounded corners = 1pt, cap=round] (51:2.04) to[bend right] (12:1.9);
\begin{scope}[xscale=1.4]
\draw[<->, line width=\arrowwidth, \sdualcol, rounded corners = 1pt, cap=round]
(75:2.60) arc (70-800:365-800:0.38);
\end{scope}
\draw[<-, line width=\arrowwidth, \compcol, rounded corners = 1pt, cap=round] (116:2.15) to[bend right] (64:2.11);

\draw 
(0,0) node {\LARGE\textbf{{M}}}
(0:2.45) node {\textbf{\underline{type IIA}}}
(60:2.3) node {\textbf{\underline{type IIB}}}
(120:2.6) node {{type I}}
(178:2.5) node {het-${\text{{SO}}(32)}$} 
(240:2.3) node {het-${\text{{E}}_8 \!\times\! \text{{E}}_8}$} 
(300:2.7) node {\textbf{\underline{{11d SUGRA}}}} 
(330:2.6) node[align=center] {\emph{compactify }\\ \emph{on} $\mathrm{S}^1$}
(279:2.4) node {compactify}
(275:2.58) node {on $\mathrm{I}$}
(201:2.23) node[\tdualcol] {T-duality} 
(23:2.18) node[\tdualcol] {\emph{T-duality}} 
(148:2.4) node[\sdualcol] {S-duality} 
(102:3.15) node[\sdualcol] {S-duality} 
(105:2.5) node[\compcol] {orientifold} 
;
\end{scope}
\end{tikzpicture}  \end{center}   \caption{A schematic of the web of dualities between the five 10-dimensional string theories and 11-dimensional supergravity (and M-theory). We will use the superspace generalization of this web to investigate the expansion in fermions of the superfields in different theories, and the expansion in fermions of the actions for the branes those theories contain. The parts of the web relevant for this work have been highlighted with thicker arrows. We begin with the superspace formulation of 11-dimensional supergravity. We find the expansion in fermions of the superfields therein, and use these to find the fermionic expansion of the M2-brane action. Compactification on $\mathrm{S}^1$ is then performed in order to obtain the fermionic expansion of the fields in type IIA, and of the D2-brane action. Finally, T-duality between type IIA and type IIB is used repeatedly to obtain expansions of the fields in type IIB, and so the expansions for D$p$-branes for all $p$.}  \label{fig:dualityweb}\end{figure}

It has been understood for quite some time that the five initially distinct-looking superstring theories are in fact limiting cases of a single fundamental theory, M-theory \cite{Townsend:1996xj}. The five string theories and M-theory are related to each other via a web of \textit{dualities} that we sketch in Fig. \ref{fig:dualityweb}.
In this work, we are going to concentrate on three of these related theories, the dualities which connect them, and the fermions on the branes that the theories contain. We will be investigating the M2-brane from M-theory and the D$p$-branes from the type IIA and type IIB superstring theories (more properly, we will be working with the low-energy supergravity limit of these theories, i.e. 11-dimensional supergravity, from M-theory, and type IIA and type IIB supergravities, from type IIA and type IIB string theories). Compactifying the  11-dimensional spacetime of  M-theory on a circle transforms the M2-brane into a D2-brane (when the circle is orthogonal to the brane), and then an arbitrary number of T-dualizations along directions wrapped by the brane, or orthogonal to it, allow us to investigate descriptions of any D$p$-brane. Our goal when it comes to these branes is to explore how to explicitly obtain the terms in the single-brane abelian actions corresponding to high-order couplings for the fermions. Of critical importance to us is the requirement that our methods are, at least in principle, applicable to arbitrary order in the fermions. As we will show, a central development consists of understanding how to dimensionally reduce and T-dualize the theories into each other in a manifestly supersymmetric way, by working in superspace.

We now outline the core details of the strategy that we follow in this work. Due to the existence of the string duality web, if we have a method for obtaining the high-order fermion couplings in one theory it can in principle be extended to the others.  We start with 11-dimensional supergravity which has a particularly simple formulation in superspace, wherein the usual dimensions of spacetime are augmented with anticommuting dimensions with Grassman-valued coordinates. In this formulation the usual fields are combined into superfields which contain both bosonic and fermionic degrees of freedom. What we then require is a way of systematically extracting information about the fermionic degrees of freedom from the superfield formulation. The technique used to do this in a complete way is called the `normal coordinate' method,\footnote{We shall see in section \ref{sec:norcor} that for our purposes this name is an anachronism and that for the physics we investigate we do not require the specific use of a normal coordinate system.} first developed in superspace in \cite{McArthur:1983fm}, and often simply referred to as NORCOR. The question of determining fermionic couplings is turned into a question of differential geometry in superspace in a way that is both elegant and powerful. In \cite{Grisaru:2000ij}, many of the results necessary for finding the expansion of the M2-brane action to fourth order in fermions were developed, this is also the order up to which we will expand in the examples which accompany our analysis. The NORCOR method can be applied to determine expansions of the superfields of 11-dimensional supergravity at all orders in fermions \cite{Tsimpis:2004gq}.  Nevertheless, we  show that the usefulness of the approach can be limited because the size of the formulae grows quickly as one computes terms of higher $\theta$-order in the superfields. This is the main obstacle we   find in our computations, and it will bring us to the conclusion that   unless one succeeds in  combing terms obtained with NORCOR together into simple and manageable formulae, it remains extremely challenging to extract  information valuable for physics.

After setting up the problem in M-theory we are going to use the web of string dualities to carry the information about fermionic expansions to the type II theories. However, in order to use the above mentioned superspace formalism when considering  the web of dualities, we promote the duality procedures to superspace as well. This circumvents some of the difficulties in applying NORCOR directly to the type II theories by instead only requiring the explicit use of NORCOR in the relatively simple world of 11-dimensional supergravity. In this way a circle compactification will provide us with the superspace formulation of the D2-brane action and T-dualities will allow us to obtain the D$p$-brane actions in superspace for an arbitrary value of $p$, in both type IIA and type IIB string theories. We will take advantage of the T-duality rules for fermions \cite{Hassan:1999bv, Hassan:1999mm, Hassan:2000kr} and express them in a convenient formalism for our superspace approach, spinor doublet notation.

While our motivations are certainly braney in origin, the techniques we investigate and develop are far more broadly applicable. The actions of the single M2-brane and for single D$p$-branes are just some examples of composite superfields that can be built from the fundamental superfields of their respective theories, although, as we have discussed, even these abelian cases are particularly relevant and interesting. We will structure our discussion, therefore, to concentrate on obtaining the $\theta$-expansions of certain  superfields in each theory, and investigate how they can be combined in order to obtain the brane action expansions in separate examples.  \\

This paper is organized as follows. In section \ref{sec:braneback}, we review background information about branes which motivates the analysis of later sections. We concentrate on viewing branes as hypersurfaces in curved superspace, and the role of the Goldstone fermions arising from the broken supersymmetry caused by the presence of a brane. In section \ref{sec:norcor}, we review the use of the `normal coordinate' method to provide an expansion in orders of fermions starting with the superspace formulation of the fields of 11-dimensional supergravity. In section \ref{sec:M2}, we consider the application of the normal coordinate expansion to the superspace formulation of the M2-brane action and we obtain expansions to quartic order in fermions. In section \ref{sec:D2}, we investigate the superspace generalization of the dimensional reduction of fields in 11-dimensional supergravity to type IIA. We use this to determine the D2-brane action to quartic order in fermions. In section \ref{sec:Tdual}, we discuss the superspace generalization of the T-duality relation between fields in type IIA and type IIB string theories. We demonstrate how this can be used in principle to move from the action for the D2-brane at a given order in fermions, to that for any D$p$-brane at the same order, and give explicit examples at second order. We end in section \ref{conclusions} with a summary of our results, our conclusions, and a discussion of future lines of inquiry. Our discussion is complemented by several appendices. Appendix \ref{spingamsec} summarizes our spinor conventions. Appendix \ref{11dconvapp} reviews 11-dimensional supergravity. Appendix \ref{order-4vielbeinexpansions} contains details about quartic-order fermionic expansions in superspace. Appendix \ref{DimRedCatApp} contains a catalogue of useful identities for the dimensional reduction from eleven to ten dimensions. Appendix \ref{Tdualconventions} is reserved for a discussion of topics related to T-duality.

\subsection*{Notes on notation}
Throughout this work we perform a large number of steps on a large number of quantities. Making our full discussion as clear as possible by avoiding notational clashes therefore necessitates the use of a large range of notation. It is worth our time to take a moment to mention a few of the most consequential choices and changes we make in this regard. \\

\noindent \emph{Indices} \\
We are going to be working with many different sets of indices through this paper. We collect details about all of these index choices here for easy reference. For easy reading we will repeat our conventions in the context of the sections when appropriate. 

In sections \ref{sec:braneback}, \ref{sec:norcor} and \ref{sec:M2}, we will be working with $(11|32)$-dimensional superspace. Our superspace conventions are the following. Superspace coordinates are $Z^M=(x^m,\theta^\mu)$, where upper-case letters in the middle of the alphabet are used to denote superspace coordinates, lower-case Latin letters denote spacetime indices, $m=0,1,\dots,10$, and lower-case Greek letters stand for Grassmann indices, $\mu=1,\dots,32$. We will use Latin and Greek indices in the beginning of the alphabet to refer to tangent space directions as $A=(a,\alpha)$, with $a=\underline{0}, \underline{1},\dots,\underline{10}$ and $\alpha = \underline{1}, \dots, \underline{32}$. We will use lower-case Latin indices like $i, j ,k$ for worldvolume directions, because we work only with the M2-brane this means that $i=0,1,2$.

In section \ref{sec:D2}, we will perform dimensional reduction from $(11|32)$-superspace to $(10|32)$-superspace. All 11-dimensional spacetime or tangent spacetime indices will now receive hats such that $\hat{m}=0,1,\dots,10$ and $\hat{a}=\underline{0}, \underline{1},\dots,\underline{10}$ whereas the 10-dimensional indices will not receive hats so that ${m}=0,1,\dots,9$ and ${a}=\underline{0}, \underline{1},\dots,\underline{9}$. Under the dimensional reduction we perform the M2-brane gets taken to the D2-brane. Therefore in section \ref{sec:D2}, where only the D2-brane is discussed, we still have $i=0,1,2$. The Grassman indices will remain unchanged.

Finally, in section \ref{sec:Tdual} we perform T-duality on $(10|32)$-superspace. This involves singling out a direction to take as a circle, which we will take to be the direction $x^\td$. We will then maintain the convention that 10-dimensional indices will not receive hats so that ${m}=0,1,\dots,9$, and we shall use a dotted index if referring only to the directions transverse to the T-duality circle so that $\dot{m}=0,1,\dots,8$. We will also shift to using double spinor notation; however a detailed explanation of this change is given in the section itself. When dealing with D$p$-branes, T-duality maps the brane content of the type IIA theory and the brane content of the type IIB theory into one another, changing the dimensionality. As such, the worldvolume indices $k, l$ run over all the $p+1$ worldvolume directions, whereas indices $m',n'$ span the complementary transverse directions, with $p$ always being clear in context. If the brane wraps the T-dual direction, we will employ a dot-notation $\dot{k}, \dot{l}$ when referring to all the worldvolume directions other than the T-dual one. \\

\noindent \emph{Hats} \\
In sections \ref{sec:norcor} and \ref{sec:M2}, we will be working in eleven spacetime dimensions. Then, in section \ref{sec:D2}, we will be reducing to ten dimensions many of the quantities from previous sections, and we will also work with them in section \ref{sec:Tdual}. In order to distinguish 11-dimensional quantities from 10-dimensional ones when performing dimensional reduction in section \ref{sec:D2} we place hats on all 11-dimensional objects and indices. However, because our use of 11-dimensions is implicit in sections \ref{sec:norcor} and \ref{sec:M2}, and to avoid swamping the notation in those sections with hats, we do not use the convention of hatting 11-dimensional quantities until section \ref{sec:D2} itself. Similarly, in appendices \ref{spingamsec} and \ref{DimRedCatApp}, where we discuss both 11-dimensional and 10-dimensional quantities, we are sure to distinguish them from one another with the hatting convention, however in appendices \ref{11dconvapp} and \ref{order-4vielbeinexpansions} where everything is implicitly 11-dimensional, we   drop them.

\section{Branes, fermions, and superspace}\label{sec:braneback}
In this section we provide some general background information about both M2-branes and D$p$-branes. This will motivate our discussion in the coming sections. For concreteness we   mostly focus on the case of a single M2-brane, but the ideas apply in a similar way for D$p$-branes as well. The  ideas in this section also hold for  the   M5-brane and the Green-Schwarz string, but as we already mentioned we will restrict ourselves to the M2-brane and Dp-brane cases.

M2-branes and D$p$-branes are solitonic solutions of M-theory and type II supergravities, respectively. `Brane-only' solutions are characterized by the breaking of the 11- or 10-dimensional Poincaré symmetry group down to the Poincaré group on the directions spanned by the brane times the group of rotations in the transverse space, i.e. $\mathrm{ISO}(1,10) \rightarrow \mathrm{ISO}(1,2) \times \mathrm{SO}(8)$ for M2-branes and $\mathrm{ISO}(1,9) \rightarrow \mathrm{ISO}(1,p) \times \mathrm{SO}(9-p)$ for D$p$-branes. The Goldstone modes associated to the breaking of the Poincaré symmetry become bosonic degrees of freedom living on the brane worldvolume \cite{Low:2001bw}. In these cases, the brane solution also triggers a spontaneous breaking of half of the bulk supersymmetries and the associated fermionic Goldstone modes turn into fermionic degrees of freedom on the brane.

In this paper we are interested in the action describing these localized branes, with a particular interest in fermionic modes living on them and their couplings in the brane worldvolumes. For this purpose it is convenient to approach branes from a slightly different perspective, that of the superspace formulation of the supergravity theories. In this formulation, branes can be regarded as extended objects in curved superspace. This is the approach taken in \cite{Bergshoeff:1987cm,Bergshoeff:1987qx} to construct the action of the M2-brane: the M2-brane is a $(2+1)$-dimensional object in $(11|32)$-dimensional superspace and its action consists of a brane worldvolume term, coupling the brane to the background metric, and a Wess-Zumino term, coupling the brane to the background gauge sector. Denoting the coordinates that span the worldvolume as $\zeta^i$, with $i=0,1,2$, this action reads 
\begin{equation}\label{eq:m2-action}
    {S}_{\text{M}2} = - T_{\text{M}2} \int \de^3 \zeta \; \sqrt{- \mathrm{det} \, (P[G](Z))} + \mu_{\text{M}2} \int P[A](Z),
\end{equation}
where $T_{\text{M}2}$ is the M2-brane tension, $\mu_{\text{M}2} = T_{\text{M}2}$ is the brane charge, and $P[G](Z)$ and $P[A](Z)$ are the pullbacks of the 11-dimensional supermetric and three-form gauge potential onto the brane worldvolume respectively, with $Z^M$ representing the superspace coordinates. The pulled-back superfields are built out of components of the supervielbein $E\ten{_M^{A}}(Z)$ and the super-three-form $A_{ABC}(Z)$.

The above action is a superspace generalization of the standard bosonic action of the M2-brane, where all fields in the latter  are replaced by their superfield counterparts. A product of superfields is a superfield itself, so what we have above is the M2-brane action superfield. Of course, since all superfields depend on superspace Grassmann coordinates $\theta^\mu$,  so does the action, and both allow for finite expansions in $\theta$.  Concretely, because the superfields in the action are the supervielbein $E\ten{_M^A}(Z)$ and the super three-form gauge potential $A_{ABC}(Z)$, if one knows the $\theta$-expansion of these superfields, one can obtain the expansion of the action superfield.  Both 11-dimensional supergravity, and the type II supergravities in ten dimensions considered in this paper, have 32 supercharges and so the fermionic expansion of the superfields goes up to order 32 in Grassmann coordinates $\theta^\mu$. Note that although we are dealing with the brane action, and the presence of the brane leads to partial supersymmetry breaking, we construct the brane action using  off-shell superfields.

We mentioned before that, from the perspective of the bulk, the presence of the brane in the brane-only solutions  only preserves half of the supersymmetries. Let us consider the bulk supercharges that are preserved in this type of solutions separately from those that are spontaneously broken. The Goldstone modes associated to the latter are fermionic degrees of freedom localized on the brane, arising from the $\theta^\mu$-directions that the broken supercharges generate on the (off-shell) superfields. The other supercharges are not affected by the presence of the brane, and so the brane action must be invariant under the shifts they generate in the corresponding Grassmann directions. Combining these ideas together, we see that the superspace Grassmann coordinates   on the brane action superfield are lifted to localized fermions living on the brane $\theta^\mu (\zeta^i)$, with only half of them (the ones generated by spontaneously broken supercharges) being physical and the other half being associated to transformations that leave the action invariant. From the brane worldvolume perspective, when we lift the Grassmann coordinates $\theta^\mu$ to fermions living on the brane, because we use the bulk off-shell  superfields to write the action, we find that half of these fermions are physical whereas the other half are not physical and instead correspond to redundancies. The existence of these redundancies implies a fermionic gauge symmetry of the action, commonly known as $\kappa$-symmetry. In \cite{Bergshoeff:1987cm} it was shown that the action \eqref{eq:m2-action} is indeed invariant under $\kappa$-symmetry transformations. More comments about the interplay between bulk supersymmetry and $\kappa$-symmetry are in section \ref{sec:M2}.

These arguments provide a clear approach for obtaining the fermion couplings of the M2-brane action. One needs to obtain the $\theta$-expansion of the superfields involved, plug them into the action \eqref{eq:m2-action}, and then lift the Grassmann coordinates to fermionic fields on the brane $\theta^\mu(\zeta^i)$. We will follow this approach in order to obtain the M2-brane action at order $(\theta)^4$, and so obtain fermionic interaction terms up to quartic order. The approach can in principle be used to obtain the action at all orders in fermions. 

Note that we used the brane-only solution to illustrate how to obtain fermion couplings on the brane worldvolume, but our interest   includes much more general solutions with the only demand being that they include  branes.  Many    points made above change when moving from the brane-only solution to more general solutions with branes, for example some of the fermions  on the brane can be massive and  not correspond to the goldstinos of the solution (points of this kind can be found in e.g. \cite{Bandos:2006wb}). Crucial for our purposes,   the fermion couplings that are obtained in the superspace formulation  are completely general and do not restrict to couplings on the brane-only solution.  

In the above analysis we focused on the M2-brane case, but the same ideas can be extended to all other branes, and in particular to D$p$-branes in type II supergravities. Hence, in order to obtain the D$p$-brane action superfields, one `only' needs to know the superfields involved. Unfortunately, there is no known \textit{simple} approach to obtain the $\theta$-expansion of superfields that appear in any of the theories in which we are interested. The method we will use, based on a normal coordinate expansion, is systematic but has  limitations in its current form. While effective for the expansion of the M2-brane action, computing the expansion of all superfields using this method turns out \textit{not} to be the best strategy for all D$p$-branes, as we will explain in more detail later. In fact, our strategy will be to use the `normal coordinate' method to obtain the $\theta$-expansion of the M2-brane action superfield, and then pursue the results for D$p$-branes using the superspace generalization of the duality web in Fig. \ref{fig:dualityweb}.

\section{The `normal coordinate' method}\label{sec:norcor}

In section \ref{sec:braneback} we explained that in order to obtain a fermionic expansion of the M2-brane action one requires the $\theta$-expansions of the superfields involved. In this section we review a systematic approach to obtain these $\theta$-expansions. Later we will specialize and apply this approach to obtain the expansion of some superfields in 11-dimensional supergravity, but the approach discussed here is completely general.

Supergravity in eleven dimensions \cite{Cremmer} has a well-established formulation in  superspace\footnote{Appendix \ref{spingamsec} reports our spinor and $\Gamma$-matrix conventions. Appendix \ref{11dconvapp} provides notes on the supergravity constraints and Bianchi identities necessary to carry out the analysis in this work.} \cite{Cremmer2, Brink:1980az}. From this perspective, the $\theta$-expansion of the superfields is just a Taylor expansion describing the dependence of the superfields on the superspace Grassmann coordinates $\theta^\mu$. We will use this geometric interpretation in order to obtain the $\theta$-expansions we are after. This approach is known as the `normal coordinate method', or NORCOR, because the normal coordinate system was very useful for performing the Taylor expansion of fields in spacetime when the method was originally proposed. We will show, however, that the superspace analysis in which we are interested does not require any special coordinate system. The normal coordinate method is a variant of the background field method to obtain covariant expressions in Taylor expansions of fields. Relevant literature in the development and application of NORCOR is \cite{AlvarezGaume:1981hn, McArthur:1983fm, Mukhi:1985vy, Atick:1986jr, Grisaru:1988jt, Grisaru:1997ub, Grisaru:2000ij}. In particular, \cite{Grisaru:2000ij} proposed the use of this method to obtain the $\theta$-expansion of the M2-brane action. In this section we provide an intuitive and self-contained description of the method.

The purpose of the NORCOR approach is to obtain the value of a (super)field at a point  $z_1^M$ in (super)space by starting from the value of the (super)field, and its derivatives, at another point, $z_0^M$, which is close to $z_1^M$, with
\begin{equation}
    z_1^M = z_0^M + \Sigma^M.
\end{equation}
In other words, we obtain the value of the superfield at points in the proximity of a point $z_0^M$ by performing a Taylor expansion around $z_0^M$. This approach is useful when we have plenty of information about the value of the superfield and its derivatives at the \textit{origin} $z_0^M$, but the information available at $z_1^M$ is much more limited.

In our case, we want to Taylor-expand superfields in the Grassmann directions $\theta^\mu$: we will take the spacetime, i.e. the subspace $z_0^M=(x^m,\theta^\mu=0)$, to be the origin, and perform the expansion along a direction $\Sigma^M$ that is purely Grassmannian. So, let $S=S(Z)$ be any superfield, and let $z_0^M=(x^m,0)$ be the starting point. In order to determine the value $S(z_1^M)$, we demand that there exists an auto-parallel curve $Z^M(t)$ with parameter $t$ connecting $z_0^M$ and $z_1^M$, such that $Z^M(t=0)=z_0^M$ and $Z^M(t=1)=z_1^M$. The tangent vector of the curve is $\y^M (t) \equiv \de Z^M (t) / \de t$. This tangent vector obeys the auto-parallel equation
\begin{equation}
    \y^M(t) \nabla_M \y^A(t) = \dfrac{\de \y^A(t)}{\de t} + \y^M(t)\, \omega\ten{_{M}^{A}_{B}} \y^B(t) = 0,
\end{equation}
where $\y^A(t) = \y^M(t) E\ten{_M^A}(Z^N(t))$ is written with the tangent superspace index because the superspace covariant derivative $\nabla$ comes with a superconnection $\omega$ generalizing the spin-connection, but nothing analogously comparable to the affine connection. We are expanding along a purely Grassmannian direction, so we want the tangent vector at the origin to point in Grassmann directions, i.e. $\y^M(t=0) = (\y^m=0, \y^\mu)$.

Before proceeding, let us explain why our approach does not need the normal coordinate system. The point of the normal coordinate system is to simplify the auto-parallel equation at the origin. This is usually achieved because the (affine) connection vanishes there. In our case of interest, however, we can use local Lorentz transformations to set some components of the superconnection to vanish at the origin of Grassmann coordinates, i.e. $\omega\ten{_{\mu}^{A}_{B}} (\theta^\mu=0)=0$. So the auto-parallel equation simplifies at $\theta=0$ regardless of the coordinate system used because the connection term vanishes there. 

Moreover, $\smash{\y^M \y^N \partial E\ten{_M^{A}} / \partial Z^N = 0}$ at $\theta=0$,\footnote{This can be checked using the $\theta$-expansion of the supervielbein components $E\ten{_\mu^{A}}(Z)$ that can be obtained e.g. using the method described in this section. The expansion of these components is only useful at this point for our purposes, so we omit the derivation and just give the necessary formulas here. They are \begin{equation}
    E\ten{_\mu^{a}} = -\dfrac{i}{2} \theta^\nu \delta^\alpha_\nu(\Gamma^a)_{\alpha\beta}\delta^\beta_\mu + \mathcal{O}(\theta^2), \quad \quad E\ten{_\mu^{\alpha}} = \delta_\mu^\alpha + \mathcal{O}(\theta^2).
\end{equation}} and so the auto-parallel equation at the origin is simply $\smash{ \de \y^M / \de t =0}$. The solution we are looking for is $\smash{Z^M(t) = z_0^M + \y^M(t=0) \, t}$ , and it is a good approximation at the origin and its surroundings. The point $z_1^M=(x^m,\theta^\mu)$ is at $t=1$ on the curve, and this allows us to  effectively identify the Grassmann coordinate and the origin tangent vector as $\y^\mu(t=0) \leftrightarrow \theta^\mu$.

We are now ready to obtain the $\theta$-expansion of any superfield $S(z_0)$. To do so, we first use the curve above to compute the Taylor expansion with  respect to the parameter $t$ around the point at $t=0$, i.e.
\begin{equation}
    S( Z^M (t) ) \bigr|_{t=0}=\sum_{k=0}^\infty \dfrac{t^k}{k!} \left(\dfrac{\delta}{\delta t} \right)^k S ( Z^M (t=0) ).
\end{equation}
Computing variations in $t$ means comparing the superfield at the origin with the superfield after dragging it along the auto-parallel curve, so we can replace the $t$ variations with Lie derivatives, denoted $\cl{L}_{\y}$, along the tangent vector field $\y^M(t)$. Because we evaluate the derivatives at $t=0$ the vector $\y$ that appears in the Lie derivatives will also be evaluated at this point. From here on we simply write it as $\y$ and drop that it is evaluated at $t=0$, where it only has components in Grassmann directions. Finally, we are interested in obtaining the value of the superfield at the point $z_1^M$, where $t=1$. Putting these things together we find that
\begin{equation}
    S ( Z^M (t=1) ) \bigr|_{t=0}=\sum_{k=0}^\infty \dfrac{\left(\mathcal{L}_{\y}\right)^k}{k!}S ( Z^M (t=0) ) = ( e^{\mathcal{L}_{\y}} S ) \bigr|_{t=0}.
\end{equation}
This means that the $\theta$-expansion of any superfield in this approach is obtained by repeatedly acting with the Lie derivative. This is effectively the approach followed in \cite{AlvarezGaume:1981hn, McArthur:1983fm, Mukhi:1985vy, Atick:1986jr, Grisaru:1988jt, Grisaru:1997ub, Grisaru:2000ij}. It is interesting  to point out that we can write the expansion using the exponential of a differential operator, because this agrees with the fact that a product of superfields  is a superfield itself: if $S$ is a product of superfields, using the Leibniz rule and the exponential expansion one finds that there will be an exponential acting on each superfield involved in the product.

For applying the NORCOR procedure it is important to note that in superspace we have the superconnection (that generalizes the spin connection), and we defined a Lorentz covariant derivative, but we did not define the notion of an affine connection or a fully covariant derivative. For this reason we are often interested in writing superfields with Lorentz indices. Regular Lie derivatives acting on Lorentz tensors do not lead to Lorentz tensors. To fix this problem, we need to replace the regular Lie derivative by the Lie-Lorentz derivative (see e.g. \cite{Figueroa-OFarrill:1999klq,Ortin:2002qb} and the original reference \cite{Kosmann}). This is a Lorentz covariantization of the regular Lie derivative, wherein partial derivatives are replaced by their Lorentz-covariant counterparts, complemented with the inclusion of an extra term that gives an infinitesimal Lorentz transformation. The effect of this Lorentz transformation is to trivialize the effect on the holonomy group driven by the inclusion of spin-connection terms in the covariantization. For practical purposes we observe that in \eqref{eq:m2-action} there are no free Lorentz indices, so the extra terms demanded by the Lie-Lorentz derivative will cancel each other in the expansions of the objects we are interested in. For this reason we can (and will) safely ignore the presence of these extra terms. Physics provides an alternative (and, dare we say it, more intuitive) description of the same idea: the Lorentz-Lie derivative above is a combination of a supersymmetry transformation and a local Lorentz transformation, and we will ignore the latter because brane actions have no free Lorentz indices. The $\theta$-expansion is therefore obtained by repeatedly taking supersymmetry variations of the fields.

Note that we have turned a problem about worldvolume couplings on  branes  into a differential geometry problem in superspace, and there is a price to pay for it. If we wish to obtain the superfield expansion systematically using this technique we are also required to do some extra work. On the one hand, we need the value of the superfield at the origin of Grassmann coordinates $\theta=0$, and on the other hand we need to be able to manipulate the outcome of the repeated application of the Lie derivative to write the results in terms of familiar objects. This is substantially easier to accomplish when we focus on computing the expansion of the individual superfields appearing in the action, rather than trying to treat the full action superfield directly. We will use some examples to illustrate these points.

As a first example consider the expansion of the 11-dimensional supervielbein that appears in the M2-brane action. We will employ the conventions and the definitions of 11-dimensional supergravity that are reviewed in appendix \ref{11dconvapp}.   For the first term in the expansion one needs the Lie derivative
\begin{equation}\label{firstvielvar}
    \cl{L}_{\y} E\ten{_M^A} = \nabla_M\y^A + \y^C E\ten{_M^B} T\ten{_{BC}^A}.
\end{equation}
This formula is obtained via integration by parts, and involves   the (Lorentz) covariant exterior derivative, $\nabla \y^A = \de \y^A - \y^B E^C \omega\ten{_{BC}^A}$ (where $\y^A=\y^M E\ten{_M^A}$), and the (superspace) torsion tensor $T$, whose definition is given in  \eqref{torsiondef}. Note that we wrote the torsion tensor with all indices in tangent space by  introducing a supervielbein for convenience. Obtaining the order-$(\theta)^1$ term in the expansion requires evaluating this expression at $\theta=0$, which in turn requires knowledge of the superspace torsion tensor and the supervielbein evaluated on this subspace. We will shortly explain how to perform this evaluation. For now let us  point out that without the notion of e.g. the superspace torsion tensor, the Lie derivative would be meaningless, and this makes manifest the need for extra structure to obtain any useful information from this approach.

Let us provide some further formulae necessary to compute  higher order terms of the supervielbein expansion. In particular, we will need
\begin{align}
    & \cl{L}_{\y} G = \y^A\nabla_A G\label{genfirstvar}, \\
    & \cl{L}_{\y} \y^A = 0, \\
    & \cl{L}_{\y} (\nabla_M\y^A) = -\y^B E\ten{_M^C} \y^D R\ten{_{DCB}^A}\label{expansionRterm}.
\end{align}
The first formula indicates how the Lie derivative acts on any Lorentz tensor $G$. For the second formula we used the previous one together with the auto-parallel equation. The last formula is also obtained by using integration by parts and the auto-parallel equation, and $R$ there is the superspace Riemann tensor defined in \eqref{curvedef}. Again, we find the need of extra structure in order to make sense of certain Lie derivatives. It turns out that the four expressions provided are enough to obtain the   $\theta$-expansion of the supervielbein at any order. We perform computations up to order four in section \ref{sec:M2} and appendix \ref{framesuperform}.

Once the necessary Lie derivatives have been computed, the next step is to evaluate them at the reference point for the Taylor expansion, that we choose to be $\theta=0$. Again, we concentrate on the 11-dimensional supervielbein for concreteness. The first object to evaluate at this point is the supervielbein itself. We use local Lorentz transformations to fix the so-called Wess-Zumino (WZ) gauge,\footnote{We previously used local Lorentz transformations to set to zero the component $\omega\ten{_{\mu}^{AB}}$ of the superconnection at the origin. These two choices are compatible with one another (see e.g. section 5.6 of \cite{Gates:1983nr}).} i.e.
 \begin{equation}\label{eq:WZ-gauge}
        E\ten{_M^A}(\theta=0) = \begin{pmatrix} e\ten{_m^a}(x) & \psi_m^\alpha(x) \\ 0 & \delta_\mu^\alpha \end{pmatrix},
    \end{equation}
where $e\ten{_m^a}(x) $ is the 11-dimensional vielbein and $ \psi_m^\alpha(x)$ the 11-dimensional gravitino. For all other terms appearing in the derivatives, there are a few steps to follow. First we must decide which component of the superfield we are assessing by choosing which of the free indices we would like to be bosonic or Grassmann. Contractions over superspace indices involve both kind of indices, upon expansion we will often find that only one of these kinds contributes. This can be for a number of the reasons including: (1) The supervielbein in the WZ-gauge has some vanishing component. (2) The vector tangent to the auto-parallel curve at the origin is constrained to be $\y^M=(0,\y^\mu)$ for our particular expansion. Note that the WZ-gauge means that this is $\y^A=(0,\y^\alpha)$. (3) The tangent space structure means no mixing between bosonic and fermionic indices in the superconnection (and so also no mixing in the superspace Riemann tensor). This means $\omega\ten{_{\gamma}^d}=\omega\ten{_{d}^\gamma} = R\ten{_{AB\gamma}^d} =R\ten{_{ABd}^\gamma} = 0$. 

For the terms that survive all of these constraints, one needs to evaluate the superspace tensors involved and  write them in terms of spacetime fields. To do this we make use of the  supergravity  constraints   and superspace Bianchi identities. It turns out that many components of superspace tensors vanish (for example $T_{\alpha b}^{\ \ c}=0$) or are constant (for example $T_{\alpha \beta}^{\ \ c}$) all over superspace \cite{Brink:1980az}. Fixing the value of the latter is a matter of conventions. The value of all other components of these superspace tensors can be obtained from superspace Bianchi identities. A list of supergravity constraints can be found in (\ref{SUGRAONE} - \ref{SUGRAFOUR}) and a list of useful formulae derived from Bianchi identities is given in  (\ref{BIANCHIONE} - \ref{BIANCHITHREE}).

As a clarifying example, we evaluate some components of \eqref{firstvielvar}, at $\theta=0$, using the ideas above. In both cases we consider that the index $M$ will be restricted to spacetime, and we evaluate the cases where the tangent space index $A$ is spacetime and Grassmann separately. We obtain,
\begin{subequations}
\begin{align}
    & \cl{L}_{\y} E\ten{_m^a} = \nabla_m \y^a + \y^\gamma E\ten{_m^B} T\ten{_{B\gamma}^a} = \nabla_m \y^a +\y^\gamma E\ten{_m^\beta} T\ten{_{\beta\gamma}^a} ~ \stackrel{\theta=0}{=}  -i \y^\gamma (\Gamma^a)_{\beta\gamma} \psi^\beta_m,\\
    & \cl{L}_{\y} E\ten{_m^\alpha} = \nabla_m \y^\alpha + \y^\gamma E\ten{_m^B} T\ten{_{B\gamma}^\alpha} = \nabla_m \y^\alpha + \y^\gamma E\ten{_m^b} T\ten{_{b\gamma}^\alpha} ~ \stackrel{\theta=0}{=} \, \nabla_m \y^\alpha + \y^\gamma e\ten{_m^b} T\ten{_{b\gamma}^\alpha}.
\end{align}
\end{subequations}
In both cases we first fixed as many indices as possible to be either spacetime or Grassmann, leaving only the contraction of $B$ with both types involved, then we got rid of vanishing contributions by using $T\ten{_{b\gamma}^a}=T\ten{_{\beta\gamma}^\alpha}=0$. Finally we evaluated the surviving terms using the WZ-gauge for the vielbein \eqref{eq:WZ-gauge} and our convention for the constant torsion component $T\ten{_{\beta\gamma}^a}=-i(\Gamma^a)_{\beta\gamma}$. We left $T\ten{_{b\gamma}^\alpha}$ untouched here, but it is a simple combination of $\Gamma$-matrices and four-form flux, as shown in \eqref{TORSIONGOOD}.

Once the formulae for the Lie derivatives have been evaluated at the origin of Grassmann coordinates, and re-written as described above, one can write the superfield expansion. In order to do so one must replace the tangent vector $\y$ by the Grassmann coordinate $\theta$ (this happens when we evaluate the auto-parallel curve at $t=1$). For the components of the supervielbein in the above example this gives the expansions up to order $(\theta)^1$, i.e.
\begin{subequations}
\begin{align}
    & E\ten{_m^a} (Z) = e\ten{_m^a} (x) - i \bar{\theta} \Gamma^a  \psi_m (x) + \dots ~ , \\
    & E\ten{_m^\alpha} (Z) = \psi_m^\alpha(x)+ \big(D_m(x)\theta\big)^\alpha + \dots ~ . \label{supcovdevref}
\end{align}
\end{subequations}
In the first formula we wrote the fermion bilinear with the Dirac conjugate $\bar{\theta} = \theta^T C$, with $C$ being the charge conjugation matrix, see appendix \ref{spingamsec} for our conventions.
For the second formula, we noted that the torsion can be manipulated and combined with the covariant derivative into the \textit{supercovariant} derivative $D_m=\nabla_m +\check{T}_m$, where $\check{T}\ten{_m}$ is related to $T\ten{_m}$ by a transposition. An alert reader will notice that the first order terms in the expansion are (unsurprisingly) the expressions that appear in the supersymmetry variations of the vielbein and the gravitino.

The above method gives rise to a superfield expansion in terms of familiar objects. This is not the end of the story, however. The method relies on writing all contractions in terms of tangent space indices. This often requires including numerous supervielbeins, and these can result in a rapidly growing number of terms when one computes higher and higher order Lie derivatives of any superfield. Higher-order terms, written in terms of spacetime fields, therefore involve an increasing number of contributions. This can cause the expansion to become enormously cumbersome unless one finds a way to put contributions at each level together into more compact and tractable combinations. As a simple example, recall that in the supervielbein expansion we combined the covariant derivative of $\theta$ together with the term related to the torsion into the supercovariant derivative. At higher orders it becomes increasingly complicated to combine terms together into manageable expressions. This will be the primary cause of the limitations we find in our computations. We will make further comments about this when we can make more precise statements.

Finally, we will concentrate on determining expressions for the case where the background is bosonic. Practically speaking,   we do this by  turning to zero all the terms involving the 11-dimensional gravitino. This means that we also turn to zero all superspace tensors with an odd number of Grassmann indices, since they involve the gravitinos when written in terms of spacetime objects. This restriction causes many more terms in the expansions to vanish. The Lie derivative applied to a bosonic field (a superspace tensor with an even number of Grassmann indices) an odd number of times will always vanish in bosonic backgrounds, as will the expression for the Lie derivative applied to a fermionic field an even number of times. In order to study completely general backgrounds, one would simply not perform this step and maintain all the gravitino terms in the discussion as well.

Now that we have explained the approach, we are ready to spell out  why it is more convenient to only use NORCOR in eleven dimensions. In ten dimensions there are more fields and more superspace tensors involved. This means that one needs to work harder in order to obtain all the supergravity constraints and useful formulae from Bianchi identities in each theory, and of course applying them to re-write the Taylor expansions requires performing even more computations.  Moreover, the `simplicity' of 11-dimensional supergravity enables us to  more clearly capture the structure of the terms involved, and we will show later that this structure is, in a sense, `inherited' by the 10-dimensional theories. We will make this statement more precise later. Nevertheless, we already mentioned that even in this `more simple' theory we encounter difficulties when manipulating higher-order terms. Clearly this problem does not improve for 10-dimensional type II theories. Computing NORCOR expansions in eleven dimensions is substantially cleaner and allows us to make insights and extract information about structure more easily. It is a better strategy, then, to obtain all expansions in this theory and then obtain expansions in ten dimensions via the superspace duality web, as we describe below.

A final compelling  reason to use the method in eleven dimensions only is that higher order expansions of the M2-brane action can be obtained with essentially just the 11-dimensional supervielbein expansion, whereas in all other cases one must compute the expansions of  more fields. In order to explain what we mean by `essentially', we can consider the M2-brane action. We can first note that in the volume term of the M2-brane action we find the (super)metric, whose expansion follows directly from the supervielbein. For the Wess-Zumino term, what we find is a combination of the supervielbein and of the super-three-form gauge potential. If we compute the Lie derivative of this combination we find
\begin{equation}\label{eq:WZ-term}
  \cl{L}_{\y} \big(E\ten{_{[m}^A}  E\ten{_n^B}  E\ten{_{p]}^C} A_{ABC} \big)  =  E\ten{_{[m}^A}  E\ten{_n^B}  E\ten{_{p]}^C} \y^D H_{D ABC} 
\end{equation}
up to total derivatives. This formula is a consequence of how the flux field-strength superfield is defined, in \eqref{Hsuperdef}. We now apply supergravity constraints (\ref{Hsugraconst1} -  \ref{SUGRAFOUR}) which tell us that the only components of the field-strength superfield that are non-vanishing are $H_{abcd}$ and $H_{\alpha\beta ab}=i(\Gamma_{ab})_{\alpha\beta}$, which is constant. This has important consequences for expansions of the above combination, and the M2-brane action as a whole, namely
\begin{equation}
(\cl{L}_{\y})^n(\y^D H_{D ABC})=0, \qquad \text{for $n\geq 1$}. 
\end{equation}
Hence, if we apply more Lie derivatives on the  combined superfield appearing in the WZ-part of the M2-brane action, only the terms with Lie derivatives acting on the supervielbeins   survive. This means that knowledge of the supervielbein expansion is sufficient for computing the expansion of the whole M2-brane action. This is the final argument supporting our general strategy. 

For ease of use, we summarize the computational steps of the strategy here:
\begin{enumerate}
    \item Compute the derivatives in the superfield expansion superfield. In practice this means using (\ref{firstvielvar} - \ref{expansionRterm}).
    
    \item Evaluate the expressions at the origin, $\theta=0$.  
    
    \item Apply the relevant supergravity constraints from (\ref{SUGRAONE} - \ref{SUGRAFOUR}) and those arising as a consequence of superspace Bianchi Identities (\ref{TORSIONGOOD} - \ref{derivedBfinal}) in order to write formulae in terms of familiar fields. 
    
    \item Apply the constraints of the bosonic background if appropriate. 
\end{enumerate}

\section{The M2-brane action}\label{sec:M2}
The expansion of the M2-brane action up to order four in fermions was first performed in \cite{Grisaru:2000ij}. In this section, with the aid of appendices, we review and correct the main results; appendix \ref{11dconvapp} contains a review of the 11-dimensional supergravity conventions and appendix \ref{order-4vielbeinexpansions} discusses useful superfield expansions up to fermionic order four. Our conventions are described in appendices \ref{spingamsec} and \ref{DimRedCatApp}.  

Let the M2-brane worldvolume coordinates be defined as $\zeta^i$, with $i=0,1,2$. The superfield action for the M2-brane in terms of the superspace embedding coordinates $Z^M(\zeta) = (x^m(\zeta),\theta^\mu(\zeta))$ is given in \eqref{eq:m2-action} which can be written as  
\begin{equation}\label{superfieldaction}
    S_{\text{M}2}(Z) = -T_{\text{M}2} \int \de^3\zeta \, \biggl[ \sqrt{- \text{det} \, ({G}_{ij}(Z))} - \frac{1}{6}\epsilonconv{A}_{ijk}(Z) \biggr],
\end{equation}
where, using the pullback of the supervielbein
\begin{equation}
    \qquad \qquad E\ten{_i^A}(Z)= \dfrac{\partial Z^M}{\partial\zeta^i} E\ten{_M^A}(Z),
\end{equation}
we wrote the Dirac-Born-Infeld (DBI) term in terms of the pullback of the metric and the Wess-Zumino (WZ) term in terms of the three-form pullback, which respectively read
\begin{align}
    & G_{ij}(Z) = E\ten{_i^a}(Z) E\ten{_j^b} (Z) \, \eta_{ab}, \\
    & A_{ijk} (Z) = E\ten{_i^A}(Z) E\ten{_j^B}(Z) E\ten{_k^C}(Z) \, A_{ABC}(Z). \label{superWZpart}
\end{align}

We explained in section \ref{sec:braneback} that in order to obtain the $\theta$-expansion of the action we need to obtain the $\theta$-expansions of the superfields involved. For the M2-brane we also showed that, because of \eqref{eq:WZ-term}, the only superfield expansion we need is that of the supervielbein. Nevertheless, it is more practical to work with Lorentz-invariant objects, so  in what follows we will  compute the expansion of   the (super)metric and the (super)three-form, that appear in the brane action. Obtaining the action expansion from these is then simple. Working with these superfields is sufficient and is a convenient middle-ground between dealing with the full action and dealing with the numerous supervielbeins individually. For a large proportion of the coming sections we will compute the Lorentz-invariant superfield expansions. 

We start by applying the method to compute the metric superfield expansion. In order to write the brane action up to order four in fermions, we need to expand the supermetric to the same order. We write the necessary Lie derivatives acting on the supermetric in terms of Lie derivatives acting on the supervielbeins involved and take into account the fact that we consider a bosonic background, which means that several terms will actually vanish. With the understanding that everything outside of Lie derivatives is evaluated at the origin, the relevant relations are
\begin{subequations}\label{metricshiftsblah}
\begin{align}
    & {(\cl{L}_{\y})}^2 {G}_{mn} = 2 \, \bigl[ {(\cl{L}_{\y})}^2 {E}\ten{_{(m}^{{a}}} \bigr] {e}\ten{_{n)}^{{b}}}{\eta}_{{a}{b}}, \\
    & {(\cl{L}_{\y})}^4 {G}_{mn} = 2 \, \bigl[ {(\cl{L}_{\y})}^4 {E}\ten{_{{(m}}^{{a}}} \bigr] {e}\ten{_{{n)}}^{{b}}}{\eta}_{{a}{b}} + 6 \, \bigl[ {(\cl{L}_{\y})}^2 {E}\ten{_{{(m}}^{{a}}} \bigr] \bigl[{(\cl{L}_{\y})}^2{E}\ten{_{{n)}}^{{b}}} \bigr] {\eta}_{{a}{b}}.
\end{align}
\end{subequations}
For the WZ-term, the analysis is slightly more involved because one has both the supervielbein and the three-form in the combination $(E\ten{_M^{A}} E\ten{_N^{B}} E\ten{_P^{C}} A_{ABC})(Z)$. We saw in the discussion around \eqref{eq:WZ-term} how to deal with this combination, so here we simply use those ideas and then follow the same procedure as we did for the metric. The relevant relations up to fermionic order four in bosonic backgrounds are
\begin{subequations} \label{threeformshiftsblah}
\begin{align}
    & {(\cl{L}_{\y})}^2{A}_{mnp} = -3i \, {\y}^\alpha \bigl[ \cl{L}_{\y} {E}\ten{_{[m}^{\beta}} \bigr] {e}\ten{_{{n}}^{{c}}} {e}\ten{_{{p]}}^{{d}}}({\Gamma}_{{c}{d}})_{\beta\alpha}, \\
    & {(\cl{L}_{\y})}^4{A}_{mnp} = -3i \, {\y}^\alpha \bigl[ {(\cl{L}_{\y} )}^3 {E}\ten{_{{[m}}^{\beta}} \bigr] {e}\ten{_{{n}}^{{c}}} {e}\ten{_{{p]}}^{{d}}}({\Gamma}_{{c}{d}})_{\beta\alpha} - 18i \, {\y}^\alpha \bigl[ \cl{L}_{\y} {E}\ten{_{{[m}}^{\beta}} \bigr] \bigl[ (\cl{L}_{\y})^2 {E}\ten{_{{n}}^{{c}}} \bigr] {e}\ten{_{{p]}}^{{d}}}({\Gamma}_{{c}{d}})_{\beta\alpha}\label{threeformfourthorder}.
\end{align}
\end{subequations}
We see that we require different components of the supervielbein expansion for the metric and the three-form. Happily, using the supergravity constraints it can be shown that in bosonic backgrounds these components are related by the condition, \cite{Tsimpis:2004gq}, 
\begin{equation}\label{eq:mix_orders}
    {(\cl{L}_{\y} )}^{2l+2}{E}\ten{_{{m}}^{{a}}} = - i {\y}^\beta ({\Gamma}^{{a}})_{\beta\gamma} \bigl[ {(\cl{L}_{\y} )}^{2l+1} {E}\ten{_{{m}}^\gamma} \bigr],
\end{equation}
where $l$ is a natural number. Therefore, in order to obtain the action at order four in fermions, we only require two terms in the expansion of the supervielbein. These are
\begin{subequations}\label{rawresults}
\begin{align}
    & \cl{L}_{\y} {E}\ten{_{{m}}^\alpha} = ({D}_{{m}})\ten{^\alpha_\gamma}{\y}^\gamma, \\
    & (\cl{L}_{\y})^3 {E}\ten{_{{m}}^\alpha} = -{\y}^\beta {e}\ten{_{{m}}^{{c}}} {\y}^\delta {\y}^\epsilon {\nabla}_\epsilon ( {R}\ten{_{\delta {c} \beta}^\alpha}-{\nabla}_\delta {T}\ten{_{{c}\beta}^\alpha} )  -{\y}^\beta ({D}_{{m}}{\y}^\gamma){\y}^\delta {R}\ten{_{\delta\gamma\beta}^\alpha} -i({\bar{\y}} {\Gamma}^{{c}} {{D}}_{{m}} {\y}){\y}^\beta {T}\ten{_{{c}\beta}^\alpha}\ .\label{cubicvielbefore}
\end{align}
\end{subequations} 
Here the first equation involves the supercovariant derivative that was discussed around \eqref{supcovdevref}.  
The supercovariant derivative will turn out to be a very important operator for our purposes. In \eqref{cubicvielbefore} we have left the expression written in terms of superspace components of the torsion and curvature tensors. Manipulating this expression using superspace Bianchi identities in order to write it in terms of spacetime fields, though important for our purposes, is a computation that does not add any insight to the present discussion. For this reason we present the details of that analysis in appendix \ref{supevielrearrapp}. The outcome of our manipulations is the expression
\begin{equation}\label{massagedresultsone}
    (\cl{L}_{\y} )^3 E\ten{_m^\alpha}  = i (\Gamma^{bc}\y)^\alpha(\bar{\y} \cl{W}_{m bc} \y) + i (\cl{\T}\ten{_b^{dfgh}}\y)^\alpha(\bar{\y} \cl{H}\ten{^b_{mdfgh}}\y)
\end{equation}
where we have defined
\begin{subequations}\label{crazyobjects}
\begin{align}
    & \cl{H}\ten{^b_{mdfgh}} = \Gamma^b H_{dfgh}D_m - 6e\ten{_m^b} \Gamma_{df} 
    [D_g,D_h],
    \\
    & \cl{W}_{m bc} =   \cl{R}\ten{_{bc}}  D_m +  \frac{1}{8}\Gamma_m  
    [D_b,D_c]
    + \frac{1}{4} \Gamma_b  
    [D_m,D_c],
    \\
    & \cl{R}\ten{_{bc}} = \frac{1}{576} \Big( \Gamma_{bc} \Gamma^{dfgh} - 8 \delta_{[c}^{[d}\Gamma_{b]}\Gamma^{fgh]} - 12 \delta_{[c}^{[d}\delta_{b]}^f\Gamma^{gh]} \Big)H_{dfgh}, \\
    & \cl{\T}\ten{_c^{dfgh}} = \frac{1}{288} \bigl( \Gamma\ten{_c}\Gamma\ten{^{dfgh}} - 12 \delta_c^{[d}\Gamma^{fgh]} \bigr).
\end{align}
\end{subequations}
There are some important points that need to be made about these formulae. First of all, manipulations lead to some terms involving commutators of supercovariant derivatives. It can be seen in appendix \ref{supevielrearrapp} that these arise from the first term in \eqref{cubicvielbefore}. There are also terms involving a single supercovariant derivative and $H^{(4)}$-flux. These contributions are the outcome of manipulating the last two terms in \eqref{cubicvielbefore}. We have so far been unable to write these parts of the expressions strictly in terms of the supercovariant derivative. Note that this problem appears for the first time at order $(\theta)^4$ for the M2-brane in bosonic backgrounds, and was therefore not observed in the order-$(\theta)^2$ analysis carried out in \cite{Martucci:2003gc, Marolf:2003vf, Marolf:2003ye, Martucci:2005rb} where everything can be packaged up in a tidy and supercovariant way. The result \eqref{massagedresultsone} agrees with \cite{MarcoSerra}, but there are strong indications that these formulae should allow for further manipulation into a more compact expression where supercovariance is made manifest. We will see later that dealing with these complicated objects is the chief source of the difficulty limiting our computational ability when performing dimensional reduction of the M2-brane action to obtain the D2-brane action. We conclude this section by stating plainly that our manipulation of the higher order expansion of the supervielbein probably needs to be completed into a manifestly supercovariant formulation that we would expect to be more compact and more manageable than the one presented above.

\subsection{M2-brane at fermionic order two}
In this section we review the M2-brane action at order two in fermions. We will use this `simple' analysis for two main purposes. First, it is a warm-up exercise that nicely illustrates how to proceed at higher orders. Second, we will use it to make more precise the relation between $\kappa$-symmetry and bulk supersymmetry discussed in section \ref{sec:braneback}. 

Recall that we decided to perform expansions of Lorentz-invariant superfields in the action since obtaining the full action expansion from these is simple. We begin with the metric expansion. We ignore order-$(\theta)^1$ terms since they involve the gravitino and we are interested in bosonic backgrounds.   For the order-$(\theta)^2$ terms, we combine \eqref{metricshiftsblah}, \eqref{eq:mix_orders}, and \eqref{rawresults} to obtain $\smash{{(\cl{L}_{\y})}^2{G}_{mn} \stackrel{\theta=0}{=} -2i \bar{\y}\Gamma_{(m}D_{n)}\y}$. We can use this to write a truncation of the metric superfield which includes only the terms relevant for the brane action. We will use a boldface notation to refer to these truncated superfields. For the metric, the combination is
\begin{equation}\label{eq:metric-shift2}
    \bs{g}_{mn}\equiv  g_{mn}(x)-i\bar{\theta}\Gamma_{(m}D_{n)}\theta.
\end{equation}
The expansion of the three-form superfield can be similarly obtained. In fact, \eqref{threeformshiftsblah} and \eqref{rawresults} combine to give the order-2 correction in the combination $A_{mnp}(Z)$, whose truncated expansion   reads
\begin{equation}\label{eq:A3-shift2}
    \bs{A}_{mnp} \equiv   A_{mnp}(x) - \dfrac{3i}{2}\bar{\theta}\Gamma_{[mn}D_{p]}\theta.
\end{equation}
These combinations of bosonic and fermionic fields  first appeared in \cite{Martucci:2003gc,Marolf:2003ye} in what was called a `superfield-like form of the action', allowing one to write the order-$(\theta)^2$ expansion of the M2-brane action in a compact way. Similar combinations appearing in other D$p$-brane actions were found and these allowed these actions to be written in a compact way as well.

Our discussion makes it manifest that the appearance of the truncated superfields is not a mere trick valid only for the action up to this order, but rather a consequence of how the action superfield is built in the superspace formulation of  supergravity. This means it is valid at \textit{any} order in fermions. Therefore in what follows our goal is to  provide a systematic approach to compute truncated  superfields of this type appearing in all brane actions. For practical purposes we will often refer to the metric and three form  without specifying if we refer to the field, the superfield, or the truncated superfield, as this will always be clear from context.

We  are now ready to write the M2-brane action at  order $(\theta)^2$. Plugging the truncated superfields \eqref{eq:metric-shift2} and \eqref{eq:A3-shift2} into the action \eqref{eq:m2-action} and then Taylor-expanding up to order $(\theta)^2$, we get 
\begin{equation}\label{onemorelabel}
\begin{split}
    \bs{S}_{\text{M}2}^{(2)} & = - T_{\text{M}2} \int \de^3\zeta \, \biggl[ \sqrt{-\mathrm{det} \, (\bs{g})} - \frac{1}{6}\epsilonconv\bs{A}_{ijk} \biggr] \\
    & = - T_{\text{M}2} \int \de^3\zeta \, \biggl[ \sqrt{- \mathrm{det} \, (g)} - \frac{1}{6}\epsilonconv{A}_{{i}{j}{k}} \biggr] + i T_{\text{M}2} \int \de^3\zeta \, \sqrt{- \mathrm{det} \, (g)} \; \Bigl[ {\bar{\theta}} P_-^{(0)}{\Gamma}^{i}{D}_{i}{\theta} \Bigr].
\end{split}
\end{equation}
In the last line, we combined the order-$(\theta)^2$ terms together forming the so-called $\kappa$-symmetry projector at order $(\theta)^0$, i.e.
\begin{equation}
    P_-^{(0)}=\dfrac{1}{2} \bigl( 1-{\Gamma}_{\text{M}2}^{(0)} \bigr),
\end{equation}
where the $\Gamma$-matrix combination defining the operator is
\begin{equation} \label{eq:something}
    \Gamma_{\text{M}2}^{(0)} = \frac{\epsilonconv{\Gamma}_{ijk}}{6\sqrt{- \mathrm{det} \, (g)}}.
\end{equation}
This allows us to see explicitly the manifestation of $\kappa$-symmetry in the M2-brane action at fermionic order $(\theta)^2$. We comment on $\kappa$-symmetry in detail now.

\subsection[Supersymmetry and $\kappa$-symmetry]{Supersymmetry and \texorpdfstring{$\boldsymbol{\kappa}$}{kappa}-symmetry}
We are now in a position to make more precise comments about   bulk supersymmetry and $\kappa$-symmetry. As we already mentioned, it is worth taking two perspectives. First, from the bulk perspective, the brane-only solution spontaneously breaks half of the supersymmetries, while the other half are preserved on-shell. The corresponding goldstinos turn into the fermionic degrees of freedom on the brane, $\theta^\mu(\zeta)$. Alternatively, from the brane worldvolume perspective, we construct the brane action using off-shell superfields with all 32 Grassmann coordinates, therefore only half of them correspond to actual degrees of freedom on the brane while the rest are redundancies. This means that there must exist a fermionic gauge symmetry, known as $\kappa$-symmetry, that gets rid of these redundant directions. The presence of such a fermionic gauge symmetry in the M2-brane action \eqref{eq:m2-action} was shown in \cite{Bergshoeff:1987cm}. The $\kappa$-symmetry variations are
\begin{subequations}
\begin{align}
    & (\delta_\kappa Z^M) E\ten{_M^a}(Z) = 0, \\
    & (\delta_\kappa Z^M) E\ten{_M^\alpha} (Z) = (1 + \Gamma_{\text{M}2}(Z))\ten{^\alpha_\beta}\kappa^\beta,
\end{align} 
\end{subequations}
where the operator
\begin{equation}\label{superkappaproj}
    \Gamma_{\text{M}2}(Z) = \frac{\epsilonconv\Gamma_{ijk}(Z)}{6 \sqrt{-\det(G(Z))}}
\end{equation}
is a hermitian traceless matrix squaring as $(\Gamma_{\text{M}2}(Z))^2=1$. In the transformations, $\kappa$ is an arbitrary 32-component Majorana fermion in 11-dimensional spacetime. Note that these expressions are valid all over superspace. If we evaluate them at the origin of Grassmann coordinates, using the WZ-gauge \eqref{eq:WZ-gauge} for the supervielbein, these variations read
\begin{subequations}
\begin{align}
    & \delta_\kappa x^m=0, \\
    & (\delta_\kappa\theta^\mu ) \delta_\mu^\alpha = (1 + \Gamma_{\text{M}2}^{(0)})\ten{^\alpha_\beta}\kappa^\beta,
\end{align}
\end{subequations}
with the matrix $\Gamma_{\text{M}2}^{(0)}$ defined as in \eqref{eq:something}. Hence it is possible to use $\kappa$-symmetry transformations to project out half of the Grassmann coordinates $\theta^\mu$. We see that the appearance of the orthogonal projector $P_-^{(0)}$ in the M2-brane action at order $(\theta)^2$ is not a coincidence, but rather it is a consequence of $\kappa$-symmetry and what we did there was to write the action in such a way as to make this symmetry manifest.

Let us now derive some bulk supersymmetry properties. We start with  the M2-brane-only solution, where the brane spontaneously breaks half of the supersymmetries. Here we use  $\kappa$-symmetry to determine whether a supersymmetry is preserved by the brane or spontaneously broken, following \cite{Bergshoeff:1997kr, Simon:2011rw}. To make this point explicit, we need some of the symmetries of the M2-brane action (see e.g. \cite{Bergshoeff:1987qx}).   To start, recall that superfields  transform under global supersymmetry variations, and so does the brane action.  Off-shell, supersymmetry variations are  shifts in any Grassmann direction(s) $\theta^\mu$. On-shell, in a background where the brane is present, only some of those shifts leave the background invariant. We denote the variation generated by the surviving killing spinors in this background $\delta_\epsilon\theta=\epsilon_{\text{M2}} $.  The combination of surviving global supersymmetry and $\kappa$-symmetry leads to a total variation (at the origin of Grassmann coordinates in order to connect with the above discussion)
\begin{equation}\label{eq:lalalalalalalaaaa}
    \delta_{\epsilon, \kappa} \theta = \epsilon_{\text{M2}}  + (1 + \Gamma_{\text{M}2}^{(0)}) \kappa. 
\end{equation} 
In order to get rid of the fermionic redundancies on the brane, we write  the $\kappa$-symmetry gauge-fixing condition as  $\mathrm{P} \theta = 0$, where $\mathrm{P}$ is a projector independent of background fields. This implies that the physical fermions on the brane are such that $\theta = (1 - \mathrm{P}) \theta$. Once the gauge is fixed, in order to preserve it, it is necessary that $\delta_{\epsilon, \kappa}(\mathrm{P}  \theta)=\mathrm{P}\delta_{\epsilon, \kappa}  \theta=0$ holds, and so $\delta_{\epsilon, \kappa}  \theta=0$. The latter formula, together with \eqref{eq:lalalalalalalaaaa} implies that the surviving global supersymmetry transformations that are compatible with this fact must satisfy
\begin{equation}\label{eq:1111111111}
  \epsilon_{\text{M2}} = - (1 + \Gamma_{\text{M}2}^{(0)}) \, \kappa  
\end{equation}
on the brane locus. Using this relation, one easily finds that any surviving supersymmetry must satisfy $\smash{P_+^{(0)}}\epsilon_{\text{M2}}=\epsilon_{\text{M2}}$ (equivalently    $\Gamma_{\text{M}2}^{(0)}\epsilon_{\text{M2}}=\epsilon_{\text{M2}}$) on the brane locus, where $\smash{P_+^{(0)} = (1 + \Gamma_{\text{M}2}^{(0)})/2}$.  On the other hand, the  orthogonal projector  $\smash{P_-^{(0)}}$ selects Grassmann coordinates generated by spontaneously broken supercharges, the  goldstinos  on the brane-only solution. This is the reason why   the combination $\smash{P_-^{(0)}\theta}$ appears on the brane action. \eqref{eq:1111111111} also shows that preserved bulk supersymmetries are of the same aspect as $\kappa$-symmetry transformations (they both involve $\smash{P_+^{(0)}}$) and so also leave the M2-brane action invariant thanks to the presence of  $\smash{P_-^{(0)} }$ in the brane action.

This physical picture is valid not only for the M2-brane, but also for all D$p$-branes. In order to study each case one must replace $\Gamma_{\text{M}2}$ by the corresponding matrix $\Gamma_{\text{D}p}$. In \cite{Marolf:2003vf, Martucci:2005rb} it was shown that all D$p$-brane actions at quadratic fermionic order can be written with the corresponding $\kappa$-symmetry projector. In   \cite{Cribiori:2020bgt} the breaking of supersymmetry by D$p$-branes was shown to correspond to a non-linear realization of supersymmetry, generalizing first results of this type \cite{Hughes:1986dn,Hughes:1986fa}.

The brane-only solution is illuminating for deriving multiple facts regarding bulk supersymmetry and $\kappa$-symmetry, but our interest is in more general setups. In the previous configuration all fermions on the brane are massless goldstinos and many fermionic couplings on the brane vanish.  In general, those couplings do not vanish and are physically   relevant.  For example, depending on the particular solution, some (or all) worldvolume fermions will become massive and will no longer correspond to goldstinos of the solution.  The superspace approach in this paper includes all such couplings and therefore captures all of the relevant physical features of these general solutions. Moreover, the argument above, telling which supersymmetries survive in the solutions involving branes, is also valid for such solutions.

Finally, it is worth noting that we evaluated our expressions at the origin of Grassmann coordinates and so formulae involved the zeroth order $\kappa$-symmetry matrix $\Gamma_{\text{M}2}^{(0)}$ and the   projectors $\smash{P_\pm^{(0)}}$, that we used to connect with what we found for the brane action at order $(\theta)^2$. Nevertheless, the above arguments work all over superspace and so the general formulas about preserved supercharges and   $\kappa$-symmetry   involve $\Gamma_{\text{M}2}(Z)$ and $\smash{P_\pm}(Z)$.

\subsection{M2-brane at fermionic order four}
In this section we apply what we learned at the second fermionic order to build the action at order four in quite a direct way. We saw that in order to do so we need to find the metric and 3-form superfield truncations up to order $(\theta)^4$.

We already provided all of the relevant formulae to write the supervielbein expansion at order $(\theta)^4 $ in \eqref{eq:mix_orders} and \eqref{massagedresultsone}. By plugging those results into \eqref{metricshiftsblah} and \eqref{threeformshiftsblah}, one finds the metric and three-form superfields  up to order  $(\theta)^4 $. The metric is
\begin{equation}\label{quarticm2met}
\begin{split}
    \bs{g}_{mn} = g_{mn} & - i(\bar{\theta}\Gamma_{(m}D_{n)}\theta) - \frac{1}{4} (\bar{\theta}\Gamma_aD_{(m}\theta) (\bar{\theta}\Gamma^{a}D_{n)}\theta) \\
    & +  \frac{1}{12}(\bar{\theta}\Gamma_{(m|} \cl{\T}\ten{_b^{dfgh}}\theta)(\bar{\theta} \cl{H}\ten{^b_{|n)dfgh}}\theta) + \frac{1}{12} (\bar{\theta}\Gamma_{(m|}\Gamma^{bc}\theta)(\bar{\theta} \cl{W}_{|n) bc} \theta) ,
\end{split}
\end{equation}
where we used the operators defined in \eqref{crazyobjects}, and, similarly, the three-form is
\begin{equation}\label{quarticm2form}
\begin{split}
    \bs{A}_{mnp} = {A}_{mnp} & - \frac{3}{2}i ({\bar{\theta}} {\Gamma}_{[{m}{n}}{D}_{{p}]} {\theta}) - \frac{3}{4}({\bar{\theta}} {\Gamma}_{{a}[{m}}{D}_{{n}} {\theta})({\bar{\theta}} {\Gamma}^{{a}}{D}_{{p}]} {\theta}) \\
    & + \frac{1}{8}(\bar{\theta}\Gamma_{[mn|} \cl{\T}\ten{_b^{dfgh}}\theta)(\bar{\theta} \cl{H}\ten{^b_{|p]dfgh}}\theta)+  \frac{1}{8} (\bar{\theta}\Gamma_{[mn}\Gamma^{bc}\theta)(\bar{\theta} \cl{W}_{p] bc} \theta).
\end{split}
\end{equation}
In the same way as we did at second order, these expressions can be plugged into the action and then we can perform a Taylor expansion to find the action at quartic order 
\begin{equation}\label{Quarticaction}
\begin{split}
    \bs{S}_{\text{M2}}^{(4)} = \, & - T_{\text{M2}}\int \de^3 \zeta \, \bigg[ \sqrt{- \mathrm{det} \, (\bs{g})} - \frac{1}{6}\epsilonconv\bs{A}_{ijk}\bigg] \\
    = \, & - T_{\text{M2}} \, \int \de^3 \zeta \, \sqrt{- \mathrm{det} \, (g)} \; \biggl[ 1 - \frac{1}{6} \dfrac{\epsilonconv}{\sqrt{- \mathrm{det} \, (g)}} {A}_{{i}{j}{k}} \biggr] \\
    & + T_{\text{M2}} \, \int \de^3 \zeta \, \begin{aligned}[t] & \sqrt{- \mathrm{det} \, (g)} \;
    \biggl[ i \bigl( \bar{\theta} P_-^{(0)} \Gamma^{i} D_{i} \theta \bigr) \\
    & + \frac{1}{8} \bigl( \bar{\theta} \Gamma^i D_i \theta \bigr)^2 - \frac{1}{8} \bigl( \bar{\theta} \Gamma_{i} D_{j} \theta \bigr) \bigl( \bar{\theta} \Gamma^{i} D^{j} \theta \bigr) - \frac{1}{8} \bigl( \bar{\theta} \Gamma_{i} D_{j} \theta \bigr) \bigl( \bar{\theta} \Gamma^{j} D^{i} \theta \bigr) \\
    & + \frac{1}{8} \bigl( \bar{\theta} \Gamma_m D^{i} \theta \bigr) \bigl( \bar{\theta} \Gamma^{m} D_{i} \theta \bigr) - \frac{1}{8} \, \dfrac{\epsilonconv}{\sqrt{- \mathrm{det} \, (g)}} \, \bigl( {\bar{\theta}} {\Gamma}_{{m}[{i}}{D}_{{j}} {\theta} \bigr) \bigl( {\bar{\theta}} {\Gamma}^{{m}}{D}_{{k}]} {\theta} \bigr) \\
    & - \frac{1}{12} \bigl( \bar{\theta} P_-^{(0)} \Gamma^{i} \cl{\T}\ten{_b^{dfgh}} \theta \bigr) \bigl( \bar{\theta} \cl{H}\ten{^b_{i dfgh}} \theta \bigr) - \frac{1}{12} \bigl( \bar{\theta} P_-^{(0)} \Gamma^{i} \Gamma^{bc} \theta \bigr) \bigl( \bar{\theta} \cl{W}_{i bc} \theta \bigr) \biggr].
    \end{aligned}
\end{split}
\end{equation}
We see that some of the fourth-order terms, like the second-order terms, may be organized around zeroth-order $\kappa$-symmetry projectors, whereas some cannot be. Those terms which cannot be (coming with a factor of $1/8$) are related to the higher-order fermionic expansion of the  $\kappa$-symmetry projector superfield. We leave the study of this for future work, and for now continue on without organising these terms around a $\kappa$-symmetry principle.  

This completes the expansion of the bosonic background M2-brane action to quartic order. In the next section we will examine the dimensional reduction of these expansions to determine the D2-brane action up to order four in fermions.

\section{Superspace dimensional reduction and the D2-brane action}\label{sec:D2}
We now know how to obtain the fermion couplings in the M2-brane action up to arbitrary order, and we have calculated them explicitly up to order four. Our plan is to use this knowledge to compute equivalent couplings on all D$p$-branes. The first step in doing this is compactifying M-theory on a circle, connecting the M2-brane in 11-dimensional supergravity to the D2-brane in type IIA supergravity. Then, by T-dualizing the theory, move to branes of arbitrary dimension in both type IIA and IIB theories.

D$p$-branes are solutions of 10-dimensional type II supergravities and it is therefore possible to construct their action using the superspace formulation of those supergravity theories. This is indeed   what we will do in this section and the next one. However, as we previously explained, the approach we will use to obtain the D$p$-brane action superfields will not be a direct application of the NORCOR approach of section \ref{sec:M2}. Instead, in this section we use a superspace generalization of the dimensional reduction relating M2-branes and D2-branes. We start by quickly reviewing the $\mathrm{S}^1$-compactification of the 11-dimensional spacetime that reproduces type IIA supergravity starting from 11-dimensional supergravity. We then consider the M2-brane and its dimensional reduction to the D2-brane. After revisiting the purely bosonic calculation,  we then extend the compactification method to superspace.  

For a detailed account of the notation employed, see appendix \ref{spingamsec}. See appendix \ref{DimRedCatApp} for an overview of the relevant dimensional reductions.

\subsection{Reduction of 11-dimensional supergravity to type IIA supergravity}

Type IIA string theory can be obtained by dimensional reduction of 11-dimensional supergravity. In this subsection we quickly review the main features of this dimensional reduction.

The notation for the dimensional reduction is as follows: 11-dimensional indices are hatted whereas 10-dimensional indices are not and 11-dimensional objects are also hatted whereas 10-dimensional objects are not; indices $a,b,...$ are tangent space and $m,n,...$ are spacetime indices, with explicit number indices underlined for tangent space and unadorned for spacetime, while $i,j,...$ are M2- and D2-brane worldvolume indices; 11-dimensional spacetime coordinates are  $\hat{x}^{\hat{m}}$ and they split as $(x^m,x^{10})$, while worldvolume coordinates are $\zeta^i$; we will leave implicit that the pull-back to the brane of an 11-dimensional object is a different operation than the pull-back to the brane of a 10-dimensional object, but we will keep track of this by observing whether the object is hatted or not, objects always being pulled back in the appropriate way. Background fields are independent of $x^{10}$.

To begin the dimensional reduction, we first deal with bosonic fields. Given the 11-dimensional metric $\smash{\hat{g}_{\hat{m} \hat{n}} = \hat{e}\ten{_{\hat{m}}^{\hat{a}}} \hat{e}\ten{_{\hat{n}}^{\hat{b}}} \hat{\eta}_{\hat{a} \hat{b}}}$, where $\smash{\hat{e}\ten{_{\hat{m}}^{\hat{a}}}}$ is the 11-dimensional vielbein, the $\mathrm{S}^1$-compactification ansatz for the vielbein leading to the type IIA action in the string frame is
\begin{equation}\label{xx1}
    \hat{e}\ten{_{\hat{m}}^{\hat{a}}} = \matr{e^{-\frac{\phi}{3}}e\ten{_m^a}}{e^{\frac{2\phi}{3}} C_m}{0}{e^{\frac{2\phi}{3}}},
\end{equation}
where $e\ten{_m^a}$ is the 10-dimensional vielbein, $\phi$ is the dilaton, and $C_m$ is the Ramond-Ramond one-form. The 10-dimensional metric is $g_{mn} = e\ten{_m^a} e\ten{_n^b} \, \eta_{ab}$. The 11-dimensional three-form gauge potential decomposes as
\begin{subequations} \label{xx2}
    \begin{align}
        & \hat{A}_{mnp} = C_{mnp}, \\
        & \hat{A}_{mn \,10} = B_{mn},
    \end{align}
\end{subequations}
where $C_3$ is the Ramond-Ramond three-form potential and $B_2$ is the Kalb-Ramond potential. Notice that our RR-field sign conventions differ from those used in \cite{Marolf:2003ye,Marolf:2003vf}. There are many objects for which we need the dimensional reduction. Those calculations are crucial, but laborious, so we provide a catalogue of the dimensional reduction results in appendix \ref{DimRedCatApp}.

Fermions are of course highly relevant for our purposes and so we   need many details from the dimensional reduction of fermionic fields. The ansatz for the 11-dimensional gravitino $\hat{\psi}_{\hat{m}}$ is
\begin{subequations}
    \begin{align}
        & \hat{\psi}_m = e^{-\phi/6} \, \biggl[ \psi_m - \dfrac{1}{6} \Gamma_m \lambda + \dfrac{1}{3} e^\phi C_m \Gamma^{\ulv} \lambda \biggr], \\
        & \hat{\psi}_{10} =  e^{-\phi/6} \, \biggl[\frac{1}{3}e^{\phi}\Gamma^{\ulv} \lambda \biggr],
    \end{align}
\end{subequations}
where ${\psi}_m$ is the 10-dimensional gravitino, $\lambda$ is the dilatino, and $\Gamma^{\ulv}$ is the 10-dimensonal chirality matrix. Recall that we start with 11-dimensional Majorana fermions. Upon dimensional reduction, these will split into pairs of 10-dimensional Majorana-Weyl fermions of opposite chiralities, so each 10-dimensional fermion above should be interpreted as a pair of Majorana-Weyl fermions of opposite chirality, e.g. $\lambda = \lambda_+ + \lambda_-$, where $\Gamma^{\ulv} \lambda_\pm = \pm \lambda_\pm$. This dimensional reduction leads to the type IIA action in the fermionic frame of \cite{Bergshoeff:2001pv}. Moreover, any 11-dimensional Majorana fermion, like the supersymmetry parameter or the fermions on the M2-brane, need to be dimensionally reduced like the gravitino, with a rescaling involving the dilaton, and further need splitting into pairs of 10-dimensional Majorana-Weyl fermions, so
\begin{equation}
    \hat{\theta} = e^{-\phi/6} \, \theta, \qquad \theta = \theta_+ + \theta_-.
\end{equation}

Next, we are interested in the type IIA gravitino and dilatino supersymmetry variations arising in the resulting 10-dimensional action. In 11-dimensional supergravity, the gravitino supersymmetry variation reads
\begin{equation}
    \delta_{\hat{\epsilon}} \hat{\psi}_{\hat{m}} = \hat{D}_{\hat{m}} \hat{\epsilon}.
\end{equation}
In the type IIA theory, the supersymmetry variations of fermionic fields are
\begin{subequations}
    \begin{align}
         \delta_\epsilon \psi_m & = D_m \epsilon, \\
         \delta_\epsilon \lambda & = \Delta \epsilon,
    \end{align}
\end{subequations}
with the 10-dimensional supercovariant derivative $D_m$ and the operator $\Delta$ being defined as,
\begin{subequations}\label{eq:IIA-operators}
    \begin{align}
        & D_m = \nabla_m + \frac{1}{4} \ul{H}_m^{(3)} \Gamma^{\ulv} - \frac{1}{8}e^\phi \bigl( \ul{F}^{(2)} \Gamma^{\ulv} +  \ul{F}^{(4)} \bigr) \Gamma_m, \\
        & \Delta =  \ul{\partial}\phi + \frac{1}{2}\ul{H}^{(3)}\Gamma^{\ulv} - \frac{1}{8}e^\phi \Gamma^m \bigl( {\ul{F}}^{(2)}  \Gamma^{\ulv}+  \ul{F}^{(4)} \bigr) \Gamma_m.
    \end{align}
\end{subequations}
Using these definitions, the 11- and 10-dimensional operators are   related as
\begin{subequations}\label{eq:dim-reduc-operators}
    \begin{align}
        & \hat{D}_m = D_m - \dfrac{1}{6} \Gamma_m \Delta + \dfrac{1}{3} e^{\phi} C_m \Gamma^{\ulv} \Delta + \dfrac{1}{6} \partial_m \phi, \\
        & \hat{D}_{10} = \dfrac{1}{3} \, e^{\phi} \Gamma^{\ulv} \Delta.
    \end{align}
\end{subequations}
We see that the 11-dimensional supercovariant derivative essentially splits in terms of the operators determining the type IIA gravitino and dilatino variations. Recall that we defined these operators from the supersymmetry variations of the type IIA gravitinos and dilatinos, which depend on the chosen fermionic frame. Therefore if one makes a different dimensional reduction ansatz for the 11-dimensional gravitino (or equivalently some redefinition in the fermionic sector of type IIA), the definition of these operators will be modified accordingly.

\subsection{Bosonic D2-brane action  }\label{D2bosonictreatment}

Once we know how to dimensionally reduce the background, we can dimensionally reduce the M2-brane action. We compactify along one direction that is not spanned by the M2-brane, therefore the result is the D2-brane of type IIA supergravity. We start from the bosonic part of the M2-brane action  \eqref{eq:m2-action}. Following our compactification ansatz, the pull-backs of the 11-dimensional metric and of the three-form can be written in terms of pullbacks of 10-dimensional fields as
\begin{align} 
    \hat{g}_{ij} &  = e^{-2\phi/3} g_{ij} + e^{4\phi/3} p_i p_j, \label{metric10211} \\
    \hat{A}_{ijk} & = C_{ijk} - 3 \, C_{[i} B_{jk]} + 3 \, p_{[i} B_{jk]}, \label{3form10211}
\end{align}
where we defined the combination $p_i = \partial_ix^{10} + \partial_i x^m C_m$. In terms of these fields, the bosonic M2-brane action becomes the D2-brane action and it reads
\begin{equation}\label{d2bosbeforeint}
    S^{(0)}_{\text{D}2} = - T_{\text{D}2} \int \de^3 \zeta \; e^{-\phi} \sqrt{- \mathrm{det} \, (g)} \; \sqrt{1 + e^{2\phi} p^2} + \frac{T_{\text{D}2}}{6} \int \de^3 \zeta \; \epsilonconv \Bigl[  C_{ijk} - 3 \, C_{i} B_{jk} + 3 \, p_{i} B_{jk} \Bigr],
\end{equation}
where $T_{\text{D}2} = T_{\text{M}2}$ is the D2-brane tension. We would like to obtain the action for the D2-brane in a fully 10-dimensional formulation. Currently, however, \eqref{d2bosbeforeint} contains factors of $p_i$ and so that formulation of the action implicitly knows about the M-theory circle. We need to get rid of $p_i$. We do this by including a Lagrange multiplier term involving the one form $p_1$ and its worldvolume dual, the exact 2-form $\sf{F}_2=\mathrm{d}A_1$, where $A_1$ is the D2-brane worldvolume gauge field. This Lagrange multiplier is
\begin{equation} \label{lagrange multiplier}
    S_{\text{LM}} = \frac{T_{\text{D}2}}{2}\int \de^3 \zeta \, \epsilonconv (p_i -  C_i) \sf{F}_{jk}.
\end{equation}
A fully 10-dimensional D2-brane action follows from including this term in the action, and then integrating out $p_i$ by plugging the solutions to its equation of motion back into the action. After doing this, and with a little massaging, we arrive at the familiar form  the bosonic D2-brane action
\begin{equation}\label{d2bosonic}
S^{(0)}_{\text{D}2} = - T_{\text{D}2} \int \de^3\zeta \, e^{-\phi} \sqrt{-\det(g + f)} +  T_{\text{D}2}\int (C_3 - C_1 \wedge f_2),
\end{equation}
where we made the definition $f_{ij} = B_{ij} + \sf{F}_{ij}$. This action, obtained from the M-theory dimensional reduction, is in string frame. It is worth noting explicitly here that the worldvolume field $f_{ij}$ is built using one field that is pulled back from the bulk, $B_{ij}$, and one that specifically lives only on the worldvolume, $\sf{F}_{ij}$.

We have  calculated a fully 10-dimensional formulation of the D2-brane \textit{bosonic} action. Our next goal is to find fermion couplings on the brane worldvolume. Therefore we turn to the superspace generalization of the $\mathrm{S}^1$-compactification we have just used.

\subsection{Superspace dimensional reduction   and fermions on the D2-brane}

In this section we obtain the fermion couplings on the D2-brane action. Following the same reasoning as in the case of the M2-brane action discussed in section \ref{sec:M2}, this can be done by moving to the superspace formulation of type IIA supergravity. One must promote fields in the bosonic action to superfields and then find the corresponding $\theta$-expansions. From the expansions of the constituent superfields, the expansion of the brane action superfield may then be determined.

A possible method to obtain the superfield expansions would be to construct all the necessary superfields using the same geometrical strategy as we applied to the M2-brane, i.e. NORCOR. However this requires more hard work than is necessary and there exists a better strategy. The key of our approach is the following  observation: the superspace formulation of M-theory is in  $(11|32)$-dimensional superspace, and the superspace formulation of type IIA strings is in $(10|32)$-superspace. It is therefore natural to expect that, as for the basic spacetime case, both superspaces are related via an $\mathrm{S}^1$-compactification of a bosonic direction. This superspace compactification and knowledge of 11-dimensional superfields in the M2-brane action are all we need to obtain the expansion of the type IIA superfields that appear in the D2-brane action.

Now we have to determine those 10-dimensional   superfields. Same as in the M2-brane case, at zeroth order in the $\theta$-expansion, the superfields are simply the bosonic fields. Those 10-dimensional bosonic fields  are related to the bosonic fields of 11-dimensional supergravity by the dimensional reduction ansatzes \eqref{xx1} and \eqref{xx2}. The spacetime dimensional reduction is described by those equations, and it is natural to interpret all fields appearing there (both  11- and  10-dimensional fields) as the leading-order terms of the corresponding superfield $\theta$-expansions. The superspace dimensional reduction must be described by the superspace generalization of those equations. Our method to compute the 10-dimensional superfields of interest will therefore be to use this superfield generalization together with knowledge of the 11-dimensional superfields we already gleaned in the previous section. Before we write the superspace compactification ansatz, recall that in eleven dimensions we did not compute the whole expansion of superfields, but rather we restricted to even $\hat{\theta}$ powers   because we were interested in bosonic backgrounds and we considered truncations to quartic order in the fermions. The same holds in ten dimensions, namely we are interested in explicitly obtaining the same type of restricted and truncated superfield expansions. We  promote \eqref{xx1} and \eqref{xx2} to the superfield level and use bold notation to indicate that in practice we will expand and truncated them. We obtain the promoted 11-dimensional metric
\begin{equation}\label{zz1}
    \bs{\hat{g}}_{\hat{m}\hat{n}} = \matr{e^{-2\bs{\phi}/3}(\bs{g}_{mn}+e^{2\bs{\phi}}\bs{C}_m\bs{C}_n)}{e^{4\bs{\phi}/3}\bs{C}_m}{e^{4\bs{\phi}/3} \bs{C}_n}{e^{4\bs{\phi}/3}}
\end{equation}
and the promoted 11-dimensional three-form\footnote{For future convenience, we place a prime on the 10-dimensional RR three-form superfield here. We ask that the reader indulges us in doing this for the time being and promise that the reason will be made clear. The motivation of this choice is explained in (\ref{trick}).}
\begin{align}
    & \bs{\hat{A}}_{mnp} = \bs{C}'_{mnp}, \label{zzz1} \\
    & \bs{\hat{A}}_{mn \,10}  = \bs{B}_{mn}, \label{zzz2}
\end{align} 
where $\bs{g}_{mn}$ is the truncated 10-dimensional supermetric, $\bs{\phi}$ is the truncated dilaton superfield, $\bs{B}_{mn}$ is the truncated Kalb-Ramond superfield, and $\bs{C}_m$ and $\bs{C}'_{mnp}$ are the truncated Ramond-Ramond one- and three-form superfields, respectively.

With these relations in hand, we are ready to  obtain the $\theta$-expansions of the 10-dimensional superfields. We are going to first compute the expansions of the  10-dimensional superfields up to order $(\theta)^2$ as an illustrative example. We will do this in detail. Then we will plug the expressions we find into the expression for D2-brane action superfield, expand, and compare our findings with previous results for the D2-brane in bosonic backgrounds at second order in fermions  obtained with alternative methods. The results match, confirming the validity of our approach. Finally, we will compute the order-$(\theta)^4$ terms of the truncated superfields. We will use these results to support the point we made in previous sections, i.e. that combining the terms in $\theta$-expansions into a more compact and manifestly supercovariant formulation is crucial. We argue strongly that this is the cornerstone of plausible methods for making the calculation of high-order fermionic couplings in brane actions viable in the future.

\subsubsection[Order-$(\theta)^2$ terms]{Order-\texorpdfstring{$\boldsymbol{(\theta)^2}$}{theta2} terms}\label{ord2sec5}

The superfield relations in (\ref{zz1} - \ref{zzz2}) can be Taylor-expanded, and these expansions can be truncated at a desired fermion order. This will lead to relations between  11- and 10-dimensional fields. We will use the number of fermions (both in eleven and ten dimensions) as an ordering  principle to relate those 11- and 10-dimensional fields. At leading order, one finds the original bosonic ansatz, which does not have any new information. For the bosonic backgrounds we are considering, at next order, in eleven dimensions one finds fermion bilinears with $\hat{\Gamma}$-matrices and the 11-dimensional supercovariant derivative in which several bosonic fields appear. It is natural to expect a similar behaviour in ten dimensions, namely that at this order each superfield involves a bilinear in $\theta$ as well as $\Gamma$-matrices and operators involving  10-dimensional fields. 
We can therefore make an ansatz for each truncated superfield involving a (for now) unknown fermion  bilnear, i.e.  
\begin{subequations}\label{10dshiftsgen}
    \begin{align}
        & \boldsymbol{g}_{mn} = g_{mn} + \gamma_{mn}, \label{10dshiftsgen-metric} \\
        & \boldsymbol{\phi} = \phi + \rho, \label{10dshiftsgen-dilaton} \\
        & \boldsymbol{B}_{mn} = B_{mn} + \beta_{mn}, \\
        & \boldsymbol{C}_m = C_m + \tau_m, \label{10dshiftsgen-Cm} \\
        & \boldsymbol{C}'_{mnp} = C_{mnp} + \alpha'_{mnp}.
    \end{align}
\end{subequations}
We now need to obtain expressions for the unknown 10-dimensional bilinears. Our procedure is to take each component of the 11-dimensional fields in (\ref{zz1})  - \eqref{zzz2} and then perform a Taylor expansion in fermions. We do this by NORCOR for the 11-dimensional left-hand side and by plugging in the ansatzes \eqref{10dshiftsgen} for the 10-dimensional right-hand side. Then we identify the corresponding 11-dimensional bilinears with the unknown 10-dimensional ones. At that stage one has relations between fermion bilinears in different theories. The equations indicate expressions for the unknown 10-dimensional bilinears in terms of 11-dimensional fields. In order to write the results for the 10-dimensional bilinears in terms of 10-dimensional fields, we are required to dimensionally reduce the 11-dimensional expressions. To properly elucidate this procedure, which is critical to our overall method, we will provide several examples at varying levels of technical complexity by calculating the bilinear terms for some of fields in \eqref{10dshiftsgen}.

\subsubsection*{Example 1: dilaton}
The simplest example case is that of the dilaton, for which we will provide every detail.  We read from \eqref{zz1} that it is related to the $(10,10)$-component of the 11-dimensional supermetric as $\bs{\hat{g}}_{10 \, 10} = e^{4\bs{\phi}/3}$. We Taylor-expand both sides of this relation. For the 11-dimensional left-hand side we use the result \eqref{eq:metric-shift2} from the NORCOR procedure. For the 10-dimensional right-hand side we use the expansion ansatz for the dilaton superfield in \eqref{10dshiftsgen-dilaton}. Equating the fermion bilinear   terms from each side, we find that
\begin{equation}
  -i\hat{\bar{\theta}} \hat{\Gamma}_{10} \hat{D}_{10} \hat{\theta} = e^{4\phi/3}  \dfrac{4\rho}{3},
 \end{equation}
In order to determine an expression for $\rho$ in terms of 10-dimensional fields we are required to dimensionally reduce the 11-dimensional bilinear. All the necessary results are given in appendix \ref{DimRedCatApp}. We can eventually write
\begin{equation}
    -i \hat{\bar{\theta}} \hat{\Gamma}_{10} \hat{D}_{10} \hat{\theta} = \dfrac{-i}{3} e^{4\phi/3} \bar{\theta}\Delta \theta,
\end{equation}
which means the dilaton superfield fermion bilinear contribution is
\begin{equation}\label{rhoshiftorder2}
    \rho = -\dfrac{i}{4} \bar{\theta} \Delta \theta.
\end{equation}
We have found an expression for the bilinear $\rho$ that is associated to the operator $\Delta$ which appears in the supersymmetry variation of the dilatino. This was to be expected: recall that we obtain the $\theta$-expansion by taking supersymmetry variations. In the first supersymmetry variation of the dilaton one finds the dilatino and so   the supersymmetry variation of the dilatino   appears when we take a second variation (on the dilaton). It is also worth remembering again at this point that, in ten dimensions, $\theta$ represents a pair of Majorana-Weyl  fermions of opposite chirality.

\subsubsection*{Example 2: Ramond-Ramond one-form}
For this next example we will move through the steps a little faster. We read from \eqref{zz1} that the Ramond-Ramond one-form superfield in ten dimensions is related to the $(m, 10)$-component of the 11-dimensional supermetric as $\bs{\hat{g}}_{m\,10} = e^{4\bs{\phi}/3}\bs{C}_m$. Taylor-expanding both sides using \eqref{eq:metric-shift2} and \eqref{10dshiftsgen-Cm}, and keeping  fermion bilinear terms, we obtain
\begin{equation}
-i\hat{\bar{\theta}}\hat{\Gamma}_{(m}\hat{D}_{10)}\hat{\theta} = e^{4\phi/3} \tau_m - \frac{i}{3} e^{4\phi/3} (\bar{\theta}\Delta \theta) C_m.
\end{equation}
Note that because the superfield relation involved both the dilaton and the Ramond-Ramond one-form, we were obliged to use \eqref{rhoshiftorder2}. After a little work for the dimensional reduction of the 11-dimensional bilinear (again, all the relevant results are given in appendix \ref{DimRedCatApp}), we arrive at
\begin{equation}
    -i \hat{\bar{\theta}} \hat{\Gamma}_{(m}\hat{D}_{10)} \hat{\theta} = - \dfrac{i}{2} e^{\phi/3} \, \bar{\theta} \Gamma^{\ulv} \Big( D_m - \dfrac{1}{2} \Gamma_m \Delta \Big) \theta  - \dfrac{i}{3} e^{4\phi/3}( \bar{\theta}\Delta \theta) C_m,
\end{equation}
which indicates that the bilinear $\tau_m$ must be
\begin{equation} \label{taushiftorder2}
    \tau_m =  - \dfrac{i}{2} e^{-\phi} \, \bar{\theta} \Gamma^{\ulv} \Big(D_m - \frac{1}{2}\Gamma_m \Delta\Big) \theta.
\end{equation}

\subsubsection*{Example 3: metric}
The most complicated superfield relation is that of the $(m,n)$-component of the 11-dimensonal supermetric. We read from \eqref{zz1} that it is related to the 10-dimensional supermetric, the Ramond-Ramond one-form, and the dilaton as $\bs{\hat{g}}_{mn} = e^{-2\bs{\phi}/3} (\bs{g}_{mn} + e^{2\bs{\phi}} \bs{C}_m \bs{C}_n)$. With the ansatz (\ref{10dshiftsgen-metric}) and the previous results (\ref{rhoshiftorder2}) and (\ref{taushiftorder2}) for $\rho$ and $\tau_m$, Taylor-expanding in exactly the way we have in previous examples yields
\begin{equation}
\begin{split}
    -i \hat{\bar{\theta}} \hat{\Gamma}_{(m} \hat{D}_{n)} \hat{\theta} = - i \, e^{-2\phi/3} \biggl[ & - \dfrac{1}{6} (\bar{\theta} \Delta \theta) \big( g_{mn} + e^{2 \phi} C_m C_n \big) + i \gamma_{mn} \\
    & + e^{2 \phi} \biggl( \frac{1}{2} (\bar{\theta} \Delta \theta) C_m C_n + C_{(m} e^{-\phi} \, \bar{\theta} \Gamma^{\ulv} \Big(D_{n)} - \frac{1}{2} \Gamma_{n)} \Delta \Big) \theta \biggr) \biggr].
\end{split}
\end{equation}
Once more applying the results of appendix \ref{DimRedCatApp}, the dimensional reduction of the 11-dimensional bilinear can be determined to be
\begin{equation}
\begin{split}
    -i\hat{\bar{\theta}}\hat{\Gamma}_{(m}\hat{D}_{n)}\hat{\theta} = - i \,  e^{-2\phi/3} \, \biggl[ & - \dfrac{1}{6} (\bar{\theta} \Delta \theta) \bigl( g_{mn} + e^{2 \phi} C_m C_n \bigr) +\bar{\theta} \Gamma_{(m} D_{n)} \theta \\
    & + e^{2 \phi} \biggl( \frac{1}{2} (\bar{\theta} \Delta \theta) C_m C_n + C_{(m} e^{-\phi} \, \bar{\theta} \Gamma^{\ulv} \Bigl( D_{n)} - \frac{1}{2} \Gamma_{n)} \Delta \Bigr) \theta \biggr) \biggr].
\end{split}
\end{equation}
By comparison, we are immediately able to discern the result
\begin{equation}
    \gamma_{mn} = -i \bar{\theta} \Gamma_{(m} D_{n)} \theta.
\end{equation}
One should observe that the metric superfield expansion takes on the same shape for the 11- and the 10-dimensional metrics. In each case the fields and operators involved are not the same, but equivalent objects appear in the same place. This is once again to be expected. The first-order $\theta$-expansion of the metric involves the (corresponding) gravitino, and we obtain the expansion by taking supersymmetry variations. Upon a second variation we are therefore not surprised to find the supersymmetry operator on the gravitino variation.

Application of our approach to the case of the relations (\ref{zzz1}) and (\ref{zzz2}) connecting the 11-dimensional three-form superfield to the 10-dimensional superfields is essentially straightforward and we leave the details of the calculation to the interested reader.

\subsubsection*{Full results}
At the end of the day, the expansions of the 10-dimensional superfields for type IIA supergravity up to quadratic order in fermions are
\begin{align}
    & \boldsymbol{g}_{mn} = g_{mn} - i \, \bar{\theta} \Gamma_{(m} D_{n)} \theta, \label{shiftsstart} \\
    & \boldsymbol{\phi} = \phi - \dfrac{i}{4} \, \bar{\theta} \Delta \theta, \\
    & \boldsymbol{B}_{mn} = B_{mn} - i \, \bar{\theta} \Gamma^{\ulv} \Gamma_{[m} D_{n]} \theta, \\
    & \boldsymbol{C}_{m} = C_m - \dfrac{i}{2} \, e^{-\phi} \, \bar{\theta} \Gamma^{\ulv} \Bigl( D_m - \frac{1}{2} \Gamma_m \Delta \Bigr) \theta, \label{c1shiftquad} \\
    & \boldsymbol{C}'_{mnp} = C_{mnp} - \dfrac{i}{2} \, e^{-\phi} \, \bar{\theta} \Bigl( 3\Gamma_{[mn} D_{p]} - \dfrac{1}{2} \Gamma_{mnp} \Delta \Bigr) \theta - 3 i \, C_{[m} \, \bar{\theta} \Gamma^{\ulv} \Gamma_{n} D_{p]} \theta. \label{shiftsend}
\end{align}
All bilinears involve one or both of the operators appearing in the  supersymmetry variation of the type IIA gravitino and dilatino, i.e. the supercovariant derivative $D_m$ and the operator $\Delta$, respectively. The Ramond-Ramond potential $C^{(1)}$ is also present in the expansion of $\smash{{\bs{C}'}^{(3)}}$. We asked earlier that the reader indulge us in defining the three-form superfield with a prime for the moment. The reason for this is that it will later become advantageous to consider the three-form superfield expansion restricted to that bilinear which does \textit{not} come multiplied with $C^{(1)}$, and for notational convenience it will be this restricted expansion which we shall call $\bs{C}^{(3)}$. We will say more on this in the next section.

Our results precisely match with those used in \cite{Marolf:2003ye},\footnote{What we give as $\bs{C}'_{ijk}$ here is denoted $\bs{C}_{ijk}^{\text{standard}}$ there.} where, however, the approach followed was morally quite different. They used the results of \cite{Marolf:2003vf}, where all D$p$-brane actions at order $(\theta)^2$ were computed using a brute-force approach, and noticed that all D$p$-brane actions could be written in a particularly compact and convenient way using field combinations like the ones above. The authors there labelled their observation a `superfield-like' formulation. Using our more conceptually sophisticated approach we can now confidently remove the `like'. We can see clearly that the reason these particular field combinations proved to be so useful to previous authors is that they are indeed born out of superfield considerations, namely the use of truncated superfield expansions as we have developed here. Moreover, the brute force approach is very complicated to manage at higher  orders in $\theta$. Our approach, though still somewhat complicated, does allow such computations to be performed.

\subsubsection*{D2-brane action at order $\bs{(\theta)^2}$}
What bosonic fields do, the superfields do better. Or rather, the superfields do morally the same thing but carry with them all of the information about the fermion terms. So it went for the dimensional reduction of individual (super)fields, and so it goes for manipulations of the (super)field quantities built from these constituent (super)fields. The composite quantity we are concerned with now is the D2-brane action.

In section \ref{D2bosonictreatment} we provided many details of the dimensional reduction of the bosonic M2-brane action to the bosonic D2-brane one. Now its usefulness is apparent: we are going to interpret the bosonic action as the zeroth-order fermionic expansion of the corresponding superfield. Based on this idea, we start with the M2-brane superaction \eqref{onemorelabel} and write it in terms of 10-dimensional superfields by using the superspace dimensional reduction ansatzes (\ref{zz1}) and (\ref{zzz1}, \ref{zzz2}). The appearance of pullbacks works exactly as in the bosonic case, and so the outcome is the D2-brane super action  written as
\begin{equation}
    \bs{S}_{\text{D}2}^{(2)} = - T_{\text{D}2}\int \de^3 \zeta \; e^{-\bs{\phi}} \sqrt{- \mathrm{det} \, (\bs{g})} \, \sqrt{1 + e^{2\bs{\phi}} \bs{p}^2} + \frac{T_{\text{D}2}}{6} \int \de^3 \zeta \; \epsilonconv \Bigl[  \bs{C}'_{ijk} - 3 \, \bs{C}_{i} \bs{B}_{jk} + 3 \, \bs{p}_{i} \bs{B}_{jk} \Bigr],
\end{equation}
where  $\bs{p}_i = \partial_ix^{10} + \partial_ix^m\bs{ C}_m$. Once again we would like to write this action in a fully 10-dimensional formulation, and so need to get rid of the explicit dependence on $\bs{p}_i$ (which knows about the M-theory $\mathrm{S}^1$). We do this by once again introducing the Lagrange multiplier \eqref{lagrange multiplier}. Notice that the bilinears in the truncated expansions of $\bs{p}_i$ and $\bs{C}_i$ cancel in the Lagrange multiplier which depends on the difference $(\bs{p}_i - \bs{C}_i)$ and so in effect we can promote these bosonic fields to truncated superfields for free. Integrating out $\bs{p}_i$ proceeds in formally the same way as integrating out $p_i$ did in the bosonic case. After doing so, we are arrive at the D2-brane action superfield
\begin{equation}\label{d2intout}
    \bs{S}^{(2)}_{\text{D}2} = - T_{\text{D}2}\int \de^3\zeta \; e^{-\bs{\phi}} \sqrt{- \det (\bs{g}_{ij} + \bs{f}_{ij})} + \frac{ T_{\text{D}2}}{6} \int \de^3 \zeta \; \epsilonconv (\bs{C}'_{ijk} - 3\bs{C}_{i} \bs{f}_{jk}),
\end{equation}
where we have defined $\bs{f}_{ij} = \bs{B}_{ij} + \sf{F}_{ij}$. Note the worldvolume flux $\sf{F}_2=dA_1$ remains purely bosonic because it is a brane worldvolume field, not a superfield.

Let us once again stress that this procedure is valid at \textit{any} order in $\theta$. The right-hand side of \eqref{d2intout} is the correct structure from which to obtain the D2-brane action to any order. All one needs to do is plug in the expansions of the superfields truncated at a given order in $\theta$. The problem of obtaining the D2-brane action up to a given order in $\theta$ has been reduced to the problem of determining the expansions of the individual superfields involved. Once these superfield expansions are known, the D2-brane action can be written down immediately.

To elaborate further on this claim, we reproduce the familiar form for the D2-brane action at second order in fermions. Starting with \eqref{d2intout}, in order to obtain explicit couplings we need only plug in the truncated superfields (\ref{shiftsstart} - \ref{shiftsend}). We successfully reproduce the quadratic D2-brane action
\begin{equation}\label{familiarD2quad}
    \bs{S}_{\text{D}2}^{(2)} = - T_{\text{D}2} \int \de^3 \zeta \;  e^{-\phi} \Biggl[ \sqrt{-\det(g + f)} \biggl[ 1 - i\bar{\theta} P_-^{(0)} \biggl( \mathcal{M}^{ij} \Gamma_i D_j - \frac{1}{2} \Delta \biggr) \theta \biggr] - (C_3 - C_1 \wedge f_2) \Biggr],
\end{equation}
where $\mathcal{M}^{ij}$ is the inverse of the combination $ \mathcal{M}_{ij}=(g_{ij}+\Gamma^{\clty} B_{ij})$  and we defined the (zeroth order) D2-brane $\kappa$-symmetry projector
\begin{equation}
    P_-^{(0)} \equiv \dfrac{1}{2} \bigl( 1 - {\Gamma}_{\text{D}2} \bigr),
\end{equation}
where 
\begin{equation}
    {\Gamma}_{\text{D}2} = \frac{1}{\sqrt{-\det(g + f)}} \, \epsilonconv \biggl( \dfrac{1}{6} \Gamma_{ijk} - \frac{1}{2} \Gamma^{\ulv} \Gamma_i f_{jk} \biggr).
\end{equation}
Notice that this is slightly more involved than in the M2-brane case because of the inclusion of worldvolume flux $f_2$. The outcome is the full D2-brane action at second order in fermions, and it matches exactly with the results in \cite{Marolf:2003vf, Martucci:2005rb}. This completes the fermionic second-order analysis to exemplify our alternative approach to obtain the D2-brane action at any fermion level.

\subsubsection[Order-$(\theta)^4$ terms]{Order-\texorpdfstring{$\boldsymbol{(\theta)^4}$}{theta4} terms}\label{orderfourdimred}
We have developed an improved approach for determining superfield fermionic expansions of fields in type IIA supergravity. We did this via NORCOR in 11-dimensional supergravity and the string duality that gives the type IIA theory via an $\mathrm{S}^1$-compactification. In the above subsection we demonstrated in detail how our approach can be used to obtain the known results at second order in fermions with much less hassle than previous approaches. In this section we move to use our approach to calculate the \textit{quartic} $\theta$ terms for those same type IIA superfield expansions. 

As we discussed above, using our approach, the problem of determining the D2-brane action superfield expansion gets reduced to the problem of determining the fermionic expansion of the constituent superfields. Once these expansions have been found, the D2-brane action follows immediately from plugging them into \eqref{d2intout}. All of the necessary details for this to work function at fourth order just as well as second order, and indeed at every order.

Since our method is applicable at every fermion order, to find the superfield expansions of the type IIA fields we can proceed in the same way as we did for the quadratic case above. To start, we make ansatzes for the order-four   terms in the truncated expansions of the 10-dimensional superfields. We have already determined the bilinear terms and so can include them immediately. We use the same symbols as we did for the ansatzes in the order-two case, but now label the unknown quantities with their fermion order. We have
\begin{subequations}
    \begin{align}
        & \boldsymbol{g}_{mn} = g_{mn} - i \, \bar{\theta} \Gamma_{(m} D_{n)} \theta + \gamma_{mn}^{(4)}, \label{shiftsstart2}\\
        & \boldsymbol{\phi} = \phi - \dfrac{i}{4} \, \bar{\theta} \Delta \theta + \rho^{(4)}, \label{rhoquart} \\
        & \boldsymbol{B}_{mn} = B_{mn} - i \, \bar{\theta} \Gamma^{\ulv} \Gamma_{[m} D_{n]} \theta + \beta_{mn}^{(4)}, \\
        & \boldsymbol{C}_{m} = C_m - \dfrac{i}{2} e^{-\phi} \, \bar{\theta} \Gamma^{\ulv} \Big(D_m - \frac{1}{2}\Gamma_m \Delta\Big) \theta + \tau_m^{(4)}, \\
        & \boldsymbol{C}'_{mnp} = C_{mnp} - \dfrac{i}{2} e^{-\phi} \, \bar{\theta} \Bigl( 3\Gamma_{[mn} D_{p]} - \dfrac{1}{2} \Gamma_{ijk} \Delta \Bigr) \theta - 3 i C_{[m} \, \bar{\theta} \Gamma^{\ulv} \Gamma_{n} D_{p]} \theta + \alpha'^{(4)}_{mnp}. \label{shiftsend2}
    \end{align}
\end{subequations}
Once again, we must determine the expressions for these unknown shifts by Taylor-expanding both sides of (\ref{zz1}) and (\ref{zzz1}, \ref{zzz2}), now to quartic order in $\theta$. Again, we appeal to the results of the NORCOR procedure to Taylor-expand the left-hand side, whereas we plug our quartic ansatzes in to Taylor-expand the right-hand side. Upon rearrangement, this will result in expressions for the unknowns which contain both 10- and 11-dimensional fields. We must then once again dimensionally reduce the 11-dimensional quantities that appear in order to determine expressions for the unknowns that are entirely in terms of 10-dimensional quantities. 

The mixing of the 10-dimensional metric, the dilaton and the Ramond-Ramond one-form in \eqref{zz1} causes the expressions for the quartic ansatzes to be quite complicated to deal with practically. For ease of notation, let us denote the quartic terms in the truncated expansion of the 11-dimensional supermetric \eqref{quarticm2met} as $\smash{\hat{\gamma}^{(4)}_{\hat{m}\hat{n}}}$. Now, Taylor-expanding the relation $\smash{\bs{\hat{g}}_{10 \, 10} = e^{4\bs{\phi}/3}}$ and keeping only the terms up to quartic order in fermions allows us to find that
\begin{equation}\label{dilatonexpansion}
    \rho^{(4)} = \frac{1}{24} (\bar{\theta} \Delta \theta)^2  + \frac{3}{4} e^{-4\phi/3} \hat{\gamma}_{10 \, 10}^{(4)}.
\end{equation}
Determining a 10-dimensional expression for $\smash{\rho^{(4)}}$ now requires us to perform dimensional reduction on $\smash{\hat{\gamma}_{10 \, 10}^{(4)}}$. Before that though we also note the results of Taylor-expanding and rearranging the relations that allow us to determine expressions for $\smash{\tau_m^{(4)}}$ and $\smash{\gamma_{mn}^{(4)}}$. First, expanding both sides of the equation $\bs{\hat{g}}_{m\,10} = e^{4\bs{\phi}/3}\bs{C}_m$ yields that the quartic shift on the 10-dimensional Ramond-Ramond one-form superfield is given by
\begin{equation}
\begin{split}
    e^\phi \tau^{(4)}_m = \frac{1}{6} (\bar{\theta}\Delta \theta) (\bar{\theta} \Gamma^\clty D_m\theta) - \frac{1}{12}(\bar{\theta} \Delta \theta) (\bar{\theta} \Gamma^\clty \Gamma_m\Delta \theta) + \frac{1}{18} e^\phi (\bar{\theta} \Delta \theta)^2 C_m \\
    - \frac{4}{3} e^\phi \rho^{(4)} C_m + e^{-{\phi}/{3}} \hat{\gamma}^{(4)}_{m\,10} & .
\end{split}
\end{equation}
As with the quadratic case, the mixing of 10-dimensional superfields in the right-hand side of the supermetric relation in \eqref{zz1} means we are required to use the expressions for the dilaton superfield expansion in this calculation. Second, we Taylor-expand the relation $\bs{\hat{g}}_{mn} = e^{-2\bs{\phi}/3} (\bs{g}_{mn} + e^{2\bs{\phi}}\bs{C}_m\bs{C}_n)$ in order to determine an expression for the quartic fermion term of the expansion of the 10-dimensional supermetric, obtaining
\begin{equation}
\begin{split}
    \gamma^{(4)}_{mn} = \frac{2}{3} g_{mn} \rho^{(4)}
    & + \frac{1}{4} \bigl( \bar{\theta} \Gamma^{\ulv} D_{(m} \theta \bigr) \bigl( \bar{\theta} \Gamma^{\ulv}D_{n)} \theta \bigr) - \frac{1}{4} \bigl( \bar{\theta} \Gamma^{\ulv} D_{(m} \theta \bigr) \bigl( \bar{\theta} \Gamma^{\ulv} \Gamma_{n)} \Delta \theta \bigr) \\
    & + \frac{1}{16} \bigl( \bar{\theta} \Gamma^{\ulv} \Gamma_{(m} \Delta \theta \bigr) \bigl( \bar{\theta} \Gamma^{\ulv} \Gamma_{n)} \Delta \theta \bigr) - \frac{1}{6} \bigl( \bar{\theta} \Delta \theta \bigr) \bigl( \bar{\theta} \Gamma_{(m} D_{n)} \theta \bigr) + \frac{1}{72} g_{mn} \bigl( \bar{\theta} \Delta \theta \bigr)^2 \\
    & - 2 \, e^{2\phi} C_{(m} \tau^{(4)}_{n)} + \frac{1}{3} e^{\phi} C_{(m} \bigl( \bar{\theta} \Delta \theta \bigr) \bigl( \bar{\theta} \Gamma^{\ulv} D_{n)} \theta \bigr) - \frac{1}{6} e^{\phi} C_{(m} \bigl( \bar{\theta} \Delta \theta \bigr) \bigl( \bar{\theta} \Gamma^{\ulv} \Gamma_{n)} \Delta \theta \bigr) \\
    & - \frac{4}{3} e^{2\phi} C_{(m} C_{n)} \rho^{(4)} + \frac{1}{18} e^{2\phi} C_{(m} C_{n)} \bigl( \bar{\theta} \Delta \theta \bigr)^2 + e^{{2\phi}/{3}} {\hat{\gamma}}^{(4)}_{mn}.
\end{split}
\end{equation}
The mixing of 10-dimensional superfields in the relation for the 11-dimensional supermetric has again meant that we must include the previously calculated quartic terms for the dilaton and the Ramond-Ramond one-form when making this expansion. Already we can see that the relative complexity of the relation of the 11-dimensional supermetric to the 10-dimensional superfields results in expressions of some length even before we turn our attention to the dimensional reduction step of our procedure.

As with the quadratic case, the initial Taylor expansion and rearrangement of the relations in (\ref{zzz1}, \ref{zzz2}) concerning the 11-dimensional three-form at quartic order are essentially straightforward. Denoting the quartic terms in the NORCOR expansion of the 11-dimensional super three-form \eqref{quarticm2form} as $\smash{\hat{\alpha}'^{(4)}_{\hat{m}\hat{n}\hat{p}}}$, it is clear that the quartic terms in the truncated expansion of the 10-dimensional Ramond-Ramond three-form superfield is given by $\smash{{\alpha}'^{(4)}_{{m}{n}{p}} = \hat{\alpha}'^{(4)}_{{m}{n}{p}}}$, and for the 10-dimensional Kalb-Ramond form superfield we have $\smash{{\beta}^{(4)}_{{m}{n}} = \hat{\alpha}'^{(4)}_{{m}{n}\,10}}$.

At this point, `all' that is left to do in order to obtain expressions for the quartic terms in the expansions of the 10-dimensional superfields is to dimensionally reduce the 11-dimensional quantities that appear, namely the components of $\hat{\gamma}^{(4)}_{\hat{m}\hat{n}}$ and $\hat{\alpha}^{(4)}_{\hat{m}\hat{n}\hat{p}}$. The calculation is very lengthy, so we provide all of the necessary tools and results in appendix \ref{DimRedCatApp}. Despite their cumulative length, all the steps are the simple application of the dimensional reduction procedure we are now very familiar with. For this reason, we place an example of the calculation in the case of the dilaton in appendix \ref{dimredquarApp}, but otherwise just report the results of the calculations here.

\subsubsection*{Quartic $\boldsymbol{\theta}$-terms for type IIA superfield expansions}
In order to simplify the statement of the results, it is convenient to first make a few definitions. Along with the familiar $D_m$ and $\Delta$, we will use the combinations
\begin{align}
    & \cl{D}_m \equiv D_m - \frac{1}{6} \Gamma_m \Delta, \\
    & \cmttr_q \equiv \big[\cl{D}_q, \Gamma^{\ulv} \Delta \big] + (\partial_q \phi) \Gamma^{\ulv}\Delta, \\
    & \cmttr_{pq} \equiv \big[\cl{D}_p, \cl{D}_q \big] + \frac{1}{3} e^{\phi} F^{(2)}_{pq} \Gamma^{\ulv} \Delta,
\end{align}
as well as
\begin{equation}
    \begin{split}
        & \flux_{mn} \equiv \frac{1}{24} \Bigr[ \Gamma_{mn} \bigl(e^\phi {\ul{F}}^{(4)} + \ul{H}^{(3)} \Gamma^{\ulv} \bigr) - 2 \, \Gamma_{m} \bigl( e^\phi {\ul{F}}^{(4)}_n + \ul{H}^{(3)}_n \Gamma^{\ulv} \bigr) + \bigl( e^\phi {\ul{F}}^{(4)}_{mn} + \ul{H}^{(3)}_{mn} \Gamma^{\ulv} \bigr) \Bigr], \\
        & \flux_m \equiv \frac{1}{24} \Bigl[ \Gamma_m \Gamma^{\ulv} e^\phi {\ul{F}}^{(4)} + \Gamma^{\ulv} e^\phi {\ul{F}}^{(4)}_m  \Bigr].
    \end{split}
\end{equation}
We are now ready to list the quartic terms in the superfield expansion of the 10-dimensional superfields using only 10-dimensional operators. The quartic fermionic terms in the dilaton are given by
\begin{equation}\label{quarticdil}
\begin{split}
    \rho^{(4)} = & \, - \frac{1}{768} (\bar{\theta} \Gamma^{mnpq} \theta) (\bar{\theta}  \Gamma_{mn}\cmttr_{pq} \theta) +  \frac{1}{576} (\bar{\theta} \Gamma^{\ulv}\Gamma^{mnp}\theta)\Big[\bar{\theta}\big[3 \Gamma^{\ulv}\Gamma_m  \cmttr_{np} - \Gamma_{mn} \cmttr_{p} \big]\theta\Big] \\
    & + \frac{1}{384} (\bar{\theta} \Gamma^{\ulv} \Gamma^{mn} \theta)\Big[\bar{\theta} \big[ 3 \Gamma^{\ulv}\cmttr_{mn} - 2 \Gamma_m \cmttr_n\big] {\theta}\Big] + \frac{1}{48}  (\bar{\theta}  \Gamma^{\ulv} \Gamma^{mn} \theta)(\bar{\theta} \flux_{mn} \Gamma^{\ulv} \Delta{\theta}) \\
    & - \frac{1}{48} (\bar{\theta}\Gamma_m\Gamma^{\ulv} \Delta \theta)(\bar{\theta} \Gamma^m \Gamma^{\ulv} \Delta \theta) - \frac{1}{576} \Big[\bar{\theta} \big[2\Gamma^{\ulv}e^{\phi} \ul{F}^{(4)}_{m}-\Gamma_m \ul{H}^{(3)} \big]\theta\Big](\bar{\theta} \Gamma^m  \Gamma^{\ulv} \Delta \theta) \\
    & + \frac{1}{48}(\bar{\theta}\Delta\theta)^2 + \frac{1}{576} \Big[\bar{\theta}\big[ e^{\phi} \ul{F}^{(4)} - 2\ul{H}^{(3)}\Gamma^{\ulv} \big]\theta\Big] (\bar{\theta} \Delta \theta).
\end{split}
\end{equation}
The quartic fermionic terms in the Ramond-Ramond one-form superfield are \def\y{\theta}
\begin{equation}
\begin{split}
    e^\phi \tau_{m}^{(4)} = & + \frac{1}{576} \bigl[ \bar{\y} \Gamma_m \Gamma^{npq} \y \bigr] \Big[\bar{\y} \bigl[  3 \,  \Gamma^{\ulv} \Gamma_n  \cmttr_{pq}-{\Gamma}_{np} \cmttr_q \big] \y \Bigr] \\
    & + \frac{1}{192} \bigl[ \bar{\y} \Gamma^{\ulv}\Gamma^{npq}\y \bigr] \Bigl[ \bar{\y} \big[  \Gamma_{mn}\cmttr_{pq} +   \Gamma_{np}\cmttr_{mq}\big] \y \Bigr] \\
    & + \frac{1}{576} \bigl[ \bar{\y} \Gamma_{m} \Gamma^{np} \y \bigr] \Bigl[ \bar{\y}\big[3 \Gamma^{\ulv}\cmttr_{np} - 2 \Gamma_n \cmttr_p\big]{\y} \Bigr] + \frac{1}{72} \bigl[ \bar{\y} \Gamma_{m} \Gamma^{np} \y \bigr] \bigl[ \bar{\y} \flux_{np} \Gamma^{\ulv} \Delta{\y} \bigr] \\
    &+ \frac{1}{192} \bigl[ \bar{\y} \Gamma^{\ulv} \Gamma^{np} \y \bigr] \Bigl[ \bar{\y} \big[ {\Gamma}_{{m}}\cmttr_{np} + 2 {\Gamma}_{{n}}\cmttr_{mp}\big] \y \Bigr] +  \frac{1}{24} \bigl[ \bar{\y} \Gamma^{\ulv} \Gamma^{np} \y \bigr] \bigl[ \bar{\y} \flux_{np} \cl{D}_m \y \bigr] \\
    &+ \frac{1}{144} \bigl[ \bar{\y} \Gamma_{m} \Gamma^{n} \Gamma^{\ulv} \y \bigr] \bigl[ \bar{\y} {\Gamma}^{\ulv} \cmttr_n \y \bigr]+ \frac{1}{36} \bigl[ \bar{\y} \Gamma_{m} \Gamma^{n} \Gamma^{\ulv} \y \bigr] \bigl[ \bar{\y} \flux_n \Gamma^{\ulv} \Delta\y \bigr] \\
    &+ \frac{1}{12} [\bar{\y}\Delta\y][\bar{\y} \Gamma^{\ulv} \cl{D}_{m} \y] - \frac{1}{18} [\bar{\y} \Delta \y] [\bar{\y} \Gamma^{\ulv} \Gamma_{m} \Delta \y] - \frac{1}{12} \bigl[ \bar{\y} \Gamma_n \Gamma^{\ulv} \Delta \y \bigr] \bigl[ \bar{\y} \Gamma^n \cl{D}_{m} \y \bigr] \\
    &+ \frac{1}{288} \Bigl[ \bar{\y} \big[  e^{\phi} \ul{F}^{(4)} - 2\ul{H}^{(3)}\Gamma^{\ulv} \big] \y \Bigr] \bigl[ \bar{\y} \Gamma^{\ulv} \cl{D}_m  \y \bigr] + \frac{1}{864} \Bigl[ \bar{\y} \Gamma_{m} \Gamma^{\ulv} \big[ e^{\phi} \ul{F}^{(4)} - 2\ul{H}^{(3)} \Gamma^{\ulv} \big] \y \Bigr] \bigl[ \bar{\y}  \Delta \y \bigr] \\
    &+ \frac{1}{288} \Bigl[ \bar{\y} \big[\Gamma^{\ulv} \Gamma_n e^{\phi} \ul{F}^{(4)} +  \Gamma_n \ul{H}^{(3)} - 3 \, \Gamma^{\ulv} e^{\phi} \ul{F}^{(4)}_n \big] \y \Bigr] \bigl[ \bar{\y} \Gamma^n \cl{D}_m \y \bigr] \\
    &- \frac{1}{864} \Bigl[ \bar{\y} \Gamma_{m} \big[ 3\ul{H}^{(3)}_n \Gamma^{\ulv} -\Gamma_n e^{\phi} \ul{F}^{(4)} + 3 \, e^{\phi} \ul{F}^{(4)}_n - \Gamma_n \ul{H}^{(3)} \Gamma^{\ulv}\big] \y \Bigr] \bigl[ \bar{\y}   \Gamma^n \Gamma^{\ulv}\Delta \y \bigr].
\end{split}
\end{equation}
The quartic fermionic terms for the 10-dimensional metric expansion read
\begin{equation}\label{10dmetricquartic}
    \begin{split}
        \gamma^{(4)}_{mn} = &  -  \frac{1}{384} g_{mn} (\bar{\y} \Gamma^{pqrs} \y) (\bar{\y}  \Gamma_{pq}\cmttr_{rs} \y) + \frac{1}{96} \bigl[ \bar{\y} \Gamma_{(m|} \Gamma^{pqr} \y \bigr] \Bigl[\bar{\y}\big[  \Gamma_{{|n)}p}\cmttr_{qr}+ \Gamma_{pq}\cmttr_{{|n)}r} \big]\y \Bigr] \\
        & + \frac{1}{288} \bigl[ \bar{\y} \Gamma_{(m|} \Gamma^{pq}\Gamma^{\ulv} \y \bigr] \Bigl[ \bar{\y}\big[2{\Gamma}_{{|n)}p}\cmttr_{q} + {\Gamma}_{pq}\cmttr_{{|n)}}  +  3\Gamma_{{|n)}} \Gamma^{\ulv} \cmttr_{pq} - 6\Gamma_{p} \Gamma^{\ulv} \cmttr_{{|n)}q}\big] \y \Bigr] \\
        & + \frac{1}{576} g_{mn} (\bar{\y} \Gamma^{\ulv} \Gamma^{pq} \y) \Big[\bar{\y} \big[ 3 \Gamma^{\ulv}\cmttr_{pq} - 2 \Gamma_p \cmttr_q\big] {\y}\Big] + \frac{1}{72} g_{mn} (\bar{\y} \Gamma^{\ulv} \Gamma^{pq} \y)(\bar{\y}   \flux_{pq} \Gamma^{\ulv} \Delta{\y}) \\
        & + \frac{1}{96} \bigl[ \bar{\y} \Gamma_{(m|}  \Gamma^{pq} \y \bigr] \Bigl[ \bar{\y} \big[  {\Gamma}_{{{|n)}}}\cmttr_{pq} + 2{\Gamma}_{{p}}\cmttr_{{|n)}q}\big] {\y} \Bigr] + \frac{1}{12} \bigl[ \bar{\y} \Gamma_{(m|}  \Gamma^{pq} \y \bigr] \bigl[ \bar{\y}\flux_{pq} \cl{D}_{|n)} {\y} \bigr] \\
        & + \frac{1}{144} \bigl[ \bar{\y} \Gamma_{(m|} \Gamma^{p} \Gamma^{\ulv} \y \bigr] \Bigl[ \bar{\y}\big[  {\Gamma}_{{|n)}}\cmttr_p +  {\Gamma}_{{p}} \cmttr_{|n)} - 3 {\Gamma}^{\ulv}\cmttr_{{|n)}p}\big] \y \Bigr] + \frac{1}{6} \bigl[ \bar{\y} \Gamma_{(m|} \Gamma^{p} \Gamma^{\ulv} \y \bigr] \bigl[ \bar{\y} \flux_p \cl{D}_{|n)} \y \bigr] \\
        & - \frac{1}{72} g_{mn} (\bar{\y}\Gamma_p\Gamma^{\ulv} \Delta \y)(\bar{\y} \Gamma^p \Gamma^{\ulv} \Delta \y) - \frac{1}{4} \bigl[ \bar{\y} \Gamma_p \cl{D}_{(m}\y \bigr] \bigl[ \bar{\y} \Gamma^p \cl{D}_{n)} \y \bigr] \\
        &  + \frac{1}{36} (\bar{\y} \Gamma^{\ulv} \Gamma_{(m} \Delta \y) (\bar{\y} \Gamma^{\ulv} \Gamma_{n)} \Delta \y) - \frac{1}{6} (\bar{\y} \Gamma^{\ulv} \Gamma_{(m} \Delta \y) (\bar{\y} \Gamma^{\ulv} \cl{D}_{n)} \y) - \frac{1}{6}  (\bar{\y} \Delta \y) (\bar{\y} \Gamma_{(m} \cl{D}_{n)} \y) \\
        & - \frac{1}{864} g_{mn} \Big[\bar{\y} \big[2\Gamma^{\ulv}e^{\phi} \ul{F}^{(4)}_{p} - \Gamma_p \ul{H}^{(3)} \big]\y\Big](\bar{\y}   \Gamma^p  \Gamma^{\ulv} \Delta \y) \\
        & - \frac{1}{144} \Bigl[ \bar{\y} \Gamma_{(m|} \big[ 3\ul{H}^{(3)}_p\Gamma^{\ulv}  -\Gamma_p e^{\phi} \ul{F}^{(4)}   -\Gamma_p \ul{H}^{(3)} \Gamma^{\ulv}  +3 e^{\phi} \ul{F}^{(4)}_p \big]\y \Bigr] \bigl[ \bar{\y}\Gamma^p \cl{D}_{|n)} \y \bigr] \\
        & - \frac{1}{144} \Bigl[ \bar{\y} \Gamma_{(m|}\Gamma^{\ulv}\big[2\ul{H}^{(3)}\Gamma^{\ulv} - e^{\phi} \ul{F}^{(4)} \big] \y \Bigr] \bigl[\bar{\y}  \Gamma^{\ulv}  \cl{D}_{|n)}\y \bigr] 
        \\
        & + \frac{1}{864} g_{mn} (\bar{\y} \Delta \y) \Big[\bar{\y}\big[ e^{\phi} \ul{F}^{(4)} - 2 \ul{H}^{(3)}\Gamma^{\ulv} \big]\y\Big].
    \end{split}
\end{equation}
The quartic fermionic terms for the Kalb-Ramond two-form are
\begin{equation}\label{KBquarticshift}
    \begin{split}
        \beta^{(4)}_{mn}  = & \, -\frac{1}{384} (\bar{\y} \Gamma^{\ulv}\Gamma_{mn}  \Gamma^{pqrs} \y) \bigl[ \bar{\y}  \Gamma_{pq}\cmttr_{rs} \y\bigr] 
        - \frac{1}{96} (\bar{\y} \Gamma_{[m|} \Gamma^{\ulv} \Gamma^{pqr} \y) \Bigl[ \bar{\y} \big[  \Gamma_{{|n]}p}\cmttr_{qr}+   \Gamma_{pq}\cmttr_{{|n]}r}\big] \y \Bigr] \\
        &+ \frac{1}{576} (\bar{\y} \Gamma_{mn} \Gamma^{pq} \y) \Bigl[ \bar{\y} \big[3 \Gamma^{\ulv}\cmttr_{pq} - 2\Gamma_p \cmttr_q\big] {\y} \Bigr] + \frac{1}{72} (\bar{\y} \Gamma_{mn} \Gamma^{pq} \y) \bigl[ \bar{\y} \flux_{pq} \Gamma^{\ulv} \Delta {\y} \bigr] \\
        & - \frac{1}{96} (\bar{\y} \Gamma_{[m|} \Gamma^{\ulv} \Gamma^{pq} \y) \Bigl[\bar{\y} \big[ {\Gamma}_{{{|n]}}}\cmttr_{pq} + 2 {\Gamma}_{{p}}\cmttr_{{|n]}q}\big] {\y} \Bigr] - \frac{1}{12} (\bar{\y} \Gamma_{[m|} \Gamma^{\ulv} \Gamma^{pq} \y) \bigl[\bar{\y} \flux_{pq} \cl{D}_{|n]} {\y} \bigr] \\
        & - \frac{1}{288} (\bar{\y} \Gamma_{[m|}\Gamma^{pq} \y)\Big[\bar{\y} \big[2{\Gamma}_{{|n]}p}\cmttr_{q} + {\Gamma}_{pq}\cmttr_{{|n]}}  +  3\Gamma_{{|n]}} \Gamma^{\ulv} \cmttr_{pq} - 6\Gamma_{p} \Gamma^{\ulv} \cmttr_{{|n]}q}\big]\y\Big] \\
        & + \frac{1}{144} (\bar{\y} \Gamma_{mn}  \Gamma^{p} \Gamma^{\ulv} \y) \bigl[ \bar{\y} {\Gamma}^{\ulv} \cmttr_p {\y} \bigr] + \frac{1}{36} (\bar{\y} \Gamma_{mn}  \Gamma^{p} \Gamma^{\ulv} \y) \bigl[ \bar{\y} \check{R}_p \Gamma^{\ulv} \Delta {\y} \bigr] + \frac{1}{6} (\bar{\y} \Gamma_{[m|} \Gamma^p \y) \bigl[\bar{\y} \flux_p \cl{D}_{|n]} {\y} \bigr] \\
        & + \frac{1}{144} (\bar{\y} \Gamma_{[m|} \Gamma^p \y) \Bigl[\bar{\y} \big[ {\Gamma}_{{|n]}}\cmttr_p +  {\Gamma}_{{p}} \cmttr_{|n]} - 3 {\Gamma}^{\ulv}\cmttr_{{|n]}p}\big] {\y} \Bigr] - \frac{1}{12} (\bar{\y} \Gamma_{p}\Gamma_{[m} \cl{D}_{n]} \y) (\bar{\y} \Gamma^p \Gamma^{\ulv} \Delta \y) \\
        & - \frac{1}{4} (\bar{\y} \Gamma_p \Gamma^{\ulv} \cl{D}_{[m} \y) (\bar{\y} \Gamma^p \cl{D}_{n]} \y)  + \frac{1}{12} (\bar{\y} \Gamma_{p} \Gamma_{[m} \Gamma^{\ulv} \Delta \y) (\bar{\y} \Gamma^p \cl{D}_{n]} \y) - \frac{1}{12} (\bar{\y} \Gamma^{\ulv} \Gamma_{[m} \cl{D}_{n]} \y) (\bar{\y} \Delta \y)  \\
        & - \frac{1}{12} (\bar{\y} \Gamma_{[m} \Delta \y) (\bar{\y} \Gamma^{\ulv} \cl{D}_{n]} \y) - \frac{1}{12} (\bar{\y}  \Gamma^{\ulv} \Delta \y) (\bar{\y} \Gamma_{[m} \cl{D}_{n]} \y)  - \frac{1}{12} (\bar{\y}  \cl{D}_{[m} \y) (\bar{\y} \Gamma_{n]} \Gamma^{\ulv} \Delta \y) \\
        & - \frac{1}{144} \Big[\bar{\y} \Gamma_{[m|} \big[ \Gamma^{\ulv} \Gamma_p e^{\phi} \ul{F}^{(4)}+ \Gamma_p \ul{H}^{(3)} - 3 \Gamma^{\ulv} e^{\phi} \ul{F}^{(4)}_p - 3\ul{H}^{(3)}_p \big] \y\Big] \bigl[ \bar{\y} \Gamma^p \cl{D}_{|n]} \y\bigr] \\
        & + \frac{1}{864} \Big[\bar{\y} \Gamma_{mn} \big[ \Gamma_p e^{\phi} \ul{F}^{(4)}  + \Gamma_p \ul{H}^{(3)} \Gamma^{\ulv} - 3 \ul{H}^{(3)}_p\Gamma^{\ulv}  - 3 e^{\phi} \ul{F}^{(4)}_p \big] \y\Big] \bigl[ \bar{\y} \Gamma^p  \Gamma^{\ulv} \Delta \y \bigr] \\
        & - \frac{1}{144} \Big[\bar{\y}\Gamma_{[m|}\Gamma^{\ulv} \big[ e^{\phi} \Gamma^{\ulv} \ul{F}^{(4)} + 2 \ul{H}^{(3)}   \big] \y\Big] \bigl[ \bar{\y} \Gamma^{\ulv} \cl{D}_{|n]} \y \bigr] \\
        & + \frac{1}{864} \Big[\bar{\y} \Gamma_{mn} \big[ \Gamma^{\ulv} e^{\phi} \ul{F}^{(4)} + 2\ul{H}^{(3)} \big] \y\Big] \bigl[ \bar{\y}  \Delta \y \bigr]  .
    \end{split}
\end{equation}
Finally, the quartic fermionic terms for the Ramond-Ramond three-form superfield can be written as
\begin{equation}
    \alpha'^{(4)}_{mnp} = {\alpha}''^{(4)}_{mnp} + 3\beta^{(4)}_{[mn} C_{p]},
\end{equation}
where ${\alpha}''^{(4)}_{mnp}$ is given by the expression,
\begin{equation}
    \begin{split}
        e^\phi{\alpha}''^{(4)}_{mnp} = & - \frac{1}{384} (\bar{\y} \Gamma_{mnp} \Gamma^{qrst} \y) \bigl[ \bar{\y} \Gamma_{qr}\cmttr_{st} \y \bigr]  - \frac{1}{576} (\bar{\y} \Gamma_{mnp}  \Gamma^{qrs}\Gamma^{\ulv} \y) \Bigl[ \bar{\y}\big[ {\Gamma}_{qr} \cmttr_s - 3 \, \Gamma^{\ulv}\Gamma_q  \cmttr_{rs}\big] \y \Bigr] \\
        & + \frac{1}{64} (\bar{\y} \Gamma_{[mn|}\Gamma^{qrs}\y) \Bigl[\bar{\y} \big[ \Gamma_{|p]q}\cmttr_{rs}+  \Gamma_{qr}\cmttr_{{|p]} s}\big]\y \Bigr] \\
        & + \frac{1}{192} (\bar{\y} \Gamma_{[mn|} \Gamma^{qr}\Gamma^{\ulv} \y) \Bigl[ \bar{\y} \big[ 2{\Gamma}_{{|p]} q}\cmttr_{r} + {\Gamma}_{qr}\cmttr_{{|p]} }  +  3\Gamma_{{|p]} } \Gamma^{\ulv} \cmttr_{qr} - 6\Gamma_{q} \Gamma^{\ulv} \cmttr_{{|p]} r}\big] \y\Bigr] \\
        & + \frac{1}{64} (\bar{\y} \Gamma_{[mn|} \Gamma^{qr} \y) \Bigl[ \bar{\y} \big[ {\Gamma}_{{{|p]} }}\cmttr_{qr} + 2 {\Gamma}_{{q}}\cmttr_{{|p]} r}\big] {\y} \Bigr] + \frac{1}{8} (\bar{\y} \Gamma_{[mn|} \Gamma^{qr} \y) \bigl[ \bar{\y} \flux_{qr} \cl{D}_{|p]} {\y} \bigr] \\
        & + \frac{1}{96} (\bar{\y} \Gamma_{[mn|}  \Gamma^{q}\Gamma^{\ulv} \y)\Big[\bar{\y}\big[ {\Gamma}_{|p]}\cmttr_q +  {\Gamma}_{{q}} \cmttr_{|p]}  - 3 {\Gamma}^{\ulv}\cmttr_{{|p]} q}\big] {\y}\Big] + \frac{1}{4} (\bar{\y} \Gamma_{[mn|}  \Gamma^{q} \Gamma^{\ulv} \y) (\bar{\y} \flux_q \cl{D}_{|p]} {\y}) \\
        & - \frac{3}{4} (\bar{\y}\Gamma_{q}\Gamma_{[m} \cl{D}_n \y)(\bar{\y} \Gamma^q \cl{D}_{p]} \y)  - \frac{3}{4} (\bar{\y} \cl{D}_{[m} \y)(\bar{\y} \Gamma_n \cl{D}_{p]} \y) -  \frac{3}{4} (\bar{\y}\Gamma^{\ulv}\Gamma_{[m} \cl{D}_{n} \y)(\bar{\y} \Gamma^{\ulv} \cl{D}_{p]} \y) \\
        & + \frac{1}{96} \Big[\bar{\y} \Gamma_{[mn|}\big[ \Gamma_q e^{\phi} \ul{F}^{(4)}+ \Gamma_q \ul{H}^{(3)}\Gamma^{\ulv} - 3 e^{\phi} \ul{F}^{(4)}_q  - 3 \ul{H}^{(3)}_q\Gamma^{\ulv}              \big]\y\Big] \bigl[ \bar{\y} \Gamma^q \cl{D}_{|p]} \y \bigr] \\
        & + \frac{1}{96} \Big[\bar{\y} \Gamma_{[mn|}\big[2\ul{H}^{(3)}  + \Gamma^{\ulv} e^{\phi} \ul{F}^{(4)}   \big]\y\Big] \bigl[\bar{\y} \Gamma^{\ulv}  \cl{D}_{|p]}\y \bigr].
    \end{split}
\end{equation}\def\y{\sf{y}}A few comments are due, as in the above formulae the 10-dimensional quartic fermionic terms look complicated and have an enormous length. With current understanding, the quartic fermion expansions of type IIA superfields seem unavoidably lengthy, as also seen in \cite{Wulff:2013kga}. We will discuss some promising avenues for improving this quality of these results in what follows. On the other hand, the most prominent feature of these results is their completeness. The robustness and systematicity of the methods we have employed \textit{guarantee} that these are the full and complete quartic fermion terms for the type IIA superfield expansions. This is the first time that some of these terms have been calculated and our results will serve as a foundation for future understanding of such expansions.

\subsubsection*{Avenues to simplification}
Our current expressions for the results for the quartic order fermion terms in the type IIA superfield expansions are unwieldy. It is therefore worthwhile to discuss how they might be made more manageable.

The first thought that might occur is to try and tidy up the large number of `loose' flux terms in the expansions. One would do this by attempting to package these terms up using the operators $D_m$ and $\Delta$ (or combinations thereof) just as everything at second order was packaged neatly. Indeed, this idea is met with some initial success, for example, with a little effort, one can see that three of the terms appearing above in the dilaton shift come together to give
\begin{equation}
\begin{split}\label{simplifexample}
& \frac{1}{128} (\bar{\theta} \Gamma^{\ulv} \Gamma^{mn} \theta) \big( \bar{\theta} \Gamma^{\ulv} \cmttr_{mn}{\theta} \big)  + \frac{1}{48} (\bar{\theta} \Delta \theta)^2 + \frac{1}{576} \Big[ \bar{\theta} \big[ e^{\phi} \ul{F}^{(4)} - 2\ul{H}^{(3)} \Gamma^{\ulv} \big] \theta \Big] (\bar{\theta} \Delta \theta) = \\
= \, & \frac{1}{128} (\bar{\theta}  \Gamma^{\ulv} \Gamma^{mn} \theta)\Big(\bar{\theta}  \Gamma^{\ulv}\big[\cl{D}_m \, , \, \cl{D}_n \big]{\theta}\Big) + \frac{1}{72} (\bar{\theta}\Delta \theta)^2.
\end{split}
\end{equation}
However, reorganizations along these lines often require spotting tricks in the calculations, for example with $\Gamma$-matrix identities, with the symmetry properties of bilinears, and potentially with Fierz identities. It rapidly becomes utterly impractical to hope to significantly reorganize these shifts as they currently stand in this way. We must try and find a better strategy.

We can see in the quadratic and quartic cases that the process of dimensional reduction sharply increases the number and complexity of terms in the expansions. However, dimensional reduction will not generate the capacity for any significant recombination or reorganization of terms all by itself. Any game-changing reorganizational principle for the 10-dimensional quartic terms should be identifiable in the simpler quartic terms in the 11-dimensional description. The most promising line, therefore, is not to try and massage the many terms appearing in ten dimensions, but to return to 11 dimensions and fix them there. The quartic fermion terms in the expansions of the supermetric and super three-form in 11-dimensional supergravity are given in \eqref{quarticm2met} and \eqref{quarticm2form}. We saw in our discussion of the M2-brane that in actuality the only 11-dimensional superfield we need to expand using NORCOR in order to obtain the expansions required for the brane action is the supervielbein $E\ten{_M^A}(Z)$. All the components of the expansion of this superfield that we require to get to quartic fermion order for the M2-brane are given in \eqref{rawresults} in conjucture with \eqref{eq:mix_orders}. Recall that we also performed significant manipulation of the higher-order expansions using Bianchi identities until we arrived at \eqref{massagedresultsone}. We can see then that it is the relative unwieldiness of \textit{these} expressions for components of the NORCOR expansion of the 11-dimensional supervielbein where the vastness of the quartic 10-dimensional terms has its origin. Meaningful rearrangement or simplification of the quartic terms in the type IIA superfield expansion will be identifiable at the level of improvements of \eqref{massagedresultsone}. These improvements have the potential to come from a couple of different lines of reasoning. The most obvious is by improving the application of the Bianchi identities (and litany of other subtle identities that emerge in their combination) when moving from \eqref{rawresults} to \eqref{massagedresultsone}. Another direction might be to improve the NORCOR procedure itself, or making significant geometrical insight there, such that the  left-hand side of \eqref{massagedresultsone} can be made more and more amenable.

Crucial to note, however, is that even with these improvements to the treatment of the 11-dimensional supervielbein, the best subsequent method for obtaining the type IIA quartic terms is still the one we have presented here, when applied to the improved formulation. We will say some more about how the quartic results might be improved once we have explored the next step in our procedure and obtained information about both type II supergravities.

\section{Superspace T-duality and D\texorpdfstring{$\bs{p}$}{p}-brane actions}\label{sec:Tdual}

In this section we complete the task  initiated in section \ref{sec:D2} and provide a systematic method to compute fermion couplings on all D$p$-branes. The method is based on ideas analogous to the ones in section \ref{sec:D2}, and for this reason we will make reference to explanations there when possible to avoid repetition.

Let us briefly summarize the approach. Our proposal relies on two facts. First,  D$p$-branes are solutions of type II supergravities related by T-dualities. Second, fermion couplings on D$p$-brane actions arise naturally in the superspace formulation of the corresponding supergravity theory. For reasons analogous to the ones in the previous section, here we combine those two facts and extend the relation between the T-dual geometries to the superspace level. Using  this  generalization we  find  relations between superfields in curved superspaces that are T-dual to each other.  We use those relations to find the $\theta$-expansions of   superfields appearing in D$p$-brane actions. We already explained that this is equivalent to finding fermion couplings on all D$p$-brane actions.

\subsection{T-duality toolkit}

With the general picture in mind, we can move into the details. In type II theories,\footnote{T-duality is a more general concept in String Theory and it also relates heterotic strings, but here we are interested in type II theories only.} T-duality represents the equivalence of type IIA strings on a background with an isometry along a non-trivial circle $\mathrm{S}^1$ of size $R$ and type IIB strings compactified on another background also with an isometric on a  non-trivial circle $\tilde{\mathrm{S}}^1$, this time with size $\tilde{R} = l_s^2/ R$ (in our conventions, the string length is $l_s = 2 \pi \sqrt{\alpha'}$). We are interested in this underlying structure that connects the two theories. The relations for Neveu-Schwarz fields were first given by Buscher \cite{BUSCHER198759, BUSCHER1988466} and expanded to Ramond-Ramond fields in \cite{Bergshoeff:1\td \td5as}, and then they were extended to fermionic fields in \cite{Hassan:1\td \td9bv, Hassan:1\td \td9mm, Hassan:2000kr}.

\subsubsection{Bosons}
Analogously to the dimensional reduction, we begin with a reminder of the standard T-duality relations for bosonic fields. We take the T-duality $\mathrm{S}^1$-direction to be $x^\td$. Our notation will be the following:  the indices ${m},{n} = 0,\dots,9$ run through all spacetime directions, and the indices $\dot{m},\dot{n} = 0,\dots,8$ through all but the circle $\mathrm{S}^1$, that we take to be $x^\td\sim x^\td+R$. We indicate which fields belong to each theory   by introducing a  tilde for fields in one theory and no adornment of symbols for fields in the other one. All fields are independent of the T-duality direction.  We start by providing the well-known Buscher rules\footnote{Notice that fields are dimensionless in this setup, e.g. $g_{\td \td} = (R/l_s)^2$. Forms therefore have length dimension with the string length $l_s$ as a reference length. Integrals such as $\smash{\int_0^1 \de x^\td \, \sqrt{g_{\td \td}} = R}$ give dimensionful volumes with the appropriate dimension.}
\begin{subequations}\label{eq:buscher-rules}
\begin{align}
    & {{\dual{\phi}}} = {\phi} - \frac{1}{2} \ln {g}_{\td\td}, \\
    & {{\dual{g}}}_{\dot{m}\dot{n}} = {g}_{\dot{m}\dot{n}} - {g}_{\td\td}^{-1}\big( \, {g}_{\dot{m} \td}{g}_{\dot{n} \td} - {B}_{\dot{m} \td}{B}_{\dot{n} \td} \big), \\[0.90ex]
    & {{\dual{g}}}_{\dot{m}\td} =  {g}_{\td\td}^{-1} {B}_{\dot{m}\td}, \\[0.90ex]
    & {{\dual{g}}}_{\td\td} =  {g}_{\td\td}^{-1}, \\[0.90ex]
    & {{\dual{B}}}_{\dot{m}\dot{n}} = {g}_{\dot{m}\dot{n}} - {g}_{\td\td}^{-1}\big({B}_{\dot{m} \td}{g}_{\dot{n} \td} - {g}_{\dot{m} \td}{B}_{\dot{n} \td} \big), \\[0.90ex]
    & {{\dual{B}}}_{\dot{m}\td} =  {g}_{\td \td}^{-1} {g}_{\dot{m}\td}.
\end{align}
\end{subequations}
The Ramond-Ramond gauge potentials are related by the mutually implicative expressions
\begin{subequations}\label{RRTdualrels}
\begin{align}
    & \dual{C}^{(n)}_{\td \dot{m}_2 \dots \dot{m}_n} = {{C}}^{(n-1)}_{\dot{m}_2 \dots \dot{m}_n} - (n-1) \, g_{\td\td}^{-1} g_{\td [ \dot{m}_2} {C}^{(n-1)}_{|\td|\dot{m}_3 \dots \dot{m}_{n}]}, \\
    & \dual{C}^{(n)}_{\dot{m}_1 \dots \dot{m}_n} = {C}^{(n+1)}_{\td \dot{m}_1 \dots \dot{m}_n} - n \, B_{\td [\dot{m}_1} {C}^{(n-1)}_{\dot{m}_2 \dots \dot{m}_{n}]} + n (n-1) \, g_{\td \td}^{-1} g_{\td [\dot{m}_1|} B_{\td | \dot{m}_2} {C}^{(n-1)}_{\dot{m}_3 \dots \dot{m}_{n}]}.
\end{align}
\end{subequations}

\subsubsection{Spinors, supersymmetry operators, and spinor doublet notation}\label{spinordoubletnotation}

When fermions are involved, T-duality becomes somewhat more subtle and complicated. The groundwork for the treatment of fermions under T-duality is represented by the Hassan rules \cite{Hassan:1\td \td9bv, Hassan:1\td \td9mm, Hassan:2000kr}.

The intricate world of fermion T-duality begins with making an   observation concerning the T-duality rules for fields in the Neveu-Schwarz sector: there are two different vielbeins that are dual to the original one. Properly dealing with this fact requires  the introduction of some extra structure. We denote the `initial' vielbein as $e\ten{_a^m}$, and the two possible `final' dual vielbeins as $(\dual{e}_+)\ten{_a^m}$ and $(\dual{e}_-)\ten{_a^m}$. Both choices give the correct T-dual metric. The initial and final vielbeins are related according to the T-duality rules
\begin{subequations}\label{eq:vielbeins}
   \begin{align}
       & (\dual{e}_{\pm})\ten{_a^n} = e\ten{_a^m}(Q_{\pm})\ten{_m^n}, \\
       & (\dual{e}_{\pm})\ten{_m^a} = (Q_{\pm}^{-1})\ten{_m^n} e\ten{_n^a},
   \end{align}
\end{subequations}
where we have defined
\begin{subequations}
    \begin{align}
        & (Q_\pm)\ten{_m^n} = \matr{\delta\ten{_{\dot{m}}^{\dot{n}}}}{\mp (g_{\td \dot{m}} \pm B_{\td \dot{m}})}{0}{\mp g_{\td \td}}, \\
        & (Q_\pm^{-1})\ten{_m^n} = \matr{\delta\ten{_{\dot{m}}^{\dot{n}}}}{- g_{\td \td}^{-1} (g_{\td \dot{m}} \pm B_{\td \dot{m}})}{0}{\mp g_{\td \td}^{-1}}.
    \end{align}
\end{subequations}
Notice that $ (\dual{Q}^{-1}_\pm)\ten{_m^n} = (Q_\pm)\ten{_m^n}$. The two vielbeins $(\dual{e}_{+})\ten{_m^a}$ and $(\dual{e}_{-})\ten{_m^a}$ are related to one another by a local Lorentz transformation as $(\dual{e}_{+})\ten{_m^a} = \Lambda\ten{^a_b}(\dual{e}_{-})\ten{_m^b}$, with $\Lambda\ten{^a_b} = e\ten{_b^m} (Q_-)\ten{_m^p} (Q_+^{-1})\ten{_p^n} e\ten{_n^a}$. This is irrelevant for the Lorentz-invariant quantities in the bosonic analysis, but it plays a vital role when considering fermions. For example there are now two choices for $\Gamma$-matrices in the dual theory, i.e.
\begin{equation}\label{eq:gamma_matrices_t_duality1}
    (\dual{\Gamma}_{\pm})_m = (Q_{\pm}^{-1})\ten{_m^n} \Gamma_n = (\tilde{e}_\pm)\ten{_m^a} \Gamma_a.
\end{equation}
These are naturally related by a spinorial representation $\Omega$ of the Lorentz transformation $\Lambda$, defined via $\Omega \, \Gamma_a \, \Omega^{-1} = \Gamma_b \, (\Lambda^{-1}) \, \ten{^b_a}$, as
\begin{equation}
    \Omega \, (\dual{\Gamma}_+)_m \, \Omega^{-1} = (\dual{\Gamma}_{-})_m.
\end{equation}
It can be determined that this matrix reads (also notice it squares as $\Omega^2 = -1$)
\begin{equation}
    \Omega = \dual{\Omega} = \frac{1}{\sqrt{g_{\td\td}}} \, \Gamma^{\clty} \Gamma_\td.
\end{equation}
The extra complication when T-dualizing objects that are sensitive to the difference between the two choices of vielbein, such as spinors and $\Gamma$-matrices, is that for self-consistency it is necessary that all Lorentz tensors in the dual theory are computed with respect to the same vielbein. We will choose $\smash{(\dual{e}_{-})\ten{_n^a}}$ as our reference dual vielbein. Let us point out that this does not imply that we will write all duality relations using $Q_-$: we will often find it convenient to transform objects using $Q_+$ and then perform Lorentz transformations.

With these tools in hand, we are in principle ready to provide all of the rules for fermion T-dualization introduced by Hassan. Before doing so, however, we   introduce a new notation that  allows us to perform computations in a clean and compact way: the  spinor doublet notation. The spinor doublet notation we   introduce   has   differences to the ones found in the literature, e.g. in \cite{Marolf:2003vf,Marolf:2003ye,Martucci:2005rb,Lust:2008zd,Grana:2020hyu}. These differences will make performing the necessary T-duality computations cleaner. The motivation for this new notation is the following: in type II theories spinors come in doublets of Majorana-Weyl spinors. In type IIA theese have opposite chirality whereas in IIB they have the same chirality, which we take to be positive for the gravitinos and supersymmetry parameters, and negative for dilatinos. It is therefore convenient to use spinor doublet in the latter in order to write most combinations, such as fermion bilinears, in a compact way. We define the IIB doublets 
\begin{equation}
    \epsilon^\B = \vect{\epsilon_1}{\epsilon_2}, \qquad \psi_m^\B = \vect{\psi_{1 m}}{\psi_{2 m}}, \qquad \lambda^\B = \vect{\lambda_1}{\lambda_2}.
\end{equation}
It is also convenient to do the same in the type IIA theory. In this case, we must bear in mind that chirality plays a crucial role in organizing fermion bilinears in this theory, and so we need to use chirality as an organizing principle. Our convention will be to have positive chirality fermions on the top of type IIA fermion doublets. We can now define
\begin{equation}
    \epsilon^\A = \vect{\epsilon_+}{\epsilon_-}, \qquad \psi_m^\A = \vect{\psi_{+ m}}{\psi_{- m}}, \qquad \lambda^\A = \vect{\lambda_+}{\lambda_-}.
\end{equation}
Given these doublets, the natural matrices that act on them can always be written in terms of the 2-dimensional identity $1_2$ and the Pauli matrices $\sigma^1, \sigma^2$, $\sigma^3$. This also comes with further implications. For instance,  chirality matrices in type IIA theory can always be replaced by $\sigma^3$ in our conventions, for instance as in $\Gamma^\clty \epsilon^\A = \sigma^3 \epsilon^\A$. Also, to account for the fact that multiplications by a $\Gamma$-matrix flip chiralities, one must introduce a $\sigma^1$ matrix for each $\Gamma$-matrix when moving to the spinor doublet notation from the one in the previous section. The appearance of multiple Pauli matrices in this notation change can make fomulae more complicated to read. In order to make them more readable, we compute the product of Pauli matrices and just give the resulting one, such that all other operators appearing in the expressions now come with $1_2$. For example, the type IIA product $\Gamma_m \Gamma^{\ulv} \epsilon$ leads to $\smash{(\sigma^1 \otimes \Gamma_m) \sigma^3 \epsilon^\A = (1_2\otimes\Gamma_m) (-i\sigma^2) \epsilon^\A=(-i\sigma^2) \otimes \Gamma_m \epsilon^\A}$ in our doublet notation. We will omit `$\otimes$' symbols from now on.  Hence operators implicitly come with $1_2$.  We will also write $\mathbb{\Gamma}_m = 1_2 \otimes \Gamma_m$. In type IIB strings, chirality cannot be used as an organizing principle, instead the Pauli-matrix structure is inherited from type IIA.

As clarifying examples, and because they will be useful for later purposes, we provide here the second-order truncated superfields (\ref{shiftsstart} - \ref{shiftsend}) that appeared in the D2-brane action with fermion bilinears written in this notation. These are
\begin{align}
    & \boldsymbol{g}_{mn}  = g_{mn} - i \, \bar{\theta}^\A \sigma^1 \ds{\Gamma}_{(m} \ds{D}_{n)}^\A \theta^\A,  \label{eq:metric_in_double_spinor_nota}\\[0.75ex]
    & \boldsymbol{\phi} = \phi - \dfrac{i}{4} \, \bar{\theta}^\A \ds{\Delta}^\A \theta^\A, \\[0.95ex]
    & \boldsymbol{B}_{mn} = B_{mn} - i \, \bar{\theta}^\A (i\sigma^2) \ds{\Gamma}_{[m} \ds{D}_{n]}^\A \theta^\A, \\[0.75ex]
    & \boldsymbol{C}_{m}  = C_m - \dfrac{i}{2} \, e^{-\phi} \, \bar{\theta}^\A \sigma^3 \Bigl( \ds{D}_m^\A - \frac{1}{2} \sigma^1 \ds{\Gamma}_m \ds{\Delta}^\A \Bigr) \theta^\A, \label{eq:c1-double-spinor}\\
    & \boldsymbol{C}'_{mnp}  = C_{mnp} - \dfrac{i}{2} \, e^{-\phi} \, \bar{\theta}^\A  \Bigl( 3 \ds{\Gamma}_{[mn} \ds{D}_{p]}^\A  - \dfrac{1}{2} \sigma^1 \ds{\Gamma}_{mnp} \ds{\Delta}^\A  \Bigr) \theta^\A  - 3 i \, C_{[m} \, \bar{\theta}^\A (i \sigma^2) \ds{\Gamma}_{n} \ds{D}_{p]} ^\A \theta^\A.\label{eq:c3-prime}
\end{align}
In order to write the superfields, we used the operators appearing in the  type IIA   gravitino and dilatino supersymmetry variations, that  in the spinor doublet notation are
\begin{align}
    & \delta_\epsilon \psi_m^\A = \ds{D}_m^\A \epsilon^\A, \\
    & \delta_\epsilon \lambda^\A = \ds{\Delta}^\A \epsilon^\A,
\end{align}
with
\begin{align}
    & \ds{D}_m^\A \equiv 1_2   \nabla_m + \dfrac{1}{4} \, \sigma^3  \underline{H}_m^{(3)} - \dfrac{1}{8} \, e^{\phi} \, \bigl[ i \sigma^2  \underline{F}^{(2)} + \sigma^1  \underline{\F}^{(4)} \bigr] \Gamma_m, \\
    & \ds{\Delta}^\A \equiv \sigma^1  \underline{\der} \phi + \dfrac{1}{2} \, i \sigma^2  \underline{H}^{(3)} - \dfrac{1}{8} \, e^\phi \, \Gamma^m \bigl[ \sigma^3  \underline{F}^{(2)} + 1_2  \underline{\F}^{(4)} \bigr] \Gamma_m.
\end{align} 
We also need to define the equivalent operators in type IIB. For doing so we first give the supersymmetry variations in the spinor doublet notation
\begin{align}
    & \delta_\epsilon \psi_m^\B = \ds{D}_m^\B \epsilon^\B, \\
    & \delta_\epsilon \lambda^\B = \ds{\Delta}^\B \epsilon^\B,
\end{align}
and this time
\begin{align}
    & \ds{D}_m^\B \equiv 1_2  \nabla_m + \dfrac{1}{4} \, \sigma^3  \underline{H}_m^{(3)} + \dfrac{1}{8} \, e^{\phi} \, \Bigl[ i \sigma^2  \Bigl( \underline{F}^{(1)} + \underline{F}^{(5)} \Bigr) + \sigma^1  \underline{\F}^{(3)} \Bigr] \Gamma_m, \\
    & \ds{\Delta}^\B \equiv \sigma^1  \underline{\der} \phi + \dfrac{1}{2} \, i \sigma^2  \underline{H}^{(3)} + \dfrac{1}{8} \, e^\phi \, \Gamma^m \Bigl[ \sigma^3  \Bigl( \underline{F}^{(1)} +   \underline{F}^{(5)} \Bigr) +   1_2  \underline{\F}^{(3)} \Bigr]\Gamma_m.
\end{align}

Now we have to express the basic T-duality relations in this spinor doublet notation. The extensions of $Q_\pm$ and $\Omega$ can be simply achieved by defining
\begin{equation}
    (\ds{Q}_\pm)\ten{_m^n} = \matr{(Q_\pm)\ten{_m^n}}{0}{0}{(Q_\mp)\ten{_m^n}}
\end{equation}
and
\begin{equation}
    \Upsilon = \matr{1}{0}{0}{\Omega}. 
\end{equation}
These definitions allow us to extend the T-duality rules for many objects to the spinor doublet notation which will be used later on. For instance, once we take $(\dual{e}_{-})\ten{_m^a}$ as the reference frame in the dual theory, the $\Gamma$-matrix rule can be manipulated to give
\begin{equation}\label{eq:useful_identity_zzz}
    \tilde{\mathbb{\Gamma}}_m = (\ds{Q}_-^{-1})\ten{_m^n} \Upsilon \ds{\Gamma}_n \Upsilon^{-1}.
\end{equation}
Adapting the notation of \cite{Hassan:1\td \td9bv, Hassan:1\td \td9mm, Hassan:2000kr, Marolf:2003vf,Marolf:2003ye}  to our conventions, spinors in type IIA and type IIB theories are related to each other by the T-duality rules
\begin{align}
    & \epsilon^\B = \Upsilon \epsilon^\A,  \label{eq:epsilon_relation} \\
    & \psi_m^\B = (\ds{Q}_+^{-1})\ten{_m^n} \Upsilon \psi_n^\A, \label{eq:gravitino-t-duality} \\
    & \lambda^\B = (\sigma^1 \Upsilon \sigma^1) \bigl[ \lambda^\A - 2 \, g_{\td \td}^{-1} \sigma^1 \ds{\Gamma}_\td \, \psi_\td^\A \bigr],\label{eq:dilatino-t-duality}
\end{align}
Related to the above formaulae, it is convenient to define the Dirac conjugate doublets because these appear in fermion bilinears. Based on chirality arguments above this is $\bar{\epsilon}^\A = (\bar{\epsilon}_-, \bar{\epsilon}_+)$ for type IIA and we extend the structure to IIB by defining $\bar{\epsilon}^\B = (\bar{\epsilon}_2, \bar{\epsilon}_1)$. The T-duality relation between them is $\bar{\epsilon}^\B = \bar{\epsilon}^\A \sigma^1 \Upsilon^{-1} \sigma^1$.

A point worth making here is that if we invert the  relations above, the outcome is similar but involves $\Upsilon^{-1}$, instead of $\Upsilon$ itself, so there is a slight difference between going from type IIA to type IIB or taking the opposite route. This did not happen for bosonic fields above, where the relations found worked the same regardless of the direction taken to perform the duality. To conclude, the T-duality rules between the supersymmetry operators read
\begin{subequations}
\begin{align}
    & \ds{D}_m^\B = (\ds{Q}_+^{-1})\ten{_m^n} \Upsilon \ds{D}_n^\A \Upsilon^{-1}, \label{eq:supercovded-t-duality} \\
    & \ds{\Delta}^\B = \sigma^1 \Upsilon\sigma^1 \bigl[ \ds{\Delta}^\A - 2 \, g_{\td \td}^{-1} \sigma^1 \ds{\Gamma}_\td \ds{D}_\td^\A \bigr] \Upsilon^{-1} \label{IIA2IIBtwo},
\end{align}
\end{subequations}
The above results are in precise agreement with the existing literature. As should be apparent, the spinor doublet notation approach we have employed here is highly successful in compactly capturing the T-duality relationships for the fermions and supersymmetry variations in type IIA and type IIB supergravity.


\subsubsection[T-duality and bosonic D$p$-branes]{T-duality and bosonic D\texorpdfstring{$\bs{p}$}{p}-branes}\label{sec:formalTdual}

We will now review how bosonic D$p$-brane actions are related to each other under T-duality. This is instrumental in explaining our superspace approach below. In general, the basic idea is that T-dualising a theory with a D$p$-brane produces a theory with a D$(p\pm1)$-brane, depending on whether the original brane wraps the T-duality circle $\mathrm{S}^1$ or not. This is consistent with the fact that type IIA and type IIB theories are exchanged, as the former only admits even-$p$ branes and the latter only odd-$p$ ones. Starting from the bosonic D2-brane action,  one can repeatedly T-dualise the theory to infer that the bosonic action of a generic D$p$-brane is
\begin{equation} \label{general_bosonic_Dp-brane_action}
    S_{\text{D}p}^{(0)} = - T_{\text{D}p} \int \de^{p+1}\zeta \; e^{-\phi} \sqrt{ -\det \, (g + f)} + T_{\text{D}p} \int \; C \, e^{-f},
\end{equation}
where the brane tension is $T_{\text{D}p} = 2 \pi / l_s^{p+1}$. All bulk fields are pulled-back onto the brane worldvolume. The WZ-term contains a formal sum $\smash{C = \sum_q C^{(q)}}$ over forms of all degrees and we let the integral pick out the appropriate forms each time. 

In order to show in some detail how the machinery of T-duality works for D$p$-branes, we consider a bosonic D$p$-brane wrapping the T-duality circle $\mathrm{S}^1$ in the direction $x^\td$ and, with simple manipulations, we integrate its action over the circle $\mathrm{S}^1$ to obtain the action of the dual D$(p-1)$-brane that is localized on the dual circle. The initial D$p$-brane wraps a $(p+1)$-cycle $\Sigma_{p+1}$ that is an $\mathrm{S}^1$-fibration over $\Sigma_p$, which is the cycle wrapped by the final D$(p-1)$-brane. Indices $k = 0, \dots, p-1, 9$ span the D$p$-brane worldvolume and indices $\dot{k} = 0, \dots, p-1$ are parallel to the D$(p-1)$-brane, excluding the direction $x^\td$. For simplicity, we fix the static gauge for the brane embedding, with all fields independent of the $\mathrm{S}^1$-direction. For clarity, we manipulate the DBI- and the WZ-terms of the action separately. The presentation here is sketchy and we refer the interested reader to \cite{Bergshoeff:1996cy, Myers:1999ps} for further details. 

First, we deal with the DBI-action. Integrating over the circle $\mathrm{S}^1$ goes as
\begin{equation}
\begin{split}
    S_{\text{D}p}^{\text{DBI} } & = - T_{\text{D}p} \int_{\Sigma_{p+1}} \de^{p+1}\zeta \; e^{-{\phi}} \sqrt{- \det \, \bigl[ (g + f)_{kl} \bigr]} \\
    & = - T_{\text{D}p} \int_{\Sigma_{p}} \de^{p} \zeta \int  \de \zeta^\td \; e^{- \phi} \, \sqrt{g_{\td \td}} \, \sqrt{- \det \, \bigl[ (g + f)_{\dot{k}\dot{l}} - g_{\td \td}^{-1} (g + f)_{\dot{k} \td} (g + f)_{\td \dot{l}} \bigr]} \\
    & = - \tilde{T}_{\text{D}(p-1)} \int_{\Sigma_{p}} \de^{p} \zeta \; e^{-\tilde{\phi}} \, \sqrt{- \det \, \bigl[ (\tilde{g} + \tilde{f})_{\dot{k}\dot{l}} \bigr]}.
\end{split}
\end{equation}
To achieve this, we express the determinant of the block matrix singling out the $\mathrm{S}^1$-direction.  Organizing the resulting formula as shown, one can identify  the combinations appearing in the Buscher rules \eqref{eq:buscher-rules}, so the  integrand after this manipulations has the appropriate shape to be the DBI-part of the D$(p-1)$-brane in the dual background to the initial one. Also, the result of the integration over the circle transforms the D$p$-brane tension leading to the D$(p-1)$-brane tension, i.e. $T_{\text{D}p} l_s = \tilde{T}_{\text{D}(p-1)}$. So the outcome of these manipulations is the DBI-term in the resulting D$(p-1)$-brane action, also in the static gauge, as expected. For later purposes, we emphasize that this computation provides an alternative derivation of the Buscher rules \eqref{eq:buscher-rules}.

We can proceed analogously for the WZ-term. One can observe that the $\mathrm{S}^1$-integration gives
\begin{equation}
\begin{split}
    S_{\text{D}p}^{\text{WZ}} & = T_{\text{D}p} \int_{\mathrm{S}^1} \int_{\Sigma_p} \de \zeta^\td \wedge \de \zeta^{\dot{l}_1} \wedge \dots \wedge \de \zeta^{\dot{l}_{p}} \; \dfrac{p+1}{(p+1)!} \, \bigl(  C \wedge e^{-f} \bigr)_{\td \dot{l}_1 \dots \dot{l}_p} \\
    & = \tilde{T}_{\text{D}(p-1)} \int_{\Sigma_p} \de \zeta^{\dot{l}_1} \wedge \dots \wedge \de \zeta^{\dot{l}_p} \; \dfrac{1}{p!} \, \bigl( \tilde{C} \wedge e^{-\tilde{f} } \bigr)_{\dot{l}_1 \dots \dot{l}_p} .
\end{split}
\end{equation}
To achieve this, we expand the integrand and upon performing the integral over $\zeta^\td$ we recognize in it the D$(p-1)$-brane WZ-term with the appropriate charge. Similarly to the case of the Neveu-Schwarz sector, we note that this reduction provides an alternative approach to obtain the T-duality rules for the Ramond sector written as $\smash{(\tilde{C} \, e^{-\tilde{B}})^{(n)}_{\td \dot{m}_2 \dots \dot{m}_n} = (C \, e^{-B})^{(n-1)}_{\dot{m}_2 \dots \dot{m}_n}}$ and $\smash{(\tilde{C} \, e^{-\tilde{B}})^{(n)}_{\dot{m}_1 \dots \dot{m}_n} = (C \, e^{-B})^{(n+1)}_{\td \dot{m}_1 \dots \dot{m}_n}}$, that can be manipulated to give \eqref{RRTdualrels}. 

This completes our review of the behaviour of the bosonic brane actions under T-duality. One should notice a fundamental fact: T-duality maps the DBI- and WZ-actions of a D$p$-brane into the DBI- and WZ-actions of a D$(p-1)$-brane, respectively, and there is no mixing among the two in the transformation. A similar calculation to the one above may be engineered to move from a D$p$-brane to a D$(p+1)$-brane. 

\subsection{A useful rearrangement}
We just showed how to obtain all the bosonic D$p$-brane actions by T-dualizing the bosonic D2-brane one. Moreover, in the superspace formulation, the structure of the D2-brane action is formally the same both at zeroth order and in superspace at any fermionic order. Therefore, the structure of fermion couplings on all D$p$-branes just follows from the D2-brane one. Because our goal is to compute these fermionic couplings for all D$p$-branes, here we present a useful rearrangement that simplifies the computation of such couplings. In fact, because the fermion couplings are inherited from the superfield expansions appearing on the brane, the rearrangement is a smart manipulation of the superfields appearing on the D2-brane action that will simplify the computation of those appearing in the rest of D$p$-branes. 

In section \ref{sec:D2}, we defined the promoted Ramond-Ramond three-form field in type IIA with a prime symbol. That is the standard three-form superfield obtained from dimensional reduction of 11-dimensional supergravity. Rather than working with that superfield, it will be convenient to work with a related one. We define a new \textit{unprimed} three-form superfield as
\begin{equation} \label{trick}
    \bs{{C}}_{mnp} = \bs{C}'_{mnp} - 3 \, \bs{C}_{[m} (\bs{B}_{np]} - B_{np]}).
\end{equation}
From here on we will work using this unprimed three-form rather than the standard one. This new superfield is such that the last term in the superfield $\bs{C}'_{mnp}$ in \eqref{eq:c3-prime} is removed, and at order $(\theta)^2$ it reads
\begin{equation}
    \boldsymbol{C}_{mnp}  = C_{mnp} - \dfrac{i}{2} \, e^{-\phi} \, \bar{\theta}^\A  \Bigl( 3 \ds{\Gamma}_{[mn} \ds{D}_{p]}^\A  - \dfrac{1}{2} \sigma^1 \ds{\Gamma}_{mnp} \ds{\Delta}^\A  \Bigr) \theta^\A .
\end{equation}
The reason why we defined this rearrangement is easily explained: the super-D2-brane action now reads
\begin{equation}\label{D2finalform}
    \boldsymbol{S}_{\text{D}2} = - T_{\text{D}2} \int \de^3 \zeta \; e^{-\bs{\phi}} \, \sqrt{- \det \, (\bs{g} + \bs{f})} +  \frac{T_{\text{D}2}}{6} \int \de^3\zeta \; \epsilonconv (\bs{{C}}_{ijk} - 3 \, \bs{C}_{i} {f}_{jk}).
\end{equation}
In other words, we have engineered a superspace action where the Neveu-Schwarz fields appear as superfields in the DBI-term but only as bosonic fields in the WZ-term. Ramond-Ramond fields instead appear as superfields in the WZ-term. From the discussion in section \ref{sec:formalTdual} we conclude that this combination of fields and superfields will hold for any D$p$-brane if we obtain the brane superspace actions by T-dualizing this one.

\subsection[Superspace T-duality and fermions on D${p}$-branes]{Superspace T-duality and fermions on D\texorpdfstring{$\bs{p}$}{p}-branes  }
We will once again be following the reasoning of the example already laid out with dimensional reduction in section \ref{sec:D2}. We interpret the bosonic T-duality relations \eqref{eq:buscher-rules} and \eqref{RRTdualrels} as the zeroth-order terms in the fermionic expansions of superspace T-duality relationships and extend them to superspace relations. T-duality in the context of full superfields was also discussed in \cite{Bandos:2003bz}.

Now, since T-duality maps the DBI-action of D$p$-branes into the DBI-action of D$(p\pm1)$-branes, and since this mapping allows one to derive the Buscher rules (\ref{eq:buscher-rules}), one can simply conclude that the Buscher rules for the Neveu-Schwarz fields in superspace read
\begin{subequations}\label{NSTdual}
    \begin{align}
        & {\bs{\dual{\phi}}} = \bs{\phi} - \frac{1}{2} \ln \bs{g}_{\td\td}, \\
        & {\bs{\dual{g}}}_{\dot{m}\dot{n}} = \bs{g}_{\dot{m}\dot{n}} - \bs{g}_{\td\td}^{-1}\big( \, \bs{g}_{\dot{m} \td}\bs{g}_{\dot{n} \td} - \bs{B}_{\dot{m} \td}\bs{B}_{\dot{n} \td} \big), \\[0.90ex]
        & {\bs{\dual{g}}}_{\dot{m}\td} =  \bs{g}_{\td\td}^{-1} \bs{B}_{\dot{m}\td},  \\[0.90ex]
        & {\bs{\dual{g}}}_{\td\td} =  \bs{g}_{\td\td}^{-1}, \label{exampleoneTdual} \\[0.90ex]
        & {\bs{\dual{B}}}_{\dot{m}\dot{n}} = \bs{g}_{\dot{m}\dot{n}} - \bs{g}_{\td\td}^{-1}\big(\bs{B}_{\dot{m} \td}\bs{g}_{\dot{n} \td} - \bs{g}_{\dot{m} \td}\bs{B}_{\dot{n} \td} \big), \\[0.90ex] 
        & {\bs{\dual{B}}}_{\dot{m}\td} =  \bs{g}_{\td \td}^{-1} \bs{g}_{\dot{m}\td}.
    \end{align}
\end{subequations}
Some of these rules partially   appeared in \cite{Martucci:2003gc}, where they found    the T-duality relation between Green-Schwarz superstrings in type IIA and type IIB with fermionic expansions up to quadratic terms.

Similarly, T-duality maps the WZ-action of D$p$-branes into the WZ-action of D$(p\pm1)$-branes and this mapping allows one to derive the T-duality rules for Ramond-Ramond fields (\ref{RRTdual}). Because in the WZ-action of (\ref{D2finalform}) the Neveu-Schwarz field appear only bosonically and the Ramond-Ramond fields appear as superfields, we conclude that the Ramond-Ramond T-duality rules we will use are 
\begin{subequations} \label{RRTdual}
\begin{align}
    & {\bs{\dual{C}}}^{(n)}_{\td \dot{m}_2 \dots \dot{m}_n} = {\bs{C}}^{(n-1)}_{\dot{m}_2 \dots \dot{m}_n} - (n-1) g_{\td\td}^{-1} g_{\td [ \dot{m}_2} \bs{C}^{(n-1)}_{|\td|\dot{m}_3 \dots \dot{m}_{n}]}, \label{exampleTdualtwo} \\
    & {\bs{\dual{C}}}^{(n)}_{\dot{m}_1 \dots \dot{m}_n} = \bs{C}^{(n+1)}_{\td \dot{m}_1 \dots \dot{m}_n} - n \, B_{\td [\dot{m}_1} {\bs{C}}^{(n-1)}_{\dot{m}_2 \dots \dot{m}_{n}]} + n (n-1) \, g_{\td \td}^{-1} g_{\td [\dot{m}_1|} B_{\td | \dot{m}_2} {\bs{C}}^{(n-1)}_{\dot{m}_3 \dots \dot{m}_{n}]}.
\end{align}
\end{subequations}
This mechanism was used in \cite{Marolf:2003vf} for the quadratic fermionic action and we have extended that observation to any fermionic order. Note that without our manipulation on the super-three-form, one would have obtained similar results involving Neveu-Schwarz superfields rather than fields. Those are the actual  superspace T-duality rules for Ramond-Ramond superfields, but for our purposes it will be more convenient to use \eqref{RRTdual}. 

\subsubsection[Order-$(\theta)^2$ terms]{Order-\texorpdfstring{$\boldsymbol{(\theta)^2}$}{theta2} terms}
In the following, we will use the promoted T-duality relations  \eqref{NSTdual} and  \eqref{RRTdual} to calculate the second-order fermionic expansions of all the superfields that appear in type IIA and type IIB   under repeated T-dualizations. Just as in section \ref{ord2sec5}, we will provide illuminating examples of the necessary calculations before listing the full results.

\subsubsection*{Example: type IIB metric}
We will now use the simplest superspace T-duality relationships in order to provide an example of how to obtain the fermionic expansions of type IIB operators from type IIA description by using the conventional T-duality rules applied to quadratic fermionic quantities. We will focus on the supermetric.

Consider the superspace T-duality rule \eqref{exampleoneTdual}, i.e. ${\bs{\dual{g}}}_{\td \td} =  \bs{g}_{\td \td}^{-1}$. Starting from the type IIA supermetric $\boldsymbol{g}_{mn}$, in order to determine an expression for the quadratic fermionic expansion of the type IIB supermetric $\boldsymbol{\tilde{g}}_{mn}$, we Taylor-expand both sides, concentrating on the components of interest. On the type IIB left-hand side, we set the ansatz $\bs{\dual{g}}_{\td \td} = \dual{g}_{\td\td} + \dual{\gamma}_{\td\td}$, whereas on the type IIA right-hand side we use the result of the dimensional reduction \eqref{eq:metric_in_double_spinor_nota}. Using the spinor doublet notation and keeping only the second-order fermion terms from both sides (as the zeroth-order terms just reproduce the bosonic identities), one determines an expression for the type IIB shift $\dual{\gamma}_{\td\td}$ in terms of type IIA quantities, i.e.
\begin{equation}
    \dual{\gamma}_{\td\td} = i {g_{\td\td}^{-2}} \bar{\theta}^\A \sigma^1 \ds{\Gamma}_{\td} \ds{D}_{\td}^\A \theta^\A.
\end{equation}
We are now required to perform conventional T-duality on the term on the right-hand side in order to determine an expression for the expansion ansatz of the type IIB metric in terms of type IIB quantities.  We can use the basic T-duality rules in spinor doublet notation in subsection \ref{spinordoubletnotation} to write
\begin{equation}
    \begin{split}
        \dual{\gamma}_{\td\td} & = i \dual{g}_{\td\td}^2 \, (\bar{\theta}^\B \sigma^1 \Upsilon \sigma^1) \, \sigma^1 \bigr[ (\tilde{\ds{Q}}_-^{-1})\ten{_\td^p} \Upsilon^{-1} \tilde{\ds{\Gamma}}_p \Upsilon \bigr] \bigl[ (\tilde{\ds{Q}}_+^{-1})\ten{_\td^q} \Upsilon^{-1} {\ds{D}}_q^\B \Upsilon \bigr] \Upsilon^{-1} \theta^\B \\
        & = i \dual{g}_{\td\td}^2 \, \bar{\theta}^\B \sigma^1 (\tilde{\ds{Q}}_-^{-1})\ten{_{\td}^p} (\tilde{\ds{Q}}_+^{-1})\ten{_{\td}^q} \tilde{\ds{\Gamma}}_p {\ds{D}}_q^\B \theta^\B \\
        & = i \dual{g}_{\td\td}^2 \, \bar{\theta}^\B \sigma^1  (- \dual{g}_{\td\td}^{-1} \sigma^3) (\dual{g}_{\td\td}^{-1} \sigma^3) \tilde{\ds{\Gamma}}_{\td} {\ds{D}}_{\td}^\B \theta^\B \\
        & = - i \, \bar{\theta}^\B \sigma^1   \tilde{\ds{\Gamma}}_{\td} {\ds{D}}_{\td}^\B \theta^\B.
    \end{split}
\end{equation}
The result is exactly as expected. The quadratic terms in the expansions of the type IIB metric take precisely the same form as the type IIA metric, just with all of the operators and spinors being the type IIB ones and not the type IIA versions. One can proceed analogously to get the generic second-order shift of the   type IIB   dilaton and Kalb-Ramond superfields.

\subsubsection*{Example: Ramond-Ramond two-form}
The superspace promotion of the dimensional reduction from 11-dimensional supergravity to type IIA supergravity allowed us to determine the fermionic expansions for the Ramond-Ramond superfields of degrees one and three. Now that we are considering T-duality between type IIA and type IIB, we must confront the requirement that we calculate the fermionic expansions of Ramond-Ramond superfields of any degree. 

Our strategy will be  to take the promoted Ramond-Ramond T-duality rule \eqref{exampleTdualtwo}, expand in orders of fermions and keep only the quadratic contribution. Writing $\bs{C}^{(q)} = C^{(q)} + \chi^{(q)}$, where $\chi^{(q)}$ is the corresponding fermion bilinear, we are interested in obtaining $\chi^{(2)}$. Following our standard procedure, from \eqref{exampleTdualtwo} we find
\begin{equation}\label{preTdual}
    \begin{split}
        \dual{\chi}^{(2)}_{\td \dot{m}} & = \chi^{(1)}_{\td \dot{m}} - g_{\td \td}^{-1} g_{\td \dot{m}} \chi^{(1)}_{\td} \\
        & = - \dfrac{i}{2} \, e^{-\phi} \, \bar{\theta}^\A \sigma^3\biggl[  \bigl( \ds{D}^{\A}_{\dot{m}} - g_{\td\td}^{-1} g_{\td \dot{m}} \ds{D}_\td^{\A} \bigr) - \dfrac{1}{2} \,  \bigl( \ds{\Gamma}_{\dot{m}} - g_{\td\td}^{-1} g_{\td \dot{m}} \ds{\Gamma}_\td \bigr) \sigma^1 \ds{\Delta}^\A \biggr] \theta^\A
    \end{split}
\end{equation}
where for the one-form shift we have made use of \eqref{eq:c1-double-spinor}. We now need to manipulate the right-hand side in order to obtain an expression for the type IIB Ramond-Ramond two-form superfield in terms of type IIB operators. We will use  identities similar to \eqref{eq:useful_identity_zzz}
\begin{align}
    & \Upsilon \bigl( \ds{D}^{\A}_{\dot{m}} - g_{\td\td}^{-1} g_{\td \dot{m}} \ds{D}_\td^{\A} \bigr) \Upsilon^{-1} = \ds{D}^{\B}_{\dot{m}} - \dual{g}_{\td\td}^{-1} \dual{g}_{\td \dot{m}} {\ds{D}}_\td^{\B}, \label{usefulob1} \\
    & \Upsilon \bigl( \ds{\Gamma}_{\dot{m}} - g_{\td\td}^{-1} g_{\td \dot{m}} \ds{\Gamma}_\td \bigr) \Upsilon^{-1} = \dual{\ds{\Gamma}}_{\dot{m}} - \dual{g}_{\td\td}^{-1} \dual{g}_{\td \dot{m}} \dual{\ds{\Gamma}}_\td. \label{usefulob2}
\end{align}
and $\sqrt{\dual{g}_{\td\td}} \, \sigma^1 \Upsilon \sigma^1 \sigma^3 \Upsilon^{-1} = \Gamma^\clty \dual{\ds{\Gamma}}_{\td}$. Splitting the first and the second term in \eqref{preTdual}, using \eqref{usefulob1} and \eqref{usefulob2}, we find
\begin{align}
    & e^{-\phi} \, \bar{\theta}^\A \sigma^3  \big( \ds{D}^{\A}_{\dot{m}} - g_{\td\td}^{-1} g_{\td \dot{m}} \ds{D}_\td^{\A}\big) \theta^\A = e^{-\dual{\phi}}  \, \bar{\theta}^\B \Gamma^\clty  \big( \dual{\ds{\Gamma}}_\td {\ds{D}}^{\B}_{\dot{m}} - \dual{g}_{\td\td}^{-1} \dual{g}_{\td \dot{m}} \dual{\ds{\Gamma}}_\td {\ds{D}}_\td^{\B}\big)  \theta^\B, \label{c1c2term1} \\
    & e^{-\phi} \, \bar{\theta}^\A \sigma^3  \,  \sigma^1 \big( \ds{\Gamma}_{\dot{m}} - g_{\td\td}^{-1} g_{\td \dot{m}} \ds{\Gamma}_\td\big) \ds{\Delta}^\A \theta^\A = e^{-\dual{\phi}} \, \bar{\theta}^\B \Gamma^\clty \bigg[ \sigma^1 \dual{\ds{\Gamma}}_{\td \dot{m}} {\ds{\Delta}}^\B + 2 \, \dual{g}^{-1}_{\td\td} (\dual{\ds{\Gamma}}_\td \dual{g}_{\dot{m}\td} - \dual{\ds{\Gamma}}_{\dot{m}} \dual{g}_{\td\td}) {\ds{D}}^\B_\td \bigg] \theta^\B. \label{c1c2term2}
\end{align}
Therefore, putting together the expressions  we have
\begin{equation}
\begin{split}
    \dual{\chi}^{(2)}_{\td \dot{m}} & = \dfrac{i}{2} \, e^{-\dual{\phi}} \, \bar{\theta}^\B \bigg[ 2 \, \dual{\ds{\Gamma}}_{[\td} {\ds{D}}^{\B}_{\dot{m}]}  - \frac{1}{2}\sigma^1\dual{\ds{\Gamma}}_{\td \dot{m}} {\ds{\Delta}}^{\B} \bigg] \theta^\B.
\end{split}
\end{equation}
One can proceed analogously to obtain all the type IIA and type IIB bilinears in Ramond-Ramond superfields, going up in the degree of the T-dualized form one at a time.  Alternatively, a generalised discussion of the Ramond-Ramond superfields in appendix \ref{Tdualconventions} demonstrates that all of these expansions can be calculated together.

\subsubsection*{Full results}
To conclude, we list all the relevant superfields up to quadratic order  both in the Neveu-Schwarz and Ramond-Ramond sectors. 

The expansions    for the Neveu-Schwarz superfields   at order $(\theta)^2$ look same in both theories in our spinor doublet notation. They  are
\begin{align}
    & \boldsymbol{g}_{mn} = g_{mn} - i \, \bar{\theta}^\II \sigma^1 \mathbb{\Gamma}_{(m} \mathbb{D}_{n)}^\II \theta^\II,  \label{second-order_typeIIA_NSNSsuperfieldsstart} \\
    & \boldsymbol{\phi} = \phi - \dfrac{i}{4} \, \bar{\theta}^\II \mathbb{\Delta}^\II \theta^\II, \\
    & \boldsymbol{B}_{mn} = B_{mn} - i \, \bar{\theta}^\II (i\sigma^2)   \mathbb{\Gamma}_{[m} \mathbb{D}_{n]}^\II \theta^\II, \label{second-order_typeIIA_NSNSsuperfieldsend}
\end{align}
where the superscript '$^{\II}$' indicates that one must introduce the appropriate object in each theory. The order-$(\theta)^2$ terms in the Ramond-Ramond superfields in type IIA and type IIB theories can also be written compactly 
\begin{equation}\label{second-order_typeIIA_RRsuperfields}
    \bs{C}^{(n)}_{m_1 \dots m_{n}} = C^{(n)}_{m_1 \dots m_{n}} -  \dfrac{i}{2} \, e^{-\phi} \, \bar{\theta}^\II \, \biggl[ (-1)^n(\sigma^3)^{1 + \lfloor\! \frac{n}{2} \!\rfloor}\biggr] \biggl[ n \, \ds{\Gamma}_{[m_1} \ds{D}_{\dots m_{n}]}^\II - \dfrac{1}{2} \, \ds{\Gamma}_{m_1 \dots m_{n}} \sigma^1 \ds{\Delta}^\II \biggr] \theta^\II,
\end{equation}
where the parity of $n$ determines whether the spinor doublet and the supersymmetry operators are the type IIA or type IIB ones.

\subsubsection*{D$\bs{p}$-branes}
Now that we have determined the fermionic expansion of the all the fundamental superfields in type IIA and type IIB theories, we can turn our attention to the composite superfields of greatest interest, namely the worldvolume actions of a D$p$-brane for arbitrary $p$. 

Since the formal structure of purely bosonic D$p$-brane is equivalent to the structure of the action in superspace, the T-duality mechanism is also the same as the one leading to the bosonic action (\ref{general_bosonic_Dp-brane_action}). The only precaution one needs to take regards the fact that the starting point, i.e. the superspace D2-brane action (\ref{D2finalform}), and consequently the T-duality rules, are such that the Neveu-Scwharz T-duality rules see all fields in superspace whereas the Ramond-Ramond ones only contain the Ramond-Ramond fields in superspace, as exemplified in (\ref{NSTdual}) and (\ref{RRTdual}). At the end of the day, the action of any D$p$-brane in superspace takes the form
\begin{equation}\label{shiftedDpaction}
    \boldsymbol{S}_{\text{D}p} = - T_{\text{D}p} \int \de^{p+1} \zeta \; e^{-\bs{\phi}} \, \sqrt{-\det(\bs{g} + \bs{f})} + T_{\text{D}p} \int \; \bs{C} \, e^{-f},
\end{equation}
where $\boldsymbol{g}_{ij}$ is the supermetric pullback, $\boldsymbol{f}_{ij} = \boldsymbol{B}_{ij} + \sf{F}_{ij}$ is the natural superspace combination of the Kalb-Ramond field with the worldvolume flux term, with $f_{ij} = B_{ij} + \sf{F}_{ij}$ being its bosonic component, and where we have defined the formal sum $\smash{\bs{C} = \sum_q \bs{C}^{(q)}}$ over promoted Ramond-Ramond $q$-form pulled-back superfields $\smash{\bs{C}^{(q)}}$. Once again, this result holds at all orders in fermions. In order to determine the expansion of the D$p$-brane action superfield to an arbitrary order in fermions, one needs to plug the expansions of the fundamental superfields from the corresponding type II supergravity into \eqref{shiftedDpaction}. The second-order expansions in spinor doublet notation are in \eqref{second-order_typeIIA_NSNSsuperfieldsstart} - \eqref{second-order_typeIIA_RRsuperfields} for both type II theories.

\subsubsection[Order-$(\theta)^4$ terms]{Order-\texorpdfstring{$\boldsymbol{(\theta)^4}$}{theta4} terms}
We have already made some comments in section \ref{orderfourdimred} regarding the unwieldy size of the expressions obtained for the quartic fermionic couplings after dimensional reduction. There we also discussed how these expressions might be improved and simplified going forward, in order that they become more manageable. In their current formulation the calculation necessary for their full T-dualization is impractically lengthy. Important to note, however, is that there is no technical impediment. Just like the quadratic fermionic couplings, the quartic couplings may in principle be T-dualized using the techniques and results we have reviewed and developed in this section. Actively pursuing this full calculation is better delayed until such a time that the possible simplifying procedures for the quartic terms have been implemented.

Nevertheless there are some observations that can be made concretely at quadratic fermion level that we can fully expect to also happen at quartic level. Firstly, the NS superfield expansions take on the same shape in both type II supergravities. The same holds for the expansion of the 11-dimensional supermetric, that at order two has the same structure as the 10-dimensional supermetrics. This is not a coincidence: the supervielbein expansion looks schematically the same in all these theories (even though in each theory there is a different notion of what the gravitino or the supercovariant derivative are) and the outcome of manipulations at quadratic order makes this point manifest. Moreover, the existing relations go beyond that. The type IIA metric and $B_2$ superfield expansions came from different 11 dimensional superfields but at quadratic order turned out to be very similar. If it were not for this, it would have been impossible to find again this structure in type IIB upon T-duality. This extends to the whole NSNS sector, that allowed us to write those superfields up to quadratic order at once both for type IIA and type IIB (\ref{second-order_typeIIA_NSNSsuperfieldsstart} - \ref{second-order_typeIIA_NSNSsuperfieldsend}).  In principle there is no argument against the structure extending to all levels in $\theta$, but unfortunately, the current form of quartic terms did not quite allow us to make this point manifest. For example, the 10 dimensional metric expansion and the 11-dimensional one do not seem to allow for such comparisons in their order $(\theta)^4$ terms. On the other hand, there are indeed many similarities between the metric and the $B_2$-field order $(\theta)^4$ terms (modulo (anti)symmetry of indices and chirality matrices), which is a positive observation, but there are also differences on certain terms (that maybe could be manipulated to make them similar to each other). These ideas could also be used e.g. to obtain quartic terms of NSNS fields in type IIB  by `simply'  writing type IIA formulae \eqref{quarticdil}, \eqref{10dmetricquartic}, and \eqref{KBquarticshift} in spinor doublet notation. It would be nice to compare that with the outcome of performing the computation using the Hassan rules.

Finally, something that might be possible given the current formulation of the quartic order fermionic couplings for the D2-brane is to identify those parts of the expressions which would lead to particularly sought-after terms in D$p$-brane actions. For example, the work in \cite{Hamada:2021ryq} posits a particular quartic term in the action of the D7-brane. It could be possible to hunt for this term via T-dualization without laboriously T-dualizing everything appearing after dimensional reduction, however we leave this possibility for future study.

\section{Conclusions and future work} \label{conclusions}
In String Theory, branes are just as important as the strings themselves. The quantum field theories living on their worldvolumes teem with rich dynamics that is both mathematically intriguing and phenomenologically impactful. While the bosonic fields in these theories have received plentiful attention, the fermionic degrees of freedom are more challenging to study and are less well understood as a result. We have drawn our primary motivation from the fact that the current level of knowledge about the fermions living on branes requires significant improvement. One of the core reasons that fermions on branes are under-studied is that obtaining their couplings explicitly turns out to be surprisingly difficult. Higher-order couplings of fermions in brane actions have been invoked recently \cite{Gautason:2018gln,Hamada:2018qef,Kallosh:2019oxv,Hamada:2019ack,Gautason:2019jwq,Carta:2019rhx,Kachru:2019dvo}, however the impracticality of the existing methods used to obtain these terms limited their use. Very recently, a proposal for obtaining specific quartic couplings on D7-branes that can be pertinent for understanding KKLT has also been put forward \cite{Hamada:2021ryq}. In this work we have made significant progress in improving both the conceptual understanding and the practical techniques needed to pursue these terms. Furthermore, the insights we have had and connections we have made are applicable far beyond the calculation of specific couplings in brane worldvolume theories. In fact we have presented the calculation of these terms as a single, if pertinent, example of a place where our more general methods come into use.

\subsubsection*{Summary}
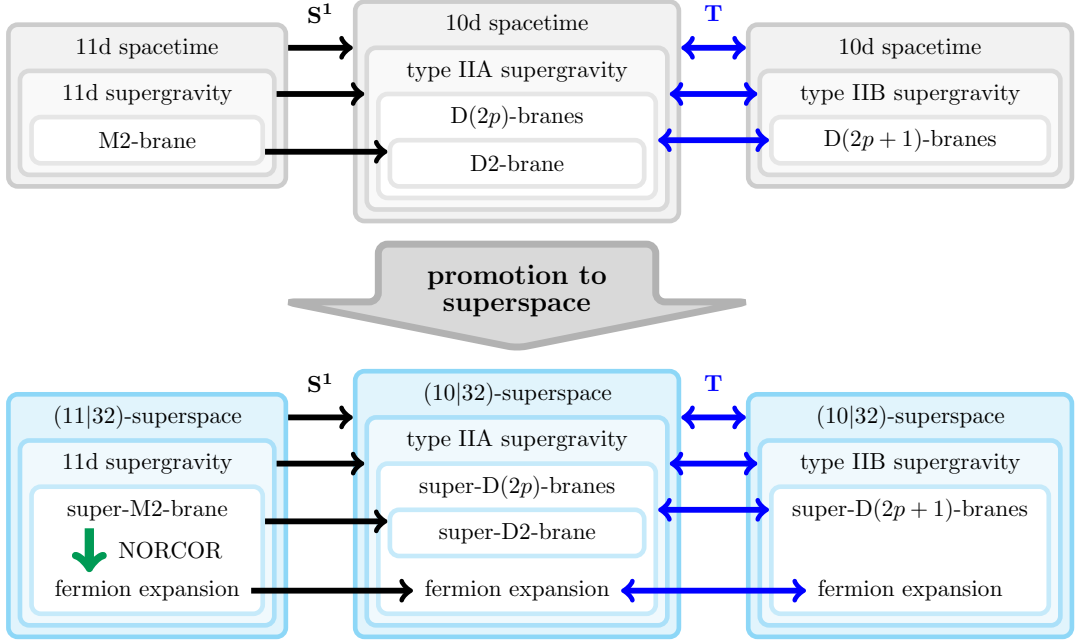
\begin{figure}  \begin{center}      \begin{tikzpicture}
    \def\topcol{gray}  
    \def\botcol{green} 
    \begin{scope}[scale=0.35]
      
    \begin{scope}[xshift=-1cm]
    \draw[line width = 3pt, rounded corners, black!30, fill=black!15] (18,9.5) -- (30,9.5) -- (30,7) -- (34,7) -- (24,5) -- (14,7) -- (18,7)  -- cycle
    ;
    \end{scope}
     
    \draw (23,7.50) node[align=center] {\large{\textbf{promotion to}} \\ \large{\textbf{superspace}}};

    \draw[line width = 2.5pt, ->, black] (13,18) -- (15.9,18);
    
    \draw[line width = 2.5pt, <->, blue] (30.1,18) -- (32.9,18);

    \draw[line width =2pt, \topcol!40, fill = \topcol!10, rounded corners] (1,19) rectangle (13,12);
     
    \draw[line width =2pt, \topcol!40, fill = \topcol!10, rounded corners] (16,20) rectangle (30,10.5);
     
    \draw[line width =2pt, \topcol!40, fill = \topcol!10, rounded corners] (33,19) rectangle (47,12);

    \draw[line width = 2.5pt, ->, black] (12.5,16) -- (16.4,16);
     
    \draw[line width = 2.5pt, <->, blue] (29.6,16) -- (33.4,16);
     
    \draw[line width =2pt, \topcol!30, fill = \topcol!5, rounded corners] (1.5,17) rectangle (12.5,12.5);
     
    \draw[line width =2pt, \topcol!30, fill = \topcol!5, rounded corners] (16.5,18) rectangle (29.5,11);
     
    \draw[line width =2pt, \topcol!30, fill = \topcol!5, rounded corners] (33.5,17) rectangle (46.5,12.5);

    \draw[line width = 2.5pt, <->, blue] (29.1,14) -- (33.9,14);
     
    \draw[line width =2pt, \topcol!20, fill = \topcol!0, rounded corners] (2,15) rectangle (12,13);
     
    \draw[line width =2pt, \topcol!20, fill = \topcol!0, rounded corners] (17,16) rectangle (29,11.5);
     
    \draw[line width =2pt, \topcol!20, fill = \topcol!0, rounded corners] (34,15) rectangle (46,13);

    \draw[line width = 2.5pt, ->, black] (12.1,13.5) -- (17.4,13.5);
     
    \draw[line width =2pt, \topcol!20, fill = cyan!0, rounded corners] (17.5,14) rectangle (28.5,12);

    \draw (7,18) node {11d spacetime};
    \draw (7,16) node {11d supergravity};
    \draw (7,14) node {M2-brane};
    
    \draw (23,19) node {10d spacetime};
    \draw (23,17) node {type IIA supergravity};
    \draw (23,15) node {D$(2p)$-branes};
    \draw (23,13) node {D2-brane};
    
    \draw (40,18) node {10d spacetime};
    \draw (40,16) node {type IIB supergravity};
    \draw (40,14) node {D$(2p+1)$-branes};
    
    \draw (14.5,19.5) node {$\boldsymbol{\mathrm{S}^1}$};
    \draw (31.5,19.5) node[blue] {\textbf{T}};
    
\end{scope}

\begin{scope}[yshift = -5.6cm, scale=0.35]
      
    \draw[line width = 2.5pt, ->, black] (13,18) -- (15.9,18);
    
     \draw[line width = 2.5pt, <->, blue] (30.1,18) -- (32.9,18);

     \draw[line width =2pt, cyan!40, fill = cyan!10, rounded corners] (1,19) rectangle (13,8.5);
     
     \draw[line width =2pt, cyan!40, fill = cyan!10, rounded corners] (16,20) rectangle (30,8.5);
     
     \draw[line width =2pt, cyan!40, fill = cyan!10, rounded corners] (33,19) rectangle (47,8.5);

    \draw[line width = 2.5pt, ->, black] (12.5,16) -- (16.4,16);
     
    \draw[line width = 2.5pt, <->, blue] (29.6,16) -- (33.4,16);
     
    \draw[line width =2pt, cyan!30, fill = cyan!5, rounded corners] (1.5,17) rectangle (12.5,9);
     
    \draw[line width =2pt, cyan!30, fill = cyan!5, rounded corners] (16.5,18) rectangle (29.5,9);
     
    \draw[line width =2pt, cyan!30, fill = cyan!5, rounded corners] (33.5,17) rectangle (46.5,9);

    \draw[line width = 2.5pt, <->, blue] (29.1,14) -- (33.9,14);
     
    \draw[line width =2pt, cyan!20, fill = cyan!0, rounded corners] (2,15) rectangle (12,9.5);
     
    \draw[line width =2pt, cyan!20, fill = cyan!0, rounded corners] (17,16) rectangle (29,9.5);
     
    \draw[line width =2pt, cyan!20, fill = cyan!0, rounded corners] (34,15) rectangle (46,9.5);

    \draw[line width = 2.5pt, ->, black] (12.1,13.5) -- (17.4,13.5);
     
    \draw[line width =2pt, cyan!20, fill = cyan!0, rounded corners] (17.5,14) rectangle (28.5,12);

    \draw (7,18) node {$(11|32)$-superspace};
    \draw (7,16) node {11d supergravity};
    \draw (7,14) node {super-M2-brane};
    
    \draw (23,19) node {$(10|32)$-superspace};
    \draw (23,17) node {type IIA supergravity};
    \draw (23,15) node {super-D$(2p)$-branes};
    \draw (23,13) node {super-D2-brane};
    
    \draw (40,18) node {$(10|32)$-superspace};
    \draw (40,16) node {type IIB supergravity};
    \draw (40,14) node {super-D$(2p+1)$-branes};
    
    \draw (14.5,19.5) node {$\boldsymbol{\mathrm{S}^1}$};
    \draw (31.5,19.5) node[blue] {\textbf{T}};
    
    \draw (7,10.5) node {fermion expansion};
    \draw (23,10.5) node {fermion expansion};
    \draw (40,10.5) node {fermion expansion};

    \draw[line width = 2.5pt, ->, black] (11.5,10.5) -- (18.5,10.5);
     
    \draw[line width = 2.5pt, <->, blue] (27.5,10.5) -- (35.5,10.5);
     
    \draw[line width = 4pt, ->, ForestGreen] (4.5,13.2) -- (4.5,11.3);
     
    \draw (8,12.25) node {NORCOR};
    
\end{scope}
\end{tikzpicture}  \end{center}  \caption{A schematic map of the procedures investigated in this work. We work in superspace, and in order to do so profitably we generalize the string duality web to superspace. We generalize the $\mathrm{S}^1$ compactification from 11-dimensional supergravity to type IIA, and we generalize the T-duality procedure connecting type IIA and type IIB to superspace. This allows us to carry the elegant geometric treatment of the `normal coordinate' (NORCOR) method in 11-dimensional supergravity over to type II supergravity, circumventing the difficulty in applying that treatment directly in those theories. This method allows us to calculate the expansion of actions of branes in orders of the worldvolume fermions. We have presented example calculations up to quartic order in fermions for the M2-brane and the D2-brane in this work, although the methods we have presented are in principle applicable to any order in fermions. }       \label{fig:processmap} \end{figure}

The structure at the heart of this work is the web of string dualities given in Fig. \ref{fig:dualityweb}. The approaches that we have developed, and used to obtain brane actions, rest upon the generalizations of the connections in this web. Such connections allowed us to take advantage of the elegance of techniques applicable to a theory in one part of the web in order to achieve progress in others. More concretely, the connections we have concentrated on are the circle compactification linking 11-dimensional to type IIA supergravity and the T-duality relating type IIA and type IIB theories to each other. The generalization we have explored is the promotion to a \textit{superspace} formalism for the connections in the web. Fig. \ref{fig:processmap} presents a map of the concepts used.

The reasons for which this particular generalization has proved to be so useful are twofold. Firstly, our starting point, 11-dimensional supergravity, has a particularly elegant formulation in $(11|32)$-superspace. Secondly, we have access to a systematic, complete, and \textit{manageable} geometrical method for determining explicit fermionic expansions of this theory's superfields, namely NORCOR. The small number of superfields in 11-dimensional supergravity in conjuction with NORCOR means we can readily obtain the fermionic expansions of all the fundamental superfields in the theory. Obtaining the fermionic expansions for composite superfields built out of these fundamental superfields is then a simple matter. The example composite superfield we have chosen to concentrate on in this case is the action for a single M2-brane. This action is constructed using the pullbacks of the supervielbein and super three-form in 11-dimensional supergravity. 

With our starting point of 11-dimensional supergravity and the M2-brane firmly in hand, we then pursued the superspace generalization of the $\mathrm{S}^1$-compactification to type IIA supergravity and the D2-brane. Our goal was to use the expansion of the 11-dimensional superfields together with this connection in the web to determine the  expansion of the type IIA superfields. The regular dimensional reduction ansatz relates the 11-dimensional vielbein and three-form to the 10-dimensional vielbein, dilaton, Ramond-Ramond one-form, Kalb-Ramond two-form and Ramond-Ramond three-form. We took the view that these bosonic relations represented the `zeroth-order' fermionic expansion of the corresponding superfield relations. As such, we promoted the dimensional reduction ansatz relations to superfields, taking the fermionic expansions of the 10-dimensional superfields (to some desired order) as unknowns to be determined. We then used the NORCOR results of the Taylor expansion of the 11-dimensional fields to determine explicit expressions for these 10-dimensional unknowns in terms of 11-dimensional fields. Finally we dimensionally reduced the 11-dimensional fermionic terms and compared the results with the expansion in terms 10-dimensional unknown fermionic terms in order to read off the desired results. We demonstrated how known second-order results for fundamental superfield expansions in type IIA can be recovered painlessly using this method. Furthermore we demonstrated how labourious manipulations of the D2-brane action can be completed almost trivially in this superfield paradigm, and how the form for the quadratic fermion terms on the D2-brane can be recovered, again relatively painlessly. Finally we calculated the fermionic expansion of the type IIA fields relevant for the D2-brane all the way up to order four in fermions. Unfortunately these terms, while systematic and complete, are unwieldy in their present formulation. We discussed some promising lines of research regarding their simplification, something we will come back to in a moment.

Finally we turned our attention to the second strand on the web of dualities that we sought to generalize to superspace. This was the T-duality relation between type IIA and type IIB theories. The structure of work mirrored that of the generalization of the dimensional reduction just discussed. We first observed what relations the T-duality demanded of the bosonic fields in either theory. These were the Buscher rules and the Ramond-Ramond field rules. We once again interpreted these relations as representing the `zeroth-order' fermionic expansion of the corresponding superfield relations, and as such promoted these T-duality rules to superfields. This required observing that the discussion of the Ramond-Ramond sector can be substantially simplified by conveniently arranging the D2-brane action. Then it was the repeated application of these promoted rules which we used to determine the fermion terms in the superfield expansions for all the superfields in both type II supergravities. When we performed the T-duality transformations, we had to become familiar with precisely how fermions behaved. This transpired to be an area of much subtle complexity, but one which we greatly streamlined by moving to spinor doublet notation. Once again, we chose as a crucial example case the calculation of the fermionic expansion of brane actions. In this case repeated T-duality transformations allowed us to leverage the knowledge we had built about the D2-brane in the previous stage to determine features of the D$p$-brane actions in general. We once again wrote down a form of the action which will yield the fermion couplings on the D$p$-brane to any order if provided with the expansions of the fundamental fields of the type II supergravity in which the brane lives. We noticed that in this formulation Ramond-Ramond fields of every degree are used implicitly, yet the first dimensional reduction step had furnished us with only degree 1 and 3. This is where the careful study of fermions under T-duality became invaluable as explicit T-dualization of these two superfield expansions allowed us to determine the expansions for all the fields we desired to quadratic order. The only remaining impediments to a full calculation at quartic order for all D$p$-branes are then of a practical nature. The expressions we have obtained, since they represent all couplings of the brane fermions to an arbitrary
bosonic background, have many terms, and the calculation for each term is non-trivial. There is no technical impediment to T-dualization and we provide all the necessary tools, however we consider it prudent to first make a proper investigation of how the expressions we have obtained for type IIA fields and D2-brane might be improved. We discuss this, and other future lines of work, next.

\subsubsection*{Future directions}
The directions in which this work will progress in the future come in two main classes: those directions that improve and build upon the work and those that use it.

The most obvious direction in which the present work might be improved is in seeking to simplify the results at quartic order in fermions. We have already discussed at the end of section \ref{orderfourdimred} how significant simplifications of the current formulation of the complete quartic order terms for the superfield expansions in type II supergravities will have their roots in a better treatment of the 11-dimensional supervielbein expansion. This might be achieved via something as simple as a more adroit rearrangement and application of the constraints imposed by Bianchi identities than we have managed here, or it could require an improvement at a higher level in the set-up of NORCOR. Pursuing such a better treatment is an obvious and tantalising direction of future study.

For the brane actions specifically, these results might be improved by getting a firmer grasp of how to arrange higher order fermionic expansions around a $\kappa$-symmetry organizational principle. As early in our process as our expression for the M2-brane action in \eqref{Quarticaction}, we neglected to explicitly organize all our terms around such a principle. When calculating the quartic terms in the M2 brane, one can interpret all of the different terms as arising from the variation of different parts of the quadratic fermionic term. Those quartic terms that came with the same, `zeroth-order' projector as in the quadratic term are interpreted as arising from varying the supercovariant derivative that appeared in the quadratic term. The remaining quartic terms (coming with a factor $\frac{1}{8}$) can be interpreted as arising from further variations of the projector, inverse metric, etc, appearing at quadratic order. The higher-order expansion of the kappa symmetry projector may be calculated directly by expanding the superfield projector \eqref{superkappaproj}. Better understanding of the structure here could then be carried over to type II theories using the duality promotion method we have presented. At second order the D$p$-brane actions were able to be organized into a similar form as the M2-brane, that is, a bilinear containing a kappa projector and some operators. The expectation would be that whatever further structure is found in the M2-brane should provide analogous arrangements of the D$p$-brane action through the promoted duality web.

With more agile control over D$p$-brane actions, it becomes natural to revisit the D7-brane quartic gaugino couplings and compare them with the existing literature, among other things. This would be instrumental in shedding further light on gaugino condensation in the stabilization of volume moduli à la KKLT. A proposal for the specific quartic gaugino terms on D7-branes necessary to achieve this was recently put forward in \cite{Hamada:2021ryq}, and hunting for the specific terms which that proposal requires within our results is a promising line of inquiry. In a different area, a further result that is now in reach is the determination of the F1-string action at arbitrary fermionic order. In fact, once the M2-brane action is known at a given order, a circle compactification along a direction wrapped by the brane (a double dimensional reduction) gives the Green-Schwarz-string action \cite{Martucci:2003gc} in a similar way to the compactification along an unwrapped direction, which gave the D2-brane action. Finally, we have worked in bosonic backgrounds. To do so we simply set to zero those terms proportional to the gravitino in the expansions of the superfields of 11-dimensional supergravity. By keeping these terms, however, the methods we employed in this article can also be used to explore more general backgrounds than purely bosonic ones. In this way, one would obtain the M2-brane couplings to the 11-dimensional gravitino and hence, upon dimensional reduction and T-dualization, the D$p$-brane couplings to the 10-dimensional gravitino and dilatino. Finally, we have concentrated in this work on obtaining the fermion couplings on brane actions in the abelian case of a single brane. Expanding this work to the non-abelian case of multiple branes, or to even more complicated brane set-ups, is yet another promising line of inquiry.  
\\

Progress in an area as central to so many discussions as the fermionic couplings on brane worldvolumes is necessarily complex. What we have presented here is both an important step in this long story, and a clear and insightful guide to what is known, and what remains to be investigated, in this exciting and consequential line of research.

\section*{Acknowledgements}
We are especially grateful to Mariana Graña, Luca Martucci, Marco Serra, and Dimitrios Tsimpis, for valuable exchanges and for patient and detailed feedback.  We also  thank Guillaume Bossard, Christopher Erickson, Nicolás Kovensky, Gabriele Lo Monaco, Severin Lüst, Ruben Minasian,  Susha Parameswaran, and Dmitri Sorokin for useful conversations and suggestions.  

A.R. is supported by the ERC Consolidator Grant 772408-Stringlandscape. 
 
J.R. is supported by an EPSRC Doctoral Training Partnership. 

R.T. is supported by the International Exchange Royal Society Grant IES/R3/170249.

\appendix

\section{Spinor conventions}\label{spingamsec}
We summarize the conventions that we use in the main text regarding spinors defined in 11- and 10-dimensional spacetime. Here we denote terms intrinsically living in 11-dimensional spacetime with a hat in order to distinguish them from the ones defined in 10-dimensional spacetime (with no hats). This is also the case in appendix \ref{DimRedCatApp}, which explains the details about dimensional reduction. In the main text we often drop hats for the sake of clarity, as the spacetime dimension is always clear from the context, only using hats for 11-dimensional objects at the point of performing dimensional reduction.

In the 11-dimensional spacetime, we use real Majorana anticommuting 32-component spinors denoted as $\hat{\theta}^\mu$, with $\mu$ representing spinor indices in the curved superspace manifold and $\alpha$ representing spinor indices on the corresponding tangent space. Spinor indices can generally be suppressed without loss of clarity. Explicitly, Dirac conjugation is defined in terms of the antisymmetric conjugation matrix $C = C_{\alpha\beta}$, with $C_{\alpha \beta} = - C_{\beta \alpha}$, as
\begin{equation}
    \hat{\bar{\theta}}_\beta = \hat{\theta}^\alpha C_{\alpha\beta}.
\end{equation}
More generally spinor indices are raised and lowered by the conjugation matrix and its inverse $C^{-1} = C^{\alpha\beta}$, with $C_{\alpha\beta} C^{\beta\gamma} = \delta_\alpha^\gamma$, according to the rule
\begin{equation}
    M\ten{_\alpha^\beta} = C_{\alpha\gamma}M\ten{^\gamma_\delta}C^{\delta\beta}.
\end{equation}
In the index-free notation, one can write $\hat{\bar{\theta}} = \hat{\theta}^T C$ and $\hat{\bar{\theta}} M \hat{\theta} = \hat{\bar{\theta}}_\alpha M\ten{^\alpha_\beta} \hat{\xi}^\beta = \hat{\theta}^\alpha M_{\alpha\beta} \hat{\xi}^\beta$. We work with the mostly-plus Minkowksi metric $\hat{\eta}_{\hat{a} \hat{b}}$, with signature $(-1,(+1)^{10})$ and indices running as $\hat{a}=\underline{0}, \dots, \underline{10}$, and employ ${\Gamma}$-matrices $\hat{\Gamma}^{\hat{a}}$ fulfilling the Clifford algebra
\begin{equation}
    \lbrace \hat{\Gamma}^{\hat{a}}, \hat{\Gamma}^{\hat{b}} \rbrace = 2 \hat{\eta}^{\hat{a} \hat{b}}.
\end{equation}
The antisymmetrized $\Gamma$-matrix products are defined as
\begin{equation}
    \hat{\Gamma}_{\hat{a}_1 \hat{a}_2 \dots \hat{a}_n} = \hat{\Gamma}_{[\hat{a}_1} \hat{\Gamma}_{\hat{a}_2} \dots \hat{\Gamma}_{\hat{a}_n]}.
\end{equation}
The combinations $(\hat{\Gamma}_{\hat{a}_1 \hat{a}_2 \dots \hat{a}_n})_{\alpha\beta}$ are symmetrical in their spinor indices for $n=1,2 \; \mathrm{mod} \, 4$ and antisymmetrical otherwise, i.e.
\begin{subequations}
    \begin{align}
        & (\hat{\Gamma}_{\hat{a}_1 \hat{a}_2 \dots \hat{a}_n})_{\alpha \beta} = + (\hat{\Gamma}_{\hat{a}_1 \hat{a}_2 \dots \hat{a}_n})_{\beta \alpha}, \qquad n = 1,2 \; \mathrm{mod} \, 4; \\
        & (\hat{\Gamma}_{\hat{a}_1 \hat{a}_2 \dots \hat{a}_n})_{\alpha \beta} = - (\hat{\Gamma}_{\hat{a}_1 \hat{a}_2 \dots \hat{a}_n})_{\beta \alpha}, \qquad n = 0,3 \; \mathrm{mod} \, 4.
    \end{align}
\end{subequations}
The Majorana nature of the anticommuting fermions $\hat{\theta}$ means that $\hat{\bar{\theta}} \, \hat{\Gamma}_{\hat{a}_1 \hat{a}_2 \dots \hat{a}_n} \, \hat{\theta} = 0$ for $n = 1,2 \; \mathrm{mod} \, 4$. The master equation for practical $\Gamma$-matrix manipulation (in any number of dimensions) is
\begin{equation}
\hat{\Gamma}^{\hat{a}_1\dots \hat{a}_m} \hat{\Gamma}_{\hat{b}_1\dots \hat{b}_n} = \sum_{r = 0}^{\text{min}(m,n)} r! \vect{m}{r} \vect{n}{r}\delta_{[\hat{b}_1}^{[\hat{a}_m} \dots \delta_{\hat{b}_r}^{\hat{a}_{m+1-r}} \hat{\Gamma}\ten{^{\hat{a}_1\dots \hat{a}_{m-r}]}_{\hat{b}_{r+1}\dots \hat{b}_n]}}.
\end{equation}

After the dimensional reduction to a 10-dimensional space spanned by indices $a=\underline{0},\dots,\underline{9}$, where the direction $x^{10}$ is compactified, it is necessary to introduce a chirality matrix. In tangent spacetime, the first ten $\Gamma$-matrices are the same because the Clifford algebra reads $\lbrace\hat{\Gamma}^{a},\hat{\Gamma}^{b} \rbrace = 2 \hat{\eta}^{ab} = 2 \eta^{ab} = \lbrace \Gamma^{a},{\Gamma}^{b} \rbrace$, so $\hat{\Gamma}^{a} = \Gamma^a$, where $\eta_{ab} = \hat{\eta}_{ab}$ is the 10-dimensional Minkowski metric; the last 11-dimensional $\Gamma$-matrix defined to be the 10-dimensional chirality matrix $\hat{\Gamma}^{{\underline{10}}} \equiv \Gamma^{\ulv}$. All the other rules on spinor indices are unchanged. Because in ten dimensions there is a notion of chirality, we split 11-dimensional Majorana spinors into pairs of 10-dimensional Majorana-Weyl spinors as $\theta = \theta_+ + \theta_-$, where $\Gamma^{\ulv} \theta_\pm = \pm \theta_\pm$. For type IIB strings, we relate the previous pair of Majorana-Weyl spinors to another pair of Majorana-Weyl spinors, but this time with equal chirality, i.e. $\theta_{1,2}$ with $\Gamma^{\ulv} \theta_{1,2} = + \theta_{1,2}$. In this case it is convenient to rearrange these fermion pairs into a Pauli matrix-valued spinor
\begin{equation}
    \theta = \vect{\theta_{1}}{\theta_{2}},
\end{equation}
which is acted on by the 2-dimensional identity $1_2$ and the three Pauli matrices $\sigma^1$, $\sigma^2$ and $\sigma^3$. All the $\Gamma$-matrices and the chirality matrix that need to act on the spinor $\theta$ can be redefined by means of a tensor product with the 2-dimensional identity $1_2$ in such a way as to act appropriately on the two spinor components $\theta_{1,2}$.

\subsection*{Note on spinor indices}\label{supcovdevnote}
In dealing with spinor contractions, we often find it useful to rearrange expressions by moving spinor indices. Given a matrix $M_{\alpha \beta}$ acting on the spinor space, we define its transpose as the matrix $\check{M}_{\alpha \beta} = M_{\beta \alpha}$. As an example, consider the torsion $\smash{\hat{T}_{\hat{a}}}$ and its transpose $\smash{\hat{\check{T}}_{\hat{a}}}$
\begin{equation*}
    \begin{split}
        \hat{T}_{\hat{a}} & = \frac{1}{288} \Big(\hat{\Gamma}\ten{_{\hat{a}}^{\hat{b}\hat{c}\hat{d}\hat{e}}} + 8 \delta_{\hat{a}}^{\hat{b}} \hat{\Gamma}^{\hat{c}\hat{d}\hat{e}}\Big) \hat{H}_{\hat{b}\hat{c}\hat{d}\hat{e}}, \\
        \hat{\check{T}}_{\hat{a}} & = \frac{1}{288} \Big(\hat{\Gamma}\ten{_{\hat{a}}^{\hat{b}\hat{c}\hat{d}\hat{e}}} - 8 \delta_{\hat{a}}^{\hat{b}} \hat{\Gamma}^{\hat{c}\hat{d}\hat{e}}\Big) \hat{H}_{\hat{b}\hat{c}\hat{d}\hat{e}}.
    \end{split}
\end{equation*}
Notice that it is not the position of the spinor indices that is used to make the distinction between $\smash{\hat{T}_{\hat{a}}}$ and $\smash{\hat{\check{T}}_{\hat{a}}}$: both are defined as in the main text and the position of the indices can be changed with the charge conjugation matrix $C_{\alpha \beta}$ and its inverse $C^{\alpha \beta}$. In fact, we can write for instance $\smash{\tensor{(\hat{T}_{\hat{m}})}{_\alpha^\beta} = - \tensor{(\hat{\T}_{\hat{m}})}{^\beta_\alpha}}$.

\section{11-dimensional supergravity  }\label{11dconvapp}
Here we summarize the set-up and conventions for 11-dimensional supergravity \cite{Cremmer, Cremmer2, Brink:1980az}, including the field content, the constraints on the torsion which are equivalent to the equations of motions, and the Bianchi identities \cite{Howe:1997he}.

In 11-dimensional supergravity, let us consider the $(11|32)$-dimensional supermanifold spanned by coordinates $Z^M=(x^m, \theta^\mu)$, where $M$ is a generalized superspace index, with $m=0,\dots,10$ representing the original spacetime directions and $\mu=1,\dots,32$ representing the corresponding spinor directions. In this formalism, one defines the supervielbein as
\begin{equation}
    E^A(x,\theta) = \de Z^M E\ten{_M^A}(x,\theta),
\end{equation}
where the index $A$ corresponds to the tangent space, with the possibility to introduce local coordinates $\y^A = (\y^a, \y^\alpha)$, with $a=0,\dots,10$ and $\alpha=1,\dots,32$. Let us also introduce a superconnection, i.e. the super-one-form $\omega\ten{_A^B}$, with Lorentzian structure group, in terms of which we define the superspace covariant derivative,
\begin{equation}
\nabla X\ten{^{A_1\dots }_{B_1 \dots }}  = \de  X\ten{^{A_1\dots }_{B_1 \dots }}  +  X\ten{^{D A_2 \dots }_{B_1 \dots }} E^C \omega\ten{_{CD}^{A_1}} \dots +  X\ten{^{A_1\dots }_{D B_2\dots }} E^C \omega\ten{_{B_1 C}^D}\dots.
\end{equation}
The superconnection is comnpatible with the structure of the tangent space Lorentz group, and it is related to the spin connection according to
\begin{equation}
    \omega\ten{_a^\beta}=\omega\ten{_\beta^a}=0\qquad , \qquad \omega\ten{_\alpha^\beta}=\dfrac{1}{4}\omega\ten{_{ab} }(\Gamma\ten{^{ab} })\ten{_\alpha^\beta}
\end{equation}

We can then define the supertorsion $T^A$ and the supercurvature $R\ten{_B^A}$ as
\begin{align}
    & T^A = \nabla E^A = \de E^A + E^B \omega\ten{_B^A} = \frac{1}{2} E^C E^B T\ten{_{CB}^A}, \label{torsiondef} \\
    & R\ten{_B^A} = \de \omega\ten{_B^A} + \omega\ten{_B^C}\omega\ten{_C^A} = \frac{1}{2}E^D E^C R\ten{_{DCB}^A}. \label{curvedef}
\end{align}
Finally, we define the super-three-form 
\begin{equation}
    A = \frac{1}{3!}E^C E^B E^A A_{CBA},
\end{equation}
along with its field-strength, i.e. the super-four-form
\begin{equation}
    H = \de A = \frac{1}{4!} E^A E^B E^C E^D H_{DCBA},
\end{equation} 
whose components explicitly read 
\begin{equation}\label{Hsuperdef}
    H_{DCBA} = \sum_{(ABCD)} \nabla_D A_{CBA} + T\ten{_{DC}^E} A_{EBA},
\end{equation}
where $\nabla_A=(E^{-1})\ten{_A^{M}}\nabla_M$.

In this formulation, 11-dimensional supergravity has only two dynamical superfields, namely the vielbein $E\ten{_M^A}(x,\theta)$ and the super-three-form $A_{MNP}(x,\theta)$. The equations of motion can be shown to be equivalent to constraints placed upon the components of the supertorsion and the super-four-form \cite{Cremmer2, Brink:1980az, Howe:1997he}. These \textit{supergravity constraints} read
\begin{subequations}
    \begin{align}
        & T\ten{_{\gamma\beta}^a} = -i(\Gamma^a)_{\gamma\beta}, \label{SUGRAONE} \\
        & T\ten{_{\gamma\beta}^\alpha} = T\ten{_{\gamma b}^a} = T\ten{_{cb}^a} = 0, \\
        & H_{\delta\gamma\beta\alpha}  = H_{\delta\gamma\beta a} = H_{\delta c b a} = 0\label{Hsugraconst1}, \\
        & H_{\delta \gamma b a} = i (\Gamma_{ba})_{\delta\gamma}. \label{SUGRAFOUR}
\end{align}
\end{subequations}
Using this superspace formulation, the physical fields of 11-dimensional supergravity
only appear through their covariant field strengths, namely the top component of the supercurvature $R\ten{_{abc}^d}$, the supertorsion component $T\ten{_{ab}^\alpha}$, and the four-form $H_{abcd}$. To see exactly how this is the case, we must use the Bianchi identities.

It is possible to observe that the supertorsion and the supercurvature obey the Bianchi identities
\begin{subequations}
    \begin{align}
        & \nabla \, T^A = E^B R\ten{_B^A}, \\
        & \nabla \, R\ten{_B^A} = 0.
    \end{align}
\end{subequations}
These, along with the closure relationship $\de H = 0$, can be expressed more explicitly as
\begin{subequations}
    \begin{align}
        & \sum_{(ABC)} \bigl( R\ten{_{ABC}^D} - \nabla_AT\ten{_{BC}^D} - T\ten{_{AB}^E}T\ten{_{EC}^D} \bigr) = 0, \label{BIANCHIONE} \\
        & \sum_{(ABCD)} \bigl( \nabla_A R\ten{_{BCD}^E} + T\ten{_{AB}^F} R\ten{_{FCD}^E} \bigr) = 0, \\
        & \sum_{(ABCDE)} \bigl( \nabla_A H_{BCDE} + T\ten{_{AB}^F}H_{FCDE} \bigr) = 0. \label{BIANCHITHREE}
\end{align}
\end{subequations}
Starting from these identities, we can determine expressions concerning the remaining components of the supertorsion, i.e.
\begin{subequations}
    \begin{align}
        & T\ten{_{c\beta}^\alpha} = \frac{1}{288} \bigl( \Gamma\ten{_c^{dfgh}} + 8 \delta_c^{d}\Gamma^{fgh} \bigr)\ten{_\beta^\alpha} H_{dfgh} = \bigl( \cl{T}\ten{_c^{dfgh}} \bigr)\ten{_\beta^\alpha} H_{dfgh}\label{TORSIONGOOD}, \\
        & T\ten{_{ab}^\alpha} = \frac{i}{42}(\Gamma^{cd})^{\alpha\beta} \nabla_\beta H_{abcd}, \\
        & (\Gamma^{abc})_{\alpha\beta} T\ten{_{bc}^\beta} = 0.
\end{align}
\end{subequations}
and the remaining components of the supercurvature, i.e.
\begin{subequations}
    \begin{align}
        & R\ten{_{\delta\gamma b a}} = - 2i \bigl( \Gamma_b \cl{\T}\ten{_a^{dfgh}} \bigr)_{(\delta\gamma)} H_{dfgh}, \label{Rtwo} \\
        & R\ten{_{\delta c b a}} = \frac{i}{2}\bigl[ (\Gamma_c)_{\delta \epsilon}T\ten{_{ba}^\epsilon} + 2 (\Gamma_{[a})\ten{_{\delta\epsilon}} T\ten{_{b]c}^\epsilon} \bigr], \label{Rthree} \\
        & R\ten{_{dc\beta}^\alpha} = 2 \nabla_{[d}T\ten{_{c]\beta}^\alpha} + 2 T\ten{_{[d|\beta}^\epsilon} T\ten{_{|c] \epsilon}^\alpha} + \nabla_{\beta}T\ten{_{dc}^\alpha}, \label{Rfour} .
    \end{align}
\end{subequations}
Note that the Riemann tensor is built from the superconnection and obeys 
\begin{equation} \label{Rrearrange}
  R\ten{_{DCa}^\beta}=R\ten{_{DC\beta}^a}=0\qquad , \qquad    R\ten{_{DC\beta}^\alpha} = \frac{1}{4} R\ten{_{DCba}}(\Gamma^{ba})\ten{_\beta^\alpha}.
\end{equation}
The $\Gamma$-matrix combination $\cl{T}$ is defined in \eqref{TORSIONGOOD} and $\check{\cl{T}}$ is its transposition. Finally, the Bianchi identities also give the expressions
\begin{subequations}
    \begin{align}
        & \nabla_\alpha H_{bcde} = - 6 i (\Gamma_{[bc})_{\alpha\beta} T\ten{_{de]}^\beta}, \label{Hmanip} \\
        & \nabla_\alpha R\ten{_{bcde}} = 2\nabla_{[b|}R\ten{_{\alpha |c]de}} + 2 T\ten{_{[b|\alpha}^\gamma} R\ten{_{\gamma |c] de}} - T\ten{_{bc}^\gamma}R\ten{_{\beta\alpha d e}}\label{derivedBfinal}.
    \end{align}
\end{subequations}

\section{Order-4 vielbein manipulations}\label{order-4vielbeinexpansions}
Expansion of the M2-brane action only requires knowledge of the expansion of the supervielbein. Therefore we record the expansion of the frame super-form to quartic order.

\subsection{Normal coordinate expansion of frame super-form}\label{framesuperform}
Using the expressions for the behaviour of the Lie derivative $\mathcal{L}_{\y}$ along the tangent field $\y = \y^M$, it can be established that the repeated action on the supervielbein $E^A$ gives \cite{Grisaru:2000ij}\footnote{Note that (\ref{disagree}) corrects (4.7, \cite{Grisaru:2000ij}), in which there is an erroneous extra term.}
\begin{equation}
    \mathcal{L}_{\y} E^A = \nabla \y^A + \y^C E^B T\ten{_{BC}^A},
\end{equation}
\begin{equation} \label{L^2 E}
    \bigl(\mathcal{L}_{\y}\bigr)^2 E^A = -\y^B E^C \y^D R\ten{_{DCB}^A} + \y^CE^B\y^D\nabla_DT\ten{_{BC}^A} + \y^C\bigl(\nabla\y^B + \y^E E^D T\ten{_{DE}^B}\bigr)T\ten{_{BC}^A},
\end{equation}
\begin{equation}
    \begin{split}\label{disagree}
    \bigl(\mathcal{L}_{\y}\bigr)^3 E^A = & - \y^D\bigl( \nabla\y^B + \y^FE^GT\ten{_{GF}^B}\bigr)\y^CR\ten{_{CBD}^A} - \y^DE^B\y^C\y^F\nabla_FR\ten{_{CBD}^A} \\
    & + 2\y^C\bigl(\nabla\y^B + \y^FE^GT\ten{_{GF}^B}\bigr)\y^D\nabla_DT\ten{_{BC}^A} + \y^CE^B\y^D\y^E\nabla_E\nabla_DT\ten{_{BC}^A} \\
    & + \y^C\y^G\bigl(\nabla\y^D+\y^EE^FT\ten{_{FE}^D}\bigr)T\ten{_{DG}^B}T\ten{_{BC}^A} - \y^C\y^DE^F\y^ER\ten{_{EFD}^B}T\ten{_{BC}^A}\\
    & +\y^C\y^FE^D\y^E\bigl(\nabla_ET\ten{_{DF}^B}\bigr)T\ten{_{BC}^A},
    \end{split}
\end{equation}
\begin{equation}
    \begin{split}
        \bigl(\mathcal{L}_{\y}\bigr)^4 E^A = & + 3\y^C\y^F\bigl(\nabla\y^E + \y^GE^HT\ten{_{HG}^E}\bigr)T\ten{_{EF}^B}\y^D\nabla_DT\ten{_{BC}^A} \\
    & + 3\y^C\bigl(\nabla\y^B + \y^FE^GT\ten{_{GF}^B}\bigr)\y^D\y^E\nabla_E\nabla_DT\ten{_{BC}^A} \\
    & -\y^C\y^D\bigl(\nabla\y^F + \y^GE^HT\ten{_{HG}^F}\bigr)\y^ER\ten{_{EFD}^B}T\ten{_{BC}^A} \\
    & + 2 \y^C\y^F\bigl(\nabla\y^D + \y^EE^GT\ten{_{GE}^D}\bigr)\y^H\bigl(\nabla_HT\ten{_{DF}^B}\bigr)T\ten{_{BC}^A} \\
    & - \y^D\y^F\bigl(\nabla\y^E + \y^HE^GT\ten{_{GH}^E}\bigr) T\ten{_{EF}^B}\y^CR\ten{_{CBD}^A} \\
    & - 2\y^D\bigl(\nabla\y^B + \y^FE^GT\ten{_{GF}^B}\bigr)\y^C\y^E\nabla_ER\ten{_{CBD}^A}\\ 
    & + \y^C\y^G\y^E\bigl(\nabla\y^F + \y^IE^HT\ten{_{HI}^F}\bigr)T\ten{_{FE}^D}T\ten{_{DG}^B}T\ten{_{BC}^A} \\
    & +\y^D\y^EE^F\y^GR\ten{_{GFE}^B}\y^CR\ten{_{CBD}^A} - \y^D\y^FE^G\y^E\bigl(\nabla_ET\ten{_{GF}^B}\bigr)\y^CR\ten{_{CBD}^A} \\
    & - \y^DE^B\y^C\y^E\y^F \nabla_F \nabla_ER\ten{_{CBD}^A} + \y^CE^B\y^D\y^E\y^F\nabla_F\nabla_E\nabla_DT\ten{_{BC}^A} \\
    & - \y^C\y^DE^F\y^E\y^G\bigl(\nabla_GR\ten{_{EFD}^B}\bigr)T\ten{_{BC}^A} - 3\y^C\y^DE^F\y^ER\ten{_{EFD}^B}\y^G\nabla_GT\ten{_{BC}^A} \\
    & - \y^C\y^E\y^FE^G\y^HR\ten{_{HGF}^D}T\ten{_{DE}^B}T\ten{_{BC}^A} + \y^C\y^G\y^EE^F\y^H\bigl(\nabla_H T\ten{_{FE}^D}\bigr)T\ten{_{DG}^B}T\ten{_{BC}^A} \\ 
    & + \y^C\y^FE^D\y^E\y^G\bigl(\nabla_G \nabla_E T\ten{_{DF}^B}\bigr)T\ten{_{BC}^A} + 3 \y^C\y^FE^D\y^E\bigl(\nabla_ET\ten{_{DF}^B}\bigr)\y^G\nabla_GT\ten{_{BC}^A}.
    \end{split}
\end{equation}
Notice that many terms can be rearranged in terms of the supercovariant derivative. However, while the order-1 variation can be written entirely in terms of this (in a bosonic bacgkround, one has $\nabla_m \y^\alpha + \y^\beta e\ten{_m^c} T\ten{_{c \beta}^\alpha} = D_m y^\alpha$), higher orders contain components of the super-Riemann tensor and operators involving the torsion that are difficult to rearrange in compact ways.

\subsection{Rearranging the expanded supervielbein using Bianchi identities}\label{supevielrearrapp}
Starting from the order-4 term in (\ref{rawresults}) and using (\ref{Rrearrange}) to perform some straightforward rearrangement while making use of $\Gamma$-matrix symmetries, we may write 
\begin{equation*}
\begin{split}
    \bigl(\mathcal{L}_{\y}\bigr)^4 E\ten{_m^a} = \frac{i}{4} (\y^\delta R_{\delta\epsilon b c} D_m \y^\epsilon) (\bar{\y}\Gamma^{abc}\y) + (\bar{\y}\Gamma^bD_m\y)(\bar{\y}\Gamma^a \cl{\T}\ten{_b^{dfgh}}\y)H_{dfgh} \\
    + \frac{i}{4} \y^\delta \y^\chi e\ten{_m^e}\y^\delta \y^\xi \nabla_{\xi} \bigl(R\ten{_{\delta e b c}}(\Gamma^{bc})\ten{_{\chi}^\beta} - 4 \nabla_\delta T\ten{_{e \chi}^\beta} \bigr)(\Gamma^a)_{\beta\gamma} & .
\end{split}
\end{equation*}
Now we will use (\ref{Rtwo}) in the first term and (\ref{TORSIONGOOD}, \ref{Rthree}, \ref{Hmanip}) in the third term, and we also split the third term. 
We also use the spinor index symmetry properties of $\Gamma$-matrices to write
\begin{equation*}
    (\Gamma_b \cl{\T}\ten{_c^{dfgh}})_{(\delta\epsilon)} = \frac{1}{288} \Bigl( \Gamma_{bc} \Gamma^{dfgh} - 8 \delta_{[c}^{[d}\Gamma_{b]}\Gamma^{fgh]} - 12 \delta_{[c}^{[d}\delta_{b]}^f\Gamma^{gh]} \Bigr)_{\delta\epsilon} H_{dfgh} \equiv 2 \, \cl{R}\ten{_{bc}^{dfgh}} H_{dfgh},
\end{equation*}
so we eventually arrive at
\begin{equation*}
    \begin{split}
    \bigl(\mathcal{L}_{\y}\bigr)^4 E\ten{_m^a} = \bigl[ \bar{\y} ( \cl{R}\ten{_{bc}^{dfgh}} H_{dfgh} D_m \y \bigr] (\bar{\y}\Gamma^{abc}\y) + (\bar{\y}H_{dfgh}\Gamma^bD_m\y)(\bar{\y}\Gamma^a \cl{\T}\ten{_b^{dfgh}}\y) \\
    + \frac{1}{8} (\bar{\y}\Gamma^{abc}\y) e\ten{_m^e} \y^\delta \bigl[ (\Gamma_e)_{\delta\sigma}\y^\xi  \nabla_\xi T\ten{_{bc}^\sigma} + 2 (\Gamma_b)_{\delta\sigma}\y^\xi  \nabla_\xi T\ten{_{e c}^\sigma}\bigr] \\
    - 6 \, (\bar{\y}\Gamma^a \cl{\T}\ten{_b^{dfgh}}\y) e\ten{_m^b} \y^\delta  (\Gamma_{df})_{\delta\sigma} \y^\xi \nabla_\xi T\ten{_{gh}^\sigma} & .
\end{split}
\end{equation*}
We see that a number of previously nasty-looking curvature and torsion terms are all reducible to expressions involving gamma matrices and the spinor derivative of the supercovariantized gravitino fields strength $\nabla_\xi T\ten{_{gh}^\sigma}$.

To assess this we step back to superspace momentarily. Using the superspace covariant derivative 
\begin{equation*}
    \nabla_M v\ten{_A^B} = \partial_M v\ten{_A^B} + v\ten{_A^C}\omega\ten{_{MC}^B} - \omega\ten{_{MA}^C}v\ten{_C^B} ,
\end{equation*}
we have $
    \big[\nabla_M,\nabla_N\big]v_A = -R\ten{_{MNA}^B}v_B,$
and so
\begin{equation*}
\begin{split}
    \big[ E\ten{_A^M}\nabla_M, E\ten{_B^N}\nabla_N\big]v_C & = \bigl( E\ten{_A^M}E\ten{_B^N}\big[\nabla_M, \nabla_{N}\big] - 2 E\ten{_{[A}^M}E\ten{_{B]}^N} (\nabla_M E\ten{_N^D}) \nabla_D\bigr) v_C \\
    & = - R\ten{_{ABC}^D}v_D - T\ten{_{AB}^D}\nabla_D v_C.
\end{split}
\end{equation*}
This means that we have $R\ten{_{ab\gamma}^\delta} = - [\nabla_a, \nabla_b]\ten{_{\gamma}^\delta} - T\ten{_{ab}^\mu}(\nabla_\mu)\ten{_\gamma^\delta}$, which in bosonic backgrounds is $R\ten{_{ab\gamma}^\delta} = - [\nabla_a, \nabla_b]\ten{_{\gamma}^\delta}$.
Using Bianchi identity results, we have in bosonic backgrounds,
\begin{equation*}
\begin{split}
    \nabla_{\gamma} T\ten{_{ab}^\delta} & = R\ten{_{ab\gamma}^\delta} - 2 \nabla_{[a}T\ten{_{b]\gamma}^\delta} - 2 T\ten{_{[a| \gamma}^\sigma} T\ten{_{\sigma|b]}^\delta} = - \big[\nabla_a + T_a, \nabla_b+T_b\big]\ten{_{\gamma}^\delta}.
\end{split}
\end{equation*}
In terms of the supercovariant derivative $D_m$ we can eventually write $\nabla_{\gamma} T\ten{_{ab}^\delta}  = e\ten{_a^m}e\ten{_b^n}\big[D_m, D_n\big]\ten{^\delta_{\gamma}}$. Applying this result we see that the two objects defined in (\ref{massagedresultsone}) arise naturally by combining terms, as we have
\begin{equation*}
\begin{split}
    \bigl(\mathcal{L}_{\y}\bigr)^4 E\ten{_m^a} = (\bar{\y}\Gamma^{abc}\y) \biggl[ \bar{\y} \biggl( \cl{R}\ten{_{bc}^{dfgh}} H_{dfgh} D_m + \frac{1}{8}\Gamma_e e\ten{_m^e} e\ten{_{b}^p}e\ten{_{c}^q}[D_p,D_q] + \frac{1}{4} \Gamma_b e\ten{_{c}^q}[D_m,D_q]\biggr) \y \biggr] \\
    + (\bar{\y}\Gamma^a \cl{\T}\ten{_b^{dfgh}}\y \bigr) \Bigl[ \bar{\y} \bigl(H_{dfgh}\Gamma^bD_m - 6e\ten{_m^b} \Gamma_{df} e\ten{_{g}^p}e\ten{_{h}^q}[D_p,D_q] \bigr) \y \Bigr] & ,
\end{split}
\end{equation*}
which means
\begin{equation}
    \bigl(\mathcal{L}_{\y}\bigr)^4 E\ten{_m^a} = (\bar{\y}\Gamma^{abc}\y)(\bar{\y} \cl{W}_{m bc} \y) +  (\bar{\y}\Gamma^a \cl{\T}\ten{_b^{dfgh}}\y)(\bar{\y} \cl{H}\ten{^b_{mdfgh}}\y).
\end{equation}

\section{Catalogue of dimensional reductions}\label{DimRedCatApp}
In this appendix we catalogue the details of the dimensional reductions of all the terms appearing in the main text.

\subsubsection*{Notation}
In the M-theory formulation, we consider the 11-dimensional spacetime to be spanned by the coordinates $x^{\hat{m}}$. This is reduced to a 10-dimensional string background via the split $x^{\hat{m}}=(x^m,x^{10})$. Unless differently stated, 11-dimensional indices are hatted whereas 10-dimensional indices are not; 11-dimensional objects are also hatted and 10-dimensional objects are not. So vectors in the 11- and 10-dimensional spacetimes read $\hat{\omega} = \hat{\omega}_{\hat{m}} \, \de x^{\hat{m}}$ and $\omega = \omega_m \, \de x^m$, respectively, and similarly for tensors of arbitrary rank. Indices $\hat{a}, \hat{b}$ and $a,b$ are 11- and 10-dimensional tangent spacetime indices, respectively, with explicit number indices being underlined for tangent space and unadorned for spacetime. Background fields are always independent of the extra M-theory coordinate $x^{10}$.

The M2- and D2-brane 3-dimensional worldvolumes are spanned by the coordinates $\xi^i$. Pulling an object back from eleven dimensions and pulling an object back from ten dimensions are different manoeuvres: for ease of notation, instead of writing these pullbacks explicitly, we shall keep track of which is being used by noting whether the object itself it hatted or not. For instance, denoting for a moment the pullback from the 11-dimensional spacetime to the 3-dimensional M2-brane worldvolume with $\varphi_\star$ and the pullback from the 10-dimensional spacetime to the 3-dimensional D2-brane worldvolume with $\phi_\star$, for two vectors $\hat{\omega}_{\hat{m}}$ and $\omega_{m}$ we will write
$\hat{\omega}_{i} = (\varphi_\star \hat{\omega})_{i} = \der_i x^{\hat{m}} \hat{\omega}_{\hat{m}}$ and $\omega_{i} = (\phi_\star \omega)_{i} = \der_i x^m \omega_{m}$.

The $n$-dimensional Levi-Civita symbol $\varepsilon_{\mu_1 \dots \mu_n}$ is normalized as $\varepsilon_{1 \dots n} = +1$ and the Levi-Civita tensor is defined as $\epsilon_{\mu_1 \dots \mu_n} = (- \mathrm{det} \, g)^{1/2} \, \varepsilon_{\mu_1 \dots \mu_n}$, where $g_{\mu_1 \mu_2}$ is the associated $n$-dimensional metric. Similarly, we define the symbol $\varepsilon^{\mu_1 \dots \mu_n} \equiv - \varepsilon_{\mu_1 \dots \mu_n}$ and $\epsilon^{\mu_1 \dots \mu_n} = (- \mathrm{det} \, g)^{-1/2} \, \varepsilon^{\mu_1 \dots \mu_n}$.

Antisymmetric and symmetric combinations of a number $n$ of indices are denoted by square brackets and parentheses, respectively, and include a normalization factor $1/n!$. For instance, we have $\Gamma_{[1} \dots \Gamma_{n]} = \sum_{\sigma \in S_n} \, {\mathrm{sgn} (\sigma)} \, \Gamma_{\sigma(1)} \dots \Gamma_{\sigma(n)}/n!$, where $\sigma \in S_n$ are the permutations of $n$ elements.

\subsection{Basic dimensional reductions}
We report details about the dimensional reductions of the essential quantities that are needed in the analysis of M2- and D2-branes.

\subsubsection*{Metric}
In terms of 10-dimensional quantities, the 11-dimensional vielbein splits according to the standard ansatz
\begin{equation} \label{metric}
\hat{e}\ten{_{\hat{m}}^{\hat{a}}} = \begin{pmatrix} e^{-\frac{\phi}{3}}e\ten{_m^a} & e^{\frac{2\phi}{3}} C_m \\
    0 & e^{\frac{2\phi}{3}} \end{pmatrix},
\end{equation}
where $e\ten{_m^a}$ is the 10-dimensional string frame vielbein, $\phi$ is the dilaton, and $C^{(1)} = \de x^m C_m$ is the Ramond-Ramond one-form. The vielbein is invertible and its inverse reads
\begin{equation}
\hat{e}\ten{_{\hat{a}}^{\hat{m}}} = \begin{pmatrix}
e^{\frac{\phi}{3}} e\ten{_a^m} & - e^{\frac{\phi}{3}} C_a \\
0 & e^{-\frac{2\phi}{3}} \end{pmatrix}.
\end{equation}
The 11-dimensional metric is defined in terms of the vielbein as $\hat{g}_{\hat{m}\hat{n}} = \hat{e}\ten{_{\hat{m}}^{\hat{a}}} \hat{e}\ten{_{\hat{n}}^{\hat{b}}} \hat{\eta}_{\hat{a}\hat{b}}$, where $\hat{\eta}_{\hat{a}\hat{b}}$ is the 11-dimensional Minkowski metric, so it reads
\begin{equation*}
\begin{split}
 \hat{g}_{\hat{m}\hat{n}} = \begin{pmatrix}
 e^{-\frac{2\phi}{3}} g_{mn} + e^{\frac{4 \phi}{3}} C_m C_n & e^{\frac{4\phi}{3}} C_m \\
e^{\frac{4\phi}{3}} C_n  & e^{\frac{4\phi}{3}} \end{pmatrix},
\end{split}
\end{equation*}
where the 10-dimensional metric is defined as $g_{mn} = e\ten{_{m}^{a}} e\ten{_n^b} \eta_{ab}$, with $\eta_{ab}$ the 10-dimensional Minkowski metric.

\subsubsection*{Three-form field}
We describe the dimensional reduction of the 11-dimensional three-form $\hat{A} = \de x^{\hat{m}} \wedge \de x^{\hat{n}} \wedge \de x^{\hat{p}} \hat{A}_{\hat{p}\hat{n}\hat{m}} / 3!$ in terms of two 10-dimensional form fields $C^{(3)} = \de x^m \wedge \de x^n \wedge \de x^p C_{pnm} / 3!$ and $B^{(2)} = \de x^m \wedge \de x^n B_{nm}/2!$ defined as
\begin{subequations}
\begin{align}
    & \hat{A}_{mnp} = C_{mnp}, \label{AC} \\
    & \hat{A}_{mn \,10} = B_{mn}, \label{AB}
\end{align}
\end{subequations}
The 11-dimensional flux is defined as $\hat{H} = \de \hat{A}$, while in the 10-dimensional formulation we have $F^{(4)} = \de C^{(3)}$ and $H^{(3)} = \de B^{(2)}$, so the 10-dimensional form field strengths are such that
\begin{subequations}
\begin{align}
    & \hat{H}_{mnpq} = F_{mnpq}, \\
    & \hat{H}_{mnp \, 10} = H_{mnp}.
\end{align}
\end{subequations}

An analysis of the dimensional-reduction ansatz shows that the tangent-space 11-dimensional flux is related to the 10-dimensional field-strength tensors as
\begin{subequations}
    \begin{align} 
    & \hat{H}_{a b c \underline{10}} = e^{\frac{\phi}{3}} \, e\ten{_a^m} e\ten{_b^n} e\ten{_c^p} H_{mnp},  \label{H ST tangent} \\
    & \hat{H}_{a b c d} = e^{\frac{4 \phi}{3}} \, e\ten{_a^m} e\ten{_b^n} e\ten{_c^p} e\ten{_d^q} (F_{mnpq} - 4 \, H_{[mnp} C_{q]}) = e^{\frac{4 \phi}{3}} \, e\ten{_a^m}e\ten{_b^n}e\ten{_c^p}e\ten{_d^q}F^{(4)}_{mnpq},
\end{align}
\end{subequations}
where we defined the combination $F^{(4)} = \de C^{(3)} - C^{(1)} \wedge H^{(3)}$.

\subsubsection*{$\boldsymbol{\hat{\Gamma}}$-matrices}
In tangent spacetime, the first ten $\hat{\Gamma}$-matrices are the same, i.e. $\hat{\Gamma}^{a} = \Gamma^a$, since the Clifford algebra is the same as a consequence of the equality $\hat{\eta}_{ab} = \eta_{ab}$; the last $\hat{\Gamma}$-matrix defined as the chirality matrix $\hat{\Gamma}^{\underline{10}} \equiv \Gamma^{\ulv}$. In curved spacetime, the 11-dimensional $\hat{\Gamma}$-matrices and 10-dimensional $\Gamma$-matrices are then related as
\begin{align}
    & \hat{\Gamma}_m = e^{-\frac{\phi}{3}} \bigl( \Gamma_{m} + e^{\phi} C_m \Gamma^{\ulv} \bigr), \label{gammaindex} \\
    & \hat{\Gamma}_{10} = e^{\frac{2 \phi}{3}} \Gamma^{\ulv}. \label{gammastar}
\end{align}
One also finds,
\begin{align}
    & \hat{\Gamma}_{mn} = e^{-\frac{2 \phi}{3}} \bigl( \Gamma_{mn} - 2 \, e^{\phi} C_{[m} \Gamma_{n]} \Gamma^{\ulv} \bigr), \\
    & \hat{\Gamma}_{mnp} = e^{-\phi} \bigl( \Gamma_{mnp} + 3 \, e^{\phi} C_{[m} \Gamma_{np]} \Gamma^{\ulv} \bigr).
\end{align}

For contractions of the components of a form field $\omega_p$ with a number $n$ of 10-dimensional curved-spacetime $\Gamma$-matrices $\Gamma^m$, we employ the underlined notation
\begin{equation}
    \ul{\omega}_{q_1 \, q_2 \dots q_m} = \frac{1}{n!} \omega_{q_1 \dots q_m p_1 \dots p_n} \Gamma^{p_1 \dots p_n}.
\end{equation}

\subsection{Supercovariant derivatives}\label{supcovapp}

\subsubsection*{Spin connection}
The 11-dimensional spin connection is defined in terms of the anhomology coefficients as
\begin{equation}
    \hat{\omega}\ten{_{\hat{a}\hat{b}}^{\hat{c}}} = \frac{1}{2}\Bigl(\hat{\Omega}\ten{_{\hat{a}\hat{b}}^{\hat{c}}} -  \hat{\Omega}\ten{_{\hat{a}\hat{d}}^{\hat{e}}}\hat{\eta}^{\hat{c}\hat{d}}\hat{\eta}_{\hat{b}\hat{e}}
    -\hat{\Omega}\ten{_{\hat{b}\hat{d}}^{\hat{e}}}\hat{\eta}^{\hat{c}\hat{d}}\hat{\eta}_{\hat{a}\hat{e}}\Bigr),
\end{equation}
where the latter read
\begin{equation*}
    \hat{\Omega}\ten{_{\hat{a}\hat{b}}^{\hat{c}}} =  \hat{e}\ten{_{\hat{a}}^{\hat{m}}} \hat{e}\ten{_{\hat{b}}^{\hat{n}}}\Bigl( \hat{\partial}_{\hat{m}} \hat{e}\ten{_{\hat{n}}^{\hat{c}}} - \hat{\partial}_{\hat{n}} \hat{e}\ten{_{\hat{m}}^{\hat{c}}}\Bigr).
\end{equation*}
These allow us to express the 11-dimensional spin connection in terms of 10-dimensional operators as
\begin{subequations}
    \begin{align}
        & \hat{\omega}\ten{_{{a}{b}}^{{c}}} = e^\frac{\phi}{3} \biggl[ \omega\ten{_{ab}^c} + \frac{1}{3} \partial_b\phi \delta_a^c - \frac{1}{3} \partial^c\phi \eta_{ba} \biggr], \\
        & \hat{\omega}\ten{_{{a}{b}}^{\underline{10}}} = \frac{1}{2}e^{\frac{4\phi}{3}}F_{ab}, \\
        & \hat{\omega}\ten{_{\underline{10} \, {a}}^{c}} = \hat{\omega}\ten{_{{a}\underline{10}}^{c}} = - \frac{1}{2} e^{\frac{4\phi}{3}} F\ten{_a^c}, \\
        & \hat{\omega}\ten{_{\underline{10} \, {a}}^{\underline{10}}} = -\frac{2}{3}e^{\frac{\phi}{3}} \partial_a\phi , \\
        & \hat{\omega}\ten{_{\underline{10} \, {\underline{10}}}^{c}} =  \frac{2}{3}e^{\frac{\phi}{3}} \partial^c\phi,
\end{align}
\end{subequations}
where all the remaining combinations are vanishing, i.e. $\hat{\omega}\ten{_{a\underline{10}}^{\underline{10}}} = \hat{\omega}\ten{_{\underline{10}~\underline{10}}^{\underline{10}}} = 0$.

\subsubsection*{Torsion}
The 11-dimensional torsion term that appears in the M2-brane action is the $\Gamma$-matrix valued term
\begin{equation}
    \hat{\T}_{\hat{a}} = \frac{1}{288} \bigl(\hat{\Gamma}\ten{_{\hat{a}}^{\hat{b}\hat{c}\hat{d}\hat{e}}} - 8 \delta_{\hat{a}}^{\hat{b}} \hat{\Gamma}^{\hat{c}\hat{d}\hat{e}}\bigr) \hat{H}_{\hat{b}\hat{c}\hat{d}\hat{e}}.
\end{equation}
In terms of 10-dimensional operators, the 11-dimensional torsion components can be seen to split as
\begin{subequations}
    \begin{align}
    & \, \hat{\T}_a = \frac{1}{12} e^{\frac{\phi}{3}} \bigl[ \Gamma_a (e^\phi \ul{F}^{(4)} + \ul{H}^{(3)}\Gamma^{\ulv}) - 3 \, e\ten{_a^m} (e^\phi \ul{F}^{(4)}_m + \ul{H}_m^{(3)} \Gamma^{\ulv}) \bigr],\label{torsiondimred} \\
    & \, \hat{\T}_{\ul{10}} = \frac{1}{12} e^{\frac{\phi}{3}} \bigl[ \Gamma^{\ulv}(e^\phi \ul{F}^{(4)} - 2\ul{H}^{(3)}\Gamma^{\ulv})\bigr]\label{tprsiondimredstar}.
\end{align}
\end{subequations}

\subsubsection*{Supercovariant derivative}
In dealing with the M2-brane action, the spinor kinetic term contains the worldvolume pullback of the 11-dimensional spacetime operator
\begin{equation}
    \hat{D}_{\hat{m}} = \hat{\nabla}_{\hat{m}} - \hat{\check{T}}_{\hat{m}},
\end{equation}
where $\hat{\nabla}_{\hat{m}}$ is the 11-dimensional spinor covariant derivative and $\hat{\T}_{\hat{m}}$ is the 11-dimensional torsion, which are defined in the tangent spacetime as
\begin{subequations}
    \begin{align}
        & \nabla_{\hat{a}} = \partial_{\hat{a}} + \dfrac{1}{4} \, \hat{\omega}\ten{_{\hat{a}}^{\hat{b} \hat{c}}} \hat{\Gamma}_{\hat{c} \hat{d}}, \\
        & \hat{\T}_{\hat{a}} = \frac{1}{288} \bigl(\hat{\Gamma}\ten{_{\hat{a}}^{\hat{b}\hat{c}\hat{d}\hat{e}}} - 8 \delta_{\hat{a}}^{\hat{b}} \hat{\Gamma}^{\hat{c}\hat{d}\hat{e}}\bigr) \hat{H}_{\hat{b}\hat{c}\hat{d}\hat{e}}.
    \end{align}
\end{subequations}
Using the above relations one can dimensionally reduce the 11 dimensional supercovariant derivative and write it in terms of 10 dimensional operators \eqref{eq:IIA-operators}, recovering the relations \eqref{eq:dim-reduc-operators}.

\subsection{Pullbacks}
We report details about the relationships between pullbacks onto M2- and D2-brane worldvolumes.

\subsubsection*{Metric}
Defining the combination 
\begin{equation}
    p_i = \partial_i x^{10} + \partial_i x^m C_m,
\end{equation}
which is the dual to the world volume flux on the D2-brane, we can express the metric pullback as
\begin{equation}
    \hat{g}_{ij} = e^{-\frac{2\phi}{3}} g_{ij} + e^{\frac{4\phi}{3}}p_i p_j.
\end{equation}
Equivalently, the pullback of the vielbein is
\begin{subequations}
    \begin{align}
    & \hat{e}\ten{_i^a} = e^{-\frac{\phi}{3}} e\ten{_i^a}, \\
    & \hat{e}\ten{_i^{\ul{10}}} = e^{\frac{2\phi}{3}} p_i.
\end{align}
\end{subequations}

Since the pulled-back metrics are 3-dimensional, using the shorthand $g^{ij} p_i p_j = p^2$, we get the exact relationship
\begin{equation}
    \det(\hat{g}_{ij}) = e^{-2\phi} \det(g_{ij}) \Bigl( 1 + e^{2\phi}p^2 \Bigr),
\end{equation}
To conclude, the relationship between the inverses of the pulled-back metrics can be seen to be
\begin{equation}
    \begin{split}
        \hat{g}^{ij} & = e^{\frac{2 \phi}{3}} \bigg( g^{ij} - \frac{e^{2\phi} p^ip^j}{1 + e^{2\phi}p^2} \bigg).
    \end{split}
\end{equation}

\subsubsection*{Three-form field}
For the three-form field, we can write
\begin{equation} \label{Adimred}
     \hat{A}_{ijk} = C_{ijk} - 3 \, C_{[i} B_{jk]} + 3 \, p_{[i} B_{jk]}.
\end{equation}

\subsubsection*{$\boldsymbol{\hat{\Gamma}}$-matrices}
The relationship between the 11-dimensional $\hat{\Gamma}$-matrix pullbacks and 10-dimensional $\Gamma$-matrix pullbacks is
\begin{equation}
    \hat{\Gamma}_i = e^{-\frac{\phi}{3}} \bigl( \Gamma_{{i}} + e^{\phi} p_i {\Gamma}^{\ulv} \bigr).
\end{equation}
Starting from this, we can then express the antisymmetric combinations of $\hat{\Gamma}$-matrices as
\begin{align}
    & \hat{\Gamma}_{ij} = e^{-\frac{2\phi}{3}} \bigl( \Gamma_{{ij}} - 2 \, e^{\phi} \, p_{{[i}}  \Gamma_{{j]}} {\Gamma}^{\ulv} \bigr), \\
    & \hat{\Gamma}_{ijk} = e^{-\phi} \bigl( \Gamma_{{ijk}} + 3 \, e^{\phi} \, p_{{[i}}  \Gamma_{{j k]}} {\Gamma}^{\ulv} \bigr).
\end{align}
Matrices with upper indices are defined by use of the metric pullback inverse, i.e. $\hat{\Gamma}^i =  \hat{g}^{ij} \hat{\Gamma}_j$ and $\Gamma^i = g^{ij} \Gamma_j$, and they are related as
\begin{equation} \label{hatGamma^i}
    \begin{split}
        \hat{\Gamma}^i = e^{\frac{\phi}{3}} \biggl[ \Gamma^i + \frac{e^\phi p^i}{1 + e^{2\phi}p^2} \Bigl( \Gamma^{\ulv} - e^{\phi}  \Gamma^k p_k \Bigr) \biggr].
    \end{split}
\end{equation}

\subsubsection*{Supercovariant derivative pullback}
The operator that appears in the M2-brane action is the 11-dimensional spinor covariant derivative pullback $\hat{D}_i\hat{\theta}$. By making use of the results above, we can determine that in terms of the D2-brane operators this reads
\begin{equation}
    \hat{D}_i \hat{\theta} = \esix \biggl[ D_i - \frac{1}{6} \Gamma_i \Delta + \frac{1}{3}e^\phi p_i \Gamma^{\ulv} \Delta \biggr] \theta.
\end{equation}

\subsection{Order-4 combinations} \label{order-4_combinations}
In the order-4 fermion expansions we find combinations of the operators that appear at second order. These are discussed in detail below.

\subsubsection*{$\boldsymbol{\hat{\Gamma}}$-matrices and fluxes}
We now treat the term
\begin{equation}
    \hat{{\cl{R}}}_{\hat{b}\hat{c}} = \frac{1}{576} \bigl( \hat{\Gamma}_{\hat{b}\hat{c}} \hat{\Gamma}^{\hat{d}\hat{f}\hat{g}\hat{h}} - 8 \delta_{[\hat{c}}^{\hat{d}}\hat{\Gamma}_{\hat{b}]}\hat{\Gamma}^{\hat{f}\hat{g}\hat{h}} - 12 \delta_{[\hat{c}}^{\hat{d}}\delta_{\hat{b}]}^{\hat{f}}\hat{\Gamma}^{\hat{g}\hat{h}} \bigr) \hat{H}_{\hat{d}\hat{f}\hat{g}\hat{h}}.
\end{equation}
Upon dimensional reduction, this allows us to define
\begin{subequations}
    \begin{align}
        & {\flux}_{bc} \equiv e^{-\frac{\phi}{3}} \hat{{\cl{R}}}_{bc} = \frac{1}{24} \Bigl[ \Gamma_{bc} \bigl( e^\phi {\ul{F}}^{(4)} \!+ \ul{H}^{(3)}\Gamma^{\ulv} \bigr) - 2 \Gamma_{[b} \bigl(e^\phi {\ul{F}}^{(4)}_{c]} \!+ \ul{H}^{(3)}_{c]} \Gamma^{\ulv}\bigr) + \bigl(e^\phi {\ul{F}}^{(4)}_{bc} \!+ \ul{H}^{(3)}_{bc}\Gamma^{\ulv} \bigr) \Bigr]\label{Rdimredone}, \\
        & {\flux}_b \equiv e^{-\frac{\phi}{3}} \hat{{\cl{R}}}_{b \, \ul{10}} = - \frac{1}{24} \, e^{\phi} \Gamma^{\ulv} \Bigl[ \Gamma_b {\ul{F}}^{(4)} \!- {\ul{F}}^{(4)}_b \Bigr]\label{Rdimeredtwo}.
    \end{align}
\end{subequations}

\subsubsection*{Supercovariant derivative commutator}
An operator appearing frequently in the order-4 fermionic expansion is the commutator of supercovariant derivatives on which we must perform dimensional reduction. First of all, we have
\begin{equation*}
    \bigl[ \hat{D}_p, \hat{D}_{10} \bigr] \hat{\y} = \frac{1}{3} \, e^{-\frac{\phi}{6}} e^\phi \cmttr_p \theta,
\end{equation*}
where we have defined the operator
\begin{equation}
    \cmttr_p \equiv \biggl[ D_{p} - \frac{1}{6} \Gamma_{p} \Delta, \Gamma^{\ulv} \Delta \biggr] + (\partial_p \phi)\Gamma^{\ulv}\Delta.
\end{equation}
The other non-zero commutator reads
\begin{equation*}
    \bigl[ \hat{D}_p, \hat{D}_q \bigr] \hat{\y} = \, e^{-\frac{\phi}{6}}\cmttr_{pq}\theta - \frac{2}{3}e^{-\frac{\phi}{6}}e^\phi C_{[p}\cmttr_{q]},
\end{equation*}
where we have defined the operator
\begin{equation}
    {\cmttr}_{pq} = \Bigl[D_p - \frac{1}{6}\Gamma_p \Delta, D_q - \frac{1}{6}\Gamma_q \Delta \Bigr] + \frac{1}{3} e^{\phi} F^{(2)}_{pq} \Gamma^{\ulv} \Delta.
\end{equation}
To conclude, we notice that $\hat{e}\ten{_{\ul{10}}^{\hat{p}}}\hat{e}\ten{_{\ul{10}}^{\hat{q}}} \big[\hat{D}_{\hat{p}}, \hat{D}_{\hat{q}}\big] \hat{\y} = \hat{e}\ten{_{\ul{10}}^{10}}\hat{e}\ten{_{\ul{10}}^{10}}\big[\hat{D}_{10} , \hat{D}_{10}\big] \hat{\y} = 0$. From these results, one can immediately derive 
\begin{subequations}
    \begin{align}
        & [\hat{D}_{a},\hat{D}_{b}]\hat{\theta} = e^{\frac{\phi}{2}} e\ten{_a^p} e\ten{_b^q} \cmttr_{pq}\theta, \label{commutedimredone} \\
        & [\hat{D}_{a},\hat{D}_{{\ul{10}}}]\hat{\theta} = \frac{1}{3} e^{\frac{\phi}{2}} e\ten{_a^p} \cmttr_{p}\theta. \label{commutedimredtwo}
    \end{align}
\end{subequations}

\def\y{\y}

\subsection{Dimensional reduction of the quartic 11-dimensional shifted fields for the dilaton}\label{dimredquarApp} \def\y{\theta}
In this appendix we provide an example of the dimensional reduction calculation for the quartic fermionic terms. We will concentrate on the dilaton as these terms are the least formidable, however the approach is fundamentally the same for the dimensional reduction for all the quartic fermonic terms in 11 dimensions. We will make heavy use of the results in appendix \ref{DimRedCatApp}.

The relationship between the quartic fermionic expansion of the 11-dimensional metric $\bs{\hat{g}}_{\hat{m}\hat{n}}$ and the quartic fermionic expansions of the 10-dimensional metric $\bs{g}_{mn}$, Ramond-Ramond one-form $\bs{C}^{(1)}_m$, and dilaton $\bs{\phi}$ is \eqref{zz1}. The expansion of the 11-dimensional metric is \eqref{quarticm2met}. Plugging in \eqref{crazyobjects} we can write the 11-dimensional shifted metric as
\begin{equation}\label{pluggedin}
\begin{split}
    \hat{\gamma}_{\hat{m}\hat{n}} = & \, - \frac{1}{4} \bigl( \hat{\bar{\theta}} \hat{\Gamma}_{\hat{a}} \hat{D}_{(\hat{m}} \hat{\theta} \bigr) \bigl( \hat{\bar{\theta}}\hat{\Gamma}^{\hat{a}}\hat{D}_{\hat{n})}\hat{\theta} \bigr) + \frac{1}{12} \bigl( \hat{\bar{\theta}} \hat{\Gamma}_{(\hat{m}|} \hat{\check{{T}}}\ten{_{\hat{a}}} \hat{\theta} \bigr) \bigl( \hat{\bar{\theta}} \hat{\Gamma}^{\hat{a}} \hat{D}_{|\hat{n})} \hat{\theta} \bigr) \\
    & - \frac{1}{576} \, \hat{g}_{\hat{m}\hat{n}} \bigl( \hat{\bar{\theta}} \hat{\Gamma}\ten{^{\hat{a}\hat{b}\hat{c}\hat{d}}} \hat{\theta} \bigr) \bigl( \hat{\bar{\theta}} \hat{\Gamma}_{\hat{a}\hat{b}}
    [\hat{D}_{\hat{c}}, \hat{D}_{\hat{d}}]\hat{\theta} \bigr) + \frac{1}{96} \bigl( \hat{\bar{\theta}} \hat{\Gamma}_{(\hat{m}} \hat{\Gamma}^{\hat{a}\hat{b}\hat{c}}\hat{\theta} \bigr) \bigl( \hat{\bar{\theta}} \hat{\Gamma}_{\hat{n})\hat{a}} 
    [\hat{D}_{\hat{b}}, \hat{D}_{\hat{c}}]\hat{\theta} \bigr)
    \\
    & + \frac{1}{96} \bigl( \hat{\bar{\theta}} \hat{\Gamma}_{(\hat{m}} \hat{\Gamma}^{\hat{a}\hat{b}\hat{c}]} \hat{\theta} \bigr) \bigl( \hat{\bar{\theta}} \hat{\Gamma}_{\hat{a}\hat{b}} 
    [\hat{D}_{|\hat{n})}, \hat{D}_{\hat{c}}] \hat{\theta} \bigr) + \frac{1}{12} \bigl( \hat{\bar{\theta}} \hat{\Gamma}_{(\hat{m}|} \hat{\Gamma}^{\hat{a}\hat{b}}\hat{\theta} \bigr) \bigl( \hat{\bar{\theta}} \hat{\cl{R}} \ten{_{\hat{a}\hat{b}}} \hat{D}_{|\hat{n})}\hat{\theta} \bigr)
    \\
    & + \frac{1}{96} \bigl( \hat{\bar{\theta}} \hat{\Gamma}_{(\hat{m}|} \hat{\Gamma}^{\hat{a}\hat{b}} \hat{\theta} \bigr) \bigl( \hat{\bar{\theta}} \hat{\Gamma}_{\hat{n})}  
    [\hat{D}_{\hat{a}}, \hat{D}_{\hat{b}}] \hat{\theta} \bigr)
    + \frac{1}{48} \bigl( \hat{\bar{\theta}} \hat{\Gamma}_{(\hat{m}|} \hat{\Gamma}^{\hat{a}\hat{b}} \hat{\theta} \bigr) \bigl( \hat{\bar{\theta}} \hat{\Gamma}_{\hat{a}} 
    [\hat{D}_{|\hat{n})}, {\hat{D}}_{\hat{b}}] \hat{\theta} \bigr).
\end{split}
\end{equation}
Plugging this into \eqref{dilatonexpansion} allows us to write the dilaton quartic shift as
\begin{equation*}
\begin{split}
    \rho^{(4)}= & \, - \frac{1}{768} \bigl( \hat{\bar{\theta}} \hat{\Gamma}\ten{^{\hat{a}\hat{b}\hat{c}\hat{d}}}\hat{\theta} \bigr) \bigl( \hat{\bar{\theta}} \hat{\Gamma}_{\hat{a}\hat{b}} 
    [\hat{D}_{\hat{c}},\hat{D}_{\hat{d}}]\hat{\theta} \bigr) + \frac{1}{128} \, e^{-\frac{4\phi}{3}} \bigl( \hat{\bar{\theta}} \hat{\Gamma}_{10} \hat{\Gamma}^{\hat{a}\hat{b}\hat{c}} \hat{\theta} \bigr) \Bigl( \hat{\bar{\theta}} \bigl( \hat{\Gamma}_{10\hat{a}} 
    [\hat{D}_{\hat{b}},\hat{D}_{\hat{c}}] + \hat{\Gamma}_{\hat{a}\hat{b}} 
    [\hat{D}_{10},\hat{D}_{\hat{c}}] \bigr) \hat{\theta} \Bigr) 
    \\
    & + \frac{1}{128} \, e^{-\frac{4\phi}{3}} \bigl( \hat{\bar{\theta}} \hat{\Gamma}_{10} \hat{\Gamma}^{\hat{a}\hat{b}} \hat{\theta} \bigr) \Bigl( \hat{\bar{\theta}} \bigl( \hat{\Gamma}_{10}  
    [\hat{D}_{\hat{a}},\hat{D}_{\hat{b}}] + 2\hat{\Gamma}_{\hat{a}} 
    [\hat{D}_{10},{\hat{D}}_{\hat{b}}] \bigr) \hat{\theta} \Bigr)
    - \frac{3}{16} \, e^{-\frac{4\phi}{3}} \bigl( \hat{\bar{\theta}} \hat{\Gamma}_{\hat{a}} \hat{D}_{10} \hat{\theta} \bigr) \bigl( \hat{\bar{\theta}}\hat{\Gamma}^{\hat{a}}\hat{D}_{10}\hat{\theta} \bigr) 
    \\
    & + \frac{1}{16} \, e^{-\frac{4\phi}{3}} \bigl( \hat{\bar{\theta}} \hat{\Gamma}_{10} \hat{\check{{T}}}\ten{_{\hat{a}}} \hat{\theta} \bigr) \bigl( \hat{\bar{\theta}} \hat{\Gamma}^{\hat{a}} \hat{D}_{10}\hat{\theta} \bigr) + \frac{1}{16} \, e^{-\frac{4\phi}{3}} \bigl( \hat{\bar{\theta}} \hat{\Gamma}_{10} \hat{\Gamma}^{\hat{a}\hat{b}} \hat{\theta} \bigr) \bigl( \hat{\bar{\theta}} \hat{\cl{R}}\ten{_{\hat{a}\hat{b}}}  \hat{D}_{10}\hat{\theta} \bigr) + \frac{1}{24} \bigl( \bar{\y} \Delta \y \bigr)^2.
\end{split}
\end{equation*}
We will demonstrate the dimensional reduction of these terms in detail. The dimensional reduction of the terms involved in the quartic shifts of the other type IIA fields follows in a very similar way, so we will forgo spelling these out. Let us tackle the dilaton shift one term at a time. We will variously require, \eqref{gammastar}, \eqref{torsiondimred}, \eqref{tprsiondimredstar}, \eqref{Rdimredone}, \eqref{Rdimeredtwo}, \eqref{commutedimredone}, and \eqref{commutedimredtwo}, at different stages of the calculations. In the order in which the terms appear, we have from the first term
\begin{equation*}
\begin{split}
    & - \frac{1}{768} (\hat{\bar{\theta}} \hat{\Gamma}\ten{^{\hat{a}\hat{b}\hat{c}\hat{d}}}\hat{\theta})(\hat{\bar{\theta}} \hat{\Gamma}_{\hat{a}\hat{b}} [\hat{D}_{\hat{c}},\hat{D}_{\hat{d}}]\hat{\theta}) =
    \\
    = \, & - \frac{1}{768}(\hat{\bar{\theta}}{\Gamma}\ten{^{{a}{b}{c}{d}}}\hat{\theta})(\hat{\bar{\theta}}  {\Gamma}_{{a}{b}} 
    [\hat{D}_{{c}},\hat{D}_{{d}}]\hat{\theta}) -\frac{1}{192}(\hat{\bar{\theta}}{\Gamma}\ten{^{{\ul{10}}{b}{c}{d}}}\hat{\theta})(\hat{\bar{\theta}}  {\Gamma}_{[{\ul{10}}{b}} 
    [\hat{D}_{{c}},\hat{D}_{{d}]}]\hat{\theta})
    \\
    = \, & - \frac{1}{768} (\bar{\y} \Gamma^{mnpq} \y) (\bar{\y}  \Gamma_{mn}\cmttr_{pq} \y) -  \frac{1}{1152} (\bar{\y} \Gamma^{\ulv}\Gamma^{mnp}\y)\Big[\bar{\y}\big[3 \Gamma^{\ulv}\Gamma_m  \cmttr_{np} - \Gamma_{mn} \cmttr_{p} \big]\y\Big].
\end{split}
\end{equation*}
In moving to the final line we used many of the results derived previously, and we move vielbeins around in order to write everything with spacetime indices rather than tangent space. In the second term, we have
\begin{equation*}
\begin{split}
    \frac{1}{128} \, e^{-\frac{4\phi}{3}}(\hat{\bar{\theta}}\hat{\Gamma}_{10}& \hat{\Gamma}^{\hat{a}\hat{b}\hat{c}}\hat{\theta})(\hat{\bar{\theta}}  \bigl(\hat{\Gamma}_{10\hat{a}} 
    [\hat{D}_{\hat{b}},\hat{D}_{\hat{c}}]  +   \hat{\Gamma}_{\hat{a}\hat{b}} 
    [\hat{D}_{10},\hat{D}_{\hat{c}}] \bigr)\hat{\theta}) =   \frac{1}{384} (\bar{\y} \Gamma^{\ulv}\Gamma^{mnp}\y)\Big[\bar{\y}\big[3 \Gamma^{\ulv}\Gamma_m  \cmttr_{np} - \Gamma_{mn} \cmttr_{p} \big]\y\Big],
\end{split}
\end{equation*}
where the term with $\hat{a}\hat{b}\hat{c} \rightarrow 10 bc$ vanishes by symmetry of the first bilinear. Next, we have
\begin{equation*}
\begin{split}
    \frac{1}{128} \, e^{-\frac{4\phi}{3}} (\hat{\bar{\theta}}\hat{\Gamma}_{10}\hat{\Gamma}^{\hat{a}\hat{b}}\hat{\theta})(\hat{\bar{\theta}}\bigl( \hat{\Gamma}_{10}  
    [\hat{D}_{\hat{a}},\hat{D}_{\hat{b}}]    + 2\hat{\Gamma}_{\hat{a}} 
    [\hat{D}_{10},{\hat{D}}_{\hat{b}}] \bigr) \hat{\theta}) = \frac{1}{384} (\bar{\y}  \Gamma^{\ulv} \Gamma^{mn} \y)\Big[\bar{\y} \big[ 3 \Gamma^{\ulv}\cmttr_{mn} - 2 \Gamma_m \cmttr_n\big] {\y}\Big],
\end{split}
\end{equation*}
where symmetry considerations of the first bilinear causes the $\hat{a}\hat{b}\rightarrow 10b$ terms to vanish. Moving to the fourth term, we have
\begin{equation*}
\begin{split}
    - \frac{3}{16} \, e^{-\frac{4\phi}{3}} (\hat{\bar{\theta}}\hat{\Gamma}_{\hat{a}}\hat{D}_{10}\hat{\theta}) (\hat{\bar{\theta}}\hat{\Gamma}^{\hat{a}}\hat{D}_{10}\hat{\theta}) = - \frac{1}{48} (\bar{\y}\Gamma_m\Gamma^{\ulv} \Delta \y)(\bar{\y} \Gamma^m \Gamma^{\ulv} \Delta \y) -  \frac{1}{48}(\bar{\y}\Delta\y)^2.
\end{split}
\end{equation*}
The fifth term gives us
\begin{equation*}
\begin{split}
    & \frac{1}{16} \, e^{-\frac{4\phi}{3}} (\hat{\bar{\theta}} \hat{\Gamma}_{10} \hat{\check{{T}}}\ten{_{\hat{a}}} \hat{\theta}) (\hat{\bar{\theta}} \hat{\Gamma}^{\hat{a}} \hat{D}_{10} \hat{\theta}) = \\
    = \, & - \frac{1}{576} \Big[\bar{\y} \big[2\Gamma^{\ulv}e^{\phi} \ul{F}^{(4)}_{m}-\Gamma_m \ul{H}^{(3)} \big]\y\Big](\bar{\y} \Gamma^m \Gamma^{\ulv} \Delta \y) + \frac{1}{576} \Big[\bar{\y}\big[ e^{\phi} \ul{F}^{(4)} - 2\ul{H}^{(3)}\Gamma^{\ulv} \big]\y\Big] (\bar{\y} \Delta \y),
\end{split}
\end{equation*}
where we have once again been able to use the symmetries of the $\Gamma$-matrices to combine some terms together. Finally, we have
\begin{equation*}
\begin{split}
    \frac{1}{16} \, e^{-\frac{4\phi}{3}} (\hat{\bar{\theta}}\hat{\Gamma}_{10}\hat{\Gamma}^{\hat{a}\hat{b}}\hat{\theta})(\hat{\bar{\theta}} \hat{\cl{R}}\ten{_{\hat{a}\hat{b}}}  \hat{D}_{10}\hat{\theta}) =  \frac{1}{48}  (\bar{\y}  \Gamma^{\ulv} \Gamma^{mn} \y)(\bar{\y} \flux_{mn} \Gamma^{\ulv} \Delta{\y}).
\end{split}
\end{equation*}
We also must not forget the final term $(\bar{\y}\Delta\y)^2/24$ in the dilaton shift, which was already built out of 10-dimensional fields. If we combine everything together, we obtain the dilaton quartic order shift
\begin{equation}
\begin{split}
    \rho^{(4)} = & \, - \frac{1}{768} (\bar{\y} \Gamma^{mnpq} \y) (\bar{\y}  \Gamma_{mn}\cmttr_{pq} \y) +  \frac{1}{576} (\bar{\y} \Gamma^{\ulv}\Gamma^{mnp}\y)\Big[\bar{\y}\big[3 \Gamma^{\ulv}\Gamma_m  \cmttr_{np} - \Gamma_{mn} \cmttr_{p} \big]\y\Big]
    \\
    & + \frac{1}{384} (\bar{\y}  \Gamma^{\ulv} \Gamma^{mn} \y)\Big[\bar{\y} \big[ 3 \Gamma^{\ulv}\cmttr_{mn} - 2 \Gamma_m \cmttr_n\big] {\y}\Big]  +   \frac{1}{48}  (\bar{\y}  \Gamma^{\ulv} \Gamma^{mn} \y)(\bar{\y} \flux_{mn} \Gamma^{\ulv} \Delta{\y}) 
    \\
    & - \frac{1}{48} (\bar{\y}\Gamma_m\Gamma^{\ulv} \Delta \y)(\bar{\y} \Gamma^m \Gamma^{\ulv} \Delta \y) - \frac{1}{576} \Big[\bar{\y} \big[2\Gamma^{\ulv}e^{\phi} \ul{F}^{(4)}_{m}-\Gamma_m \ul{H}^{(3)} \big]\y\Big](\bar{\y} \Gamma^m \Gamma^{\ulv} \Delta \y) 
    \\
    & + \frac{1}{48}(\bar{\y}\Delta\y)^2  + \frac{1}{576} \Big[\bar{\y}\big[ e^{\phi} \ul{F}^{(4)} - 2\ul{H}^{(3)}\Gamma^{\ulv} \big]\y\Big] (\bar{\y} \Delta \y).
\end{split}
\end{equation}
This is the shift given in the main text for the dilaton. 

For the sake of completion, let us also note here that the expanded expression for the quartic terms in the expansion of the three-form superfield, obtained by plugging \eqref{crazyobjects} into \eqref{quarticm2form}, is
\begin{equation}
\begin{split}
    \hat{\alpha}_{\hat{m}\hat{n}\hat{p}} = & \, - \frac{3}{4} \bigl( \hat{\bar{\theta}} \hat{\Gamma}_{\hat{a}[\hat{m}} \hat{D}_{\hat{n}} \hat{\theta} \bigr) \bigl( \hat{\bar{\theta}} \hat{\Gamma}^{\hat{a}} \hat{D}_{\hat{p}])} \hat{\theta} \bigr) +  \frac{1}{8} \bigl( \hat{\bar{\theta}} \hat{\Gamma}_{[\hat{m}\hat{n}|} \hat{\check{{T}}}\ten{_{\hat{a}}}\hat{\theta} \bigr) \bigl( \hat{\bar{\theta}} \hat{\Gamma}^{\hat{a}} \hat{D}_{|\hat{p}]}\hat{\theta} \bigr)
    \\
    & - \frac{1}{384} \bigl( \hat{\bar{\theta}} \hat{\Gamma}_{\hat{m}\hat{n}\hat{p}} \hat{\Gamma}\ten{^{\hat{a}\hat{b}\hat{c}\hat{d}}} \hat{\theta} \bigr) \bigl( \hat{\bar{\theta}} \hat{\Gamma}_{\hat{a}\hat{b}} 
    [\hat{D}_{\hat{c}}, \hat{D}_{\hat{d}}]\hat{\theta} \bigr) + \frac{1}{64} \bigl( \hat{\bar{\theta}} \hat{\Gamma}_{[\hat{m}\hat{n}} \hat{\Gamma}^{\hat{a}\hat{b}\hat{c}} \hat{\theta} \bigr) \bigl( \hat{\bar{\theta}} \hat{\Gamma}_{\hat{p}]\hat{a}} 
    [\hat{D}_{\hat{b}}, \hat{D}_{\hat{c}}]\hat{\theta} \bigr)
    \\
    & + \frac{1}{64}\bigl(\hat{\bar{\theta}} \hat{\Gamma}_{[\hat{m}\hat{n}|} \hat{\Gamma}^{\hat{a}\hat{b}\hat{c}]} \hat{\theta} \bigr) \bigl( \hat{\bar{\theta}} \hat{\Gamma}_{\hat{a}\hat{b}} 
    [\hat{D}_{|\hat{p}]}, \hat{D}_{\hat{c}}] \hat{\theta} \bigr) + \frac{1}{8} \bigl( \hat{\bar{\theta}} \hat{\Gamma}_{[\hat{m}\hat{n}|}\hat{\Gamma}^{\hat{a}\hat{b}} \hat{\theta} \bigr) \bigl( \hat{\bar{\theta}} \hat{\cl{R}}\ten{_{\hat{a}\hat{b}}}  \hat{D}_{|\hat{p}]} \hat{\theta} \bigr)
    \\
    & + \frac{1}{64} \bigl( \hat{\bar{\theta}} \hat{\Gamma}_{[\hat{m}\hat{n}} \hat{\Gamma}^{\hat{a}\hat{b}} \hat{\theta} \bigr) \bigl( \hat{\bar{\theta}} \hat{\Gamma}_{\hat{p}]} [\hat{D}_{\hat{a}}, \hat{D}_{\hat{b}}] \hat{\theta} \bigr) + \frac{1}{32} \bigl( \hat{\bar{\theta}} \hat{\Gamma}_{[\hat{m}\hat{n}|} \hat{\Gamma}^{\hat{a}\hat{b}} \hat{\theta} \bigr) \bigl( \hat{\bar{\theta}} \hat{\Gamma}_{\hat{a}} [\hat{D}_{|\hat{p}]}, {\hat{D}}_{\hat{b}}] \hat{\theta} \bigr).
\end{split}
\end{equation}

\section{Further comments on T-duality}\label{Tdualconventions}
Here we discuss the T-duality calculation for the general Ramond-Ramond superfield expansions at second order in fermions. Notice that all three of the quadratic shifts so far calculated have been of the form
\begin{equation*}
    \bs{C}^{(n)}_{{m_1}\dots{m_n}}  = {C}^{(n)}_{{m_1}\dots{m_n}} -  \begin{dcases}
    \frac{i}{2} \, e^{-{\phi}} \, \bar{\theta}^{\A} a_n \biggl[ n {\ds{\Gamma}}_{[m_1\dots m_{n-1}} {\ds{D}}^{\A}_{m_n]}  - \frac{1}{2}{\ds{\Gamma}}_{{m_1}\dots{m_n}}\sigma^1 {\ds{\Delta}}^{\A} \biggr] \theta^{\A}, \quad & n = 2p-1, \\
    \frac{i}{2} \, e^{-{\dual{\phi}}} \, \bar{\theta}^{\B} b_n \biggl[ n \, \dual{\ds{\Gamma}}_{[m_1\dots m_{n-1}} {\ds{D}}^{\B}_{m_n]}  - \frac{1}{2}\dual{\ds{\Gamma}}_{{m_1}\dots{m_n}}\sigma^1 {\ds{\Delta}}^{\B} \biggr] \theta^{\B}, \quad & n = 2p,
    \end{dcases}
\end{equation*}
where the $a_n$ and $b_n$ are some Pauli matrix combinations that need to be determined. We will show that this is the form for all the quadratic RR shifts, and determine $a_n$ and $b_n$ for all $n$, starting from the known results for $n=1,3$. The key equations to T-dualize these superfields into each other is the Ramond-Ramond superfield T-duality rule (\ref{RRTdual}). In particular, defining the quadratic Ramond-Ramond superfield expansions as $\smash{\bs{C}^{(n)} = C^{(n)} + \chi^{(n)}}$, one can write
\begin{equation} \label{shiftTdual}
    \dual{\chi}^{(n+1)}_{\td \dot{m}_2 \dots \dot{m}_{n+1}} = \chi^{(n)}_{\dot{m}_2 \dots \dot{m}_{n+1}} - n \, g_{\td\td}^{-1} g_{\td [ \dot{m}_2} \chi^{(n)}_{|\td|\dot{m}_3 \dots \dot{m}_{n+1}]}.
\end{equation}

Let us first concentrate on the terms outside of the square brackets. For now we will neglect to write what appears inside the square brackets after applying \eqref{shiftTdual}, instead we shall just label it $[\text{IIA}]$ or $[\text{IIB}]$ to keep track of whether it has yet been T-dualized. Under T-dualization, moving from type IIA to type IIB we can write,
\begin{equation*}
\begin{split}
    - \dfrac{i}{2} \, e^{-\phi} \, \bar{\theta}^\A a_{2p-1} [\text{IIA}] \theta^\A & = - \dfrac{i}{2} \, e^{-\dual{\phi}} \sqrt{\dual{g}_{\td\td}} \, (\bar{\theta}^\B \sigma^1 \Upsilon \sigma^1) a_{2p-1} \Upsilon^{-1}\Upsilon [\text{IIA}] \Upsilon^{-1} \theta^\B \\
    & = - \dfrac{i}{2} \, e^{-\dual{\phi}} \, \bar{\theta}^\B b_{2p} \ds{\dual{\Gamma}}_{\td}\Upsilon [\text{IIA}] \Upsilon^{-1} \theta^\B.
\end{split}
\end{equation*}
We will see shortly that we will require $\dual{\ds{\Gamma}}_{\td}$ when T-dualizing the terms inside the square brackets, so write it separately in line two and treat that part in a moment. Moving from type IIB to type IIA, instead, we can write
\begin{equation*}
\begin{split}
    \frac{i}{2} \, e^{-{\dual{\phi}}} \, \bar{\theta}^{\B} b_{2p} [\text{IIB}] \theta^{\B} & = - \dfrac{i}{2} \, e^{-{\phi}} \sqrt{{g}_{\td\td}} \, (\bar{\theta}^\A \sigma^1 \Upsilon^{-1} \sigma^1) b_{2p} \Upsilon\Upsilon^{-1} [\text{IIB}] \Upsilon \theta^\A \\
    & = - \dfrac{i}{2} \, e^{-{\phi}} \, \bar{\theta}^\A a_{2p+1} \ds{\Gamma}_{\td}\Upsilon^{-1} [\text{IIB}] \Upsilon \theta^\A.
\end{split}
\end{equation*}
We know from the expansions in (\ref{second-order_typeIIA_RRsuperfields}) that $a_1 = \sigma^3$ and $a_3 = 1_2$. We also know from the definitions of the T-duality operators in section \ref{sec:Tdual} that $\smash{\sqrt{\dual{g}_{\td\td}} \, \sigma^1 \Upsilon \sigma^1 \sigma^3 \Upsilon^{-1} = \Gamma^\clty \dual{\ds{\Gamma}}_{\td}}$, which allows to conclude that $\bar{\theta}^{\B} b_2 \dual{\ds{\Gamma}}_{\td} = \bar{\theta}^{\B} \Gamma^\clty \dual{\ds{\Gamma}}_{\td}$, meaning that $b_2 = -1_2$. Similarly, from the series of conditions $\smash{\bar{\theta}^\A a_{3} \ds{\Gamma}_{\td} = \sqrt{{g}_{\td\td}} \, \bar{\theta}^\A \sigma^1 \Upsilon^{-1} \sigma^1 b_{2} \Upsilon = - \sqrt{{g}_{\td\td}} \, \bar{\theta}^\A \sigma^1 \Upsilon^{-1} \sigma^1 \Upsilon = \bar{\theta}^\A \ds{\Gamma}_{\td}}$, we recover $a_3 = 1_2$. Going on, from $\smash{\bar{\theta}^\B b_4 \dual{\ds{\Gamma}}_{\td} = \sqrt{\dual{g}_{\td\td}} \, \bar{\theta}^\B \sigma^1 \Upsilon \sigma^1 a_{3} \Upsilon^{-1} = -\sigma^3 \bar{\theta}^\B \ds{\Gamma}_{\td}}$, we find $b_4 = -\sigma^3$. Finally, from the chain of relationships $\smash{\bar{\theta}^\A a_{5} \ds{\Gamma}_{\td}  = \sqrt{{g}_{\td\td}} \, \bar{\theta}^\A \sigma^1 \Upsilon^{-1} \sigma^1 b_{4} \Upsilon = - \sqrt{{g}_{\td\td}} \, \bar{\theta}^\A \sigma^1 \Upsilon^{-1} \sigma^1 \sigma^3 \Upsilon = \sigma^3 \bar{\theta}^\A \ds{\Gamma}_9}$, we obtain $a_5 = \sigma^3$. In conclusion we have
\begin{equation}\label{anbn}
    a_1 = \sigma^3, \qquad b_2 = -1_2, \qquad a_3 =     1_2, \qquad b_4 = - \sigma^3, \qquad a_5 = \sigma^3.
\end{equation}
The pattern continues, multiplying by $-1_2$ when moving from IIB to IIA and by $-\sigma^3$ when moving from IIA to IIB. 

Now let us concentrate on the expressions $[\text{IIA}]$ and $[\text{IIB}]$ inside the square brackets. Here we will look at moving from type IIA to type IIB, however moving from type IIB to type IIA employs an essentially identical structure. More specifically, when we use \eqref{shiftTdual} to determine the shift on $\bs{C}^{(n+1)}_{{m_1}\dots{m_{n+1}}}$ from the shift on $\bs{C}^{(n)}_{{m_1}\dots{m_n}}$ to move from IIA to IIB, we have to consider $\ds{\dual{\Gamma}}_{\td}\Upsilon [\text{IIB}] \Upsilon^{-1}$, and vice versa. The explicit terms are
\begin{equation*}
\begin{split}
    [\text{IIA}]^{\left[\tilde{\chi}^{(n+1)}\right]}_{\dot{m}_2\dots \dot{m}_{n+1}} = \dual{\ds{\Gamma}}_9 \Upsilon  \biggl[ n \, {\ds{\Gamma}}_{[\dot{m}_2\dots \dot{m}_{n}} {\ds{D}}^{\A}_{\dot{m}_{n+1}]} - n(n-1) g_{\td\td}^{-1} g_{\td [ \dot{m}_2} {\ds{\Gamma}}_{|\td| \dot{m}_3 \dots \dot{m}_{n}} {\ds{D}}^{\A}_{\dot{m}_{n+1}]} \\
    - (-1)^{n-1} n g_{\td\td}^{-1} g_{\td [ \dot{m}_2} {\ds{\Gamma}}_{ \dot{m}_3 \dots \dot{m}_{n+1}]} {\ds{D}}^{\A}_{{\td}} \\
    - \frac{1}{2}\Bigl({\ds{\Gamma}}_{{\dot{m}_2}\dots{\dot{m}_{n+1}}} - n g_{\td\td}^{-1} g_{\td [ \dot{m}_2}{\ds{\Gamma}}_{|\td| \dot{m}_3 \dots{m_{n+1}}]}\Bigr)\sigma^1 {\ds{\Delta}^{\A}} \biggr] \Upsilon^{-1} & .
\end{split}
\end{equation*}
After a little work and using $ (-1)^{n-1} n g_{\td\td}^{-1} g_{\td [ \dot{m}_2} {\ds{\Gamma}}_{ \dot{m}_3 \dots \dot{m}_{n+1}]} {\ds{D}}_{{\td}} = n g_{\td\td}^{-1} g_{\td [\dot{m}_{n+1}} {\ds{\Gamma}}_{ \dot{m}_2 \dots \dot{m}_{n}]} {\ds{D}}_{{\td}} $, this can be written as
\begin{equation*}
\begin{split}
    [\text{IIA}]^{\left[\tilde{\chi}^{(n+1)}\right]}_{\dot{m}_2\dots \dot{m}_{n+1}} = \dual{\ds{\Gamma}}_9 \Upsilon \biggl[ n \, \Bigl(\ds{\Gamma}_{[\dot{m}_2} - g_{\td\td}^{-1} g_{\td [ \dot{m}_2|} \ds{\Gamma}_{\td} \Bigr) \dots \Bigl(\ds{\Gamma}_{\dot{m}_{n}} - g_{\td\td}^{-1} g_{\td | \dot{m}_{n}|} \ds{\Gamma}_{\td} \Bigr)\Bigl(\ds{D}^{\A}_{\dot{m}_{n+1}]} - g_{\td\td}^{-1} g_{\td | \dot{m}_{n+1}]} \ds{D}^{\A}_{\td} \Bigr) \\
    - \frac{1}{2} \Bigl(\ds{\Gamma}_{[\dot{m}_2} - g_{\td\td}^{-1} g_{\td [ \dot{m}_2|} \ds{\Gamma}_{\td} \Bigr) \dots \Bigl(\ds{\Gamma}_{\dot{m}_{n+1}]} - g_{\td\td}^{-1} g_{\td | \dot{m}_{n+1}]} \ds{\Gamma}_{\td} \Bigr)\sigma^1 {\ds{\Delta}^{\A}} \biggr] \Upsilon^{-1} & .
\end{split}
\end{equation*}
At this point, we can use \eqref{usefulob1} and \eqref{usefulob2} to T-dualize almost everything immediately, obtaining
\begin{equation*}
    \begin{split}
        [\text{IIA}]^{\left[\tilde{\chi}^{(n+1)}\right]}_{\dot{m}_2\dots \dot{m}_{n+1}} & = \begin{aligned}[t] \dual{\ds{\Gamma}}_9 \biggl[ \, & n \, \Bigl(\dual{\ds{\Gamma}}_{[\dot{m}_2} - \dual{g}_{\td\td}^{-1} \dual{g}_{\td [ \dot{m}_2|} \dual{\ds{\Gamma}}_{\td} \Bigr) \dots \Bigl(\dual{\ds{\Gamma}}_{\dot{m}_{n}} - \dual{g}_{\td\td}^{-1} \dual{g}_{\td | \dot{m}_{n}|} \dual{\ds{\Gamma}}_{\td} \Bigr)\Bigl({\ds{D}}^{\B}_{\dot{m}_{n+1}]} - \dual{g}_{\td\td}^{-1} \dual{g}_{\td | \dot{m}_{n+1}]} {\ds{D}}^{\B}_{\td} \Bigr) \\
        & - \frac{1}{2} \Bigl(\dual{\ds{\Gamma}}_{[\dot{m}_2} - \dual{g}_{\td\td}^{-1} \dual{g}_{\td [ \dot{m}_2|} \dual{\ds{\Gamma}}_{\td} \Bigr) \dots \Bigl(\dual{\ds{\Gamma}}_{\dot{m}_{n+1}]} - \dual{g}_{\td\td}^{-1} \dual{g}_{\td | \dot{m}_{n+1}]} \dual{\ds{\Gamma}}_{\td} \Bigr)\Bigl(\sigma^1 {\ds{\Delta}}^{\B} - 2 \dual{g}_{\td\td}^{-1}\dual{\ds{\Gamma}}_{\td} {\ds{D}}^{\B}_{\td} \Bigr) \biggr] \end{aligned} \\
        & = \begin{aligned}[t] \biggl[ \, & n \, \dual{\ds{\Gamma}}_{9}\dual{\ds{\Gamma}}_{[\dot{m}_2 \dots \dot{m}_n}\Bigl({\ds{D}}^{\B}_{\dot{m}_{n+1}]} - \dual{g}_{\td\td}^{-1} \dual{g}_{\td | \dot{m}_{n+1}]} {\ds{D}}^{\B}_{\td} \Bigr) - n(n-1) \dual{g}_{\td [ \dot{m}_2}\dual{\ds{\Gamma}}_{\dot{m}_3 \dots \dot{m}_n}{\ds{D}}^{\B}_{\dot{m}_{n+1}]} \\
        & - \frac{1}{2} \dual{\ds{\Gamma}}_{9}\dual{\ds{\Gamma}}_{\dot{m}_2 \dots \dot{m}_{n+1}} \Bigl(\sigma^1 {\ds{\Delta}}^{\B} - 2 \dual{g}_{\td\td}^{-1} \dual{\ds{\Gamma}}_{\td} {\ds{D}}_{\td}^{\B} \Bigr) \\
        & + \frac{n}{2} \dual{g}_{\td [ \dot{m}_2} \dual{\ds{\Gamma}}_{\dot{m}_3 \dots \dot{m}_{n+1}]} \Bigl(\sigma^1 {\ds{\Delta}}^{\B} - 2 \dual{g}_{\td\td}^{-1}\dual{\ds{\Gamma}}_{\td} {\ds{D}}^{\B}_{\td} \Bigr) \biggr] \end{aligned} \\
        & = \begin{aligned}[t] \biggl[ \, & n \, \dual{\ds{\Gamma}}_{9 [\dot{m}_2 \dots \dot{m}_n}{\ds{D}}^{\B}_{\dot{m}_{n+1}]} -  n \, \dual{\ds{\Gamma}}_{9 [\dot{m}_2 \dots \dot{m}_n|} \dual{g}_{\td\td}^{-1} \dual{g}_{\td | \dot{m}_{n+1}]} {\ds{D}}^{\B}_{\td} \\
        & + (-1)^{n} \dual{\ds{\Gamma}}_{\dot{m}_2 \dots \dot{m}_{n+1}9} \dual{g}_{\td\td}^{-1}\dual{\ds{\Gamma}}_{\td} {\ds{D}}^{\B}_{\td} -\frac{1}{2} \dual{\ds{\Gamma}}_{9\dot{m}_2 \dots \dot{m}_{n+1}} \sigma^1 {\ds{\Delta}}^{\B} \biggr], \end{aligned}
    \end{split}
\end{equation*}
where in the final step we combined some $\Gamma$-matrices, distributed the final term and eventually rearranged some indices. A further use of useful $\Gamma$-matrix identities and a little further massaging results in some more cancellations, to give
\begin{equation*}
    \begin{split}
        [\text{IIA}]^{\left[\tilde{\chi}^{(n+1)}\right]}_{\dot{m}_2\dots \dot{m}_{n+1}} & = \begin{aligned}[t] \biggl[ \, & n \, \dual{\ds{\Gamma}}_{9 [\dot{m}_2 \dots \dot{m}_n}{\ds{D}}^{\B}_{\dot{m}_{n+1}]} -  n \, \dual{\ds{\Gamma}}_{9 [\dot{m}_2 \dots \dot{m}_n|} \dual{g}_{\td\td}^{-1} \dual{g}_{\td | \dot{m}_{n+1}]} {\ds{D}}^{\B}_{\td} \\
        & + (-1)^{n} \bigl(\dual{\ds{\Gamma}}_{\dot{m}_2 \dots \dot{m}_{n+1}}\dual{\ds{\Gamma}}_{9} - n \, \dual{\ds{\Gamma}}_{[\dot{m}_2 \dots \dot{m}_{n}}\dual{g}_{|\td|\dot{m}_{n+1}]}\bigr) \dual{g}_{\td\td}^{-1}\dual{\ds{\Gamma}}_{\td} {\ds{D}}^{\B}_{\td} -\frac{1}{2} \dual{\ds{\Gamma}}_{9\dot{m}_2 \dots \dot{m}_{n+1}} \sigma^1 {\ds{\Delta}}^{\B}  \biggr] \end{aligned} \\
        & = \begin{aligned}[t] \biggl[ \, & n \, \dual{\ds{\Gamma}}_{9 [\dot{m}_2 \dots \dot{m}_n}{\ds{D}}^{\B}_{\dot{m}_{n+1}]} + (-1)^{n} \dual{\ds{\Gamma}}_{\dot{m}_2 \dots \dot{m}_{n+1}}\dual{\ds{\Gamma}}_{9}\dual{g}_{\td\td}^{-1}\dual{\ds{\Gamma}}_{\td} {\ds{D}}^{\B}_{\td}-\frac{1}{2} \dual{\ds{\Gamma}}_{9\dot{m}_2 \dots \dot{m}_{n+1}} \sigma^1 {\ds{\Delta}}^{\B} \\
        & -  n \, \dual{\ds{\Gamma}}_{9 [\dot{m}_2 \dots \dot{m}_n|} \dual{g}_{\td\td}^{-1} \dual{g}_{\td | \dot{m}_{n+1}]} {\ds{D}}^{\B}_{\td} - (-1)^{n} n \, \dual{\ds{\Gamma}}_{[\dot{m}_2 \dots \dot{m}_{n}}\dual{g}_{|\td|\dot{m}_{n+1}]} \dual{g}_{\td\td}^{-1}\dual{\ds{\Gamma}}_{\td} {\ds{D}}^{\B}_{\td} \biggr] \end{aligned} \\
        & = \begin{aligned}[t]
        \biggl[ \, & n \, \dual{\ds{\Gamma}}_{9 [\dot{m}_2 \dots \dot{m}_n}{\ds{D}}^{\B}_{\dot{m}_{n+1}]} + (-1)^{n} \dual{\ds{\Gamma}}_{\dot{m}_2 \dots \dot{m}_{n+1}} {\ds{D}}^{\B}_{\td}-\frac{1}{2} \dual{\ds{\Gamma}}_{9\dot{m}_2 \dots \dot{m}_{n+1}} \sigma^1 {\ds{\Delta}}^{\B} \\
        & - n \, \dual{\ds{\Gamma}}_{9 [\dot{m}_2 \dots \dot{m}_n|} \dual{g}_{\td\td}^{-1} \dual{g}_{\td | \dot{m}_{n+1}]} {\ds{D}}^{\B}_{\td} - (-1)^{2n-1} n \, \dual{\ds{\Gamma}}_{9[\dot{m}_2 \dots \dot{m}_{n}}\dual{g}_{|\td|\dot{m}_{n+1}]} \dual{g}_{\td\td}^{-1} {\ds{D}}^{\B}_{\td} \biggr] \end{aligned} \\
        & = \biggl[ n \, \dual{\ds{\Gamma}}_{9 [\dot{m}_2 \dots \dot{m}_n}{\ds{D}}^{\B}_{\dot{m}_{n+1}]} + (-1)^{n} \dual{\ds{\Gamma}}_{\dot{m}_2 \dots \dot{m}_{n+1}} {\ds{D}}^{\B}_{\td}-\frac{1}{2} \dual{\ds{\Gamma}}_{9\dot{m}_2 \dots \dot{m}_{n+1}} \sigma^1 {\ds{\Delta}}^{\B} \biggr] \\
        & = \biggl[ (n+1) \dual{\ds{\Gamma}}_{[9 \dot{m}_2 \dots \dot{m}_n}{\ds{D}}^{\B}_{\dot{m}_{n+1}]} -\frac{1}{2} \dual{\ds{\Gamma}}_{9\dot{m}_2 \dots \dot{m}_{n+1}} \sigma^1 {\ds{\Delta}}^{\B} \biggr],
    \end{split}
\end{equation*}
which is exactly the desired result. Note that while powers of $(-1)$ depending on $n$ appeared, nowhere did we rely on $n$ being odd for the specific case of moving from type IIA to type IIB, and indeed the derivation moving the other way has precisely the same structure. Thanks to this procedure, one can verify the general second-order Ramond-Ramond shifts in \eqref{second-order_typeIIA_RRsuperfields}.

\bibliographystyle{JHEP}

\bibliography{Refs.bib}

\end{document}